\documentclass[useAMS,usenatbib]{mn2e}

\usepackage{graphicx}
\usepackage{lscape}
\usepackage{longtable}
\usepackage{deluxetable}
\usepackage{ulem}

\newcommand{\Msun}{$M_{\odot}$ }
\newcommand{\M}{$M^{\rm AGB}_{\rm ini}$ }

\newcommand{\g}{$\gamma$}
\newcommand{\Can}{$^{13}$C($\alpha$, n)$^{16}$O }
\newcommand{\Nean}{$^{22}$Ne($\alpha$, n)$^{25}$Mg }
\newcommand{\Neag}{$^{22}$Ne($\alpha$, $\gamma$)$^{26}$Mg }

\title[Main $s$-process Branchings and Neutron Source Uncertainties]
{The Branchings of the Main $s$-process: Their Sensitivity
to $\alpha$-induced Reactions on $^{13}$C and $^{22}$Ne and to the Uncertainties
of the Nuclear Network}
\author[S. Bisterzo, et al.]
{S. Bisterzo$^{1,2}$\thanks{E-mail: bisterzo@ph.unito.it; sarabisterzo@gmail.com}, 
R. Gallino$^{1}$,
F. K{\"a}ppeler$^{3}$,
M. Wiescher$^{4}$,
G. Imbriani$^{5}$,  
\newauthor O. Straniero$^{6}$, 
S. Cristallo$^{6,7}$,
J. G{\"o}rres$^{4}$, 
R. J. deBoer$^{4}$\\
$^{1}$Dipartimento di Fisica, Universit\`{a} 
   di Torino, Italy\\
$^{2}$INAF - Astrophysical Observatory Turin, Turin, Italy\\
$^{3}$Karlsruhe Institute of Technology, Campus Nord, 
Institut f{\"u}r Kernphysik, Karlsruhe, Germany\\
$^{4}$Joint Institute for Nuclear Astrophysics (JINA), Department 
of Physics, University of Notre Dame, IN, USA\\
$^{5}$ Dipartimento di Scienze Fisiche, Universit\`{a} di Napoli 
Federico II, Italy\\
$^{6}$INAF - Osservatorio Astronomico di Collurania, 64100 Teramo, Italy\\
$^{7}$INFN Sezione Napoli, Napoli, Italy}
\begin{document}

\date{Accepted 1988 December 15. Received 1988 December 14; in original form 1988 October 11}

\pagerange{\pageref{firstpage}--\pageref{lastpage}} \pubyear{2002}

\maketitle

\label{firstpage}

\begin{abstract}

This paper provides a detailed analysis of the main component of the $slow$ neutron capture
process (the $s$-process), which accounts for the solar abundances of half of the nuclei with 
90 $\la$ $A$ $\la$ 208. 
We examine the impact of the uncertainties of the two neutron sources operating
in low-mass asymptotic giant branch (AGB) stars: the $^{13}$C($\alpha$, n)$^{16}$O reaction, 
which releases neutrons radiatively during interpulse periods ($kT$ $\sim$ 8 keV), and the 
$^{22}$Ne($\alpha$, n)$^{25}$Mg reaction, partially activated during the convective thermal 
pulses (TPs).
We focus our attention on the branching points that mainly influence the abundance
of $s$-only isotopes.
\\
In our AGB models, the $^{13}$C is fully consumed radiatively during interpulse. In this
case, we find that the present uncertainty associated to the $^{13}$C($\alpha$,
n)$^{16}$O reaction has marginal effects on $s$-only nuclei.
On the other hand, a reduction of this rate may increase the amount of residual (or unburned) 
$^{13}$C at the end of the interpulse: in this condition,  
the residual $^{13}$C is burned at higher 
temperature in the convective zone powered by the following TP. 
\\
The neutron burst produced by the $^{22}$Ne($\alpha$,
n)$^{25}$Mg reaction has major effects on the branches along the $s$ path.
The contributions of $s$-only isotopes with 90 $\la$ $A$ $\leq$ 204 are reproduced within 
solar and nuclear uncertainties, even if the $^{22}$Ne($\alpha$, n)$^{25}$Mg rate is varied 
by a factor of two. 
Improved $\beta$-decay and neutron capture rates of a few key radioactive nuclides
would help to attain a comprehensive understanding of the solar main component.

\end{abstract}



\section{Introduction}\label{intro}

The classical analysis of the $s$ process (the $slow$ neutron capture process)
has provided a first phenomenological approach to interpret the solar $s$ distribution
by means of analytical tools \citep{clayton74,kaeppeler89}, but  
it was soon clear that the solar $s$ process requires a multiplicity 
of neutron exposures, intuitively associated to different astrophysical sites.
\\
The existence of the $main$ component of the $s$-process was advanced by the classical analysis 
to reproduce the solar abundances of $s$ isotopes between 90 $<$ $A$ $\leq$ 204.
Low-mass Asymptotic Giant Branch (AGB) stars ($M$ $\la$ 3 $M_\odot$) were 
recognised to be a most promising site for the main component \citep{ulrich73,iben83}.
The contribution of two additional $s$-process components was required to reproduce the 
solar $s$-process distribution, the so-called $weak$ and $strong$ components.
The weak component is partly responsible for the nucleosynthesis of $s$ nuclei
with $A$ $\la$ 90 during the hydrostatic evolutionary phases of massive stars 
\citep{arnett85,limongi00,rauscher02}.
The strong component was postulated by \citet{clayton67} to explain 
about half of solar $^{208}$Pb, despite the astrophysical site was not identified. 
Successively, the strong component found a natural explanation in AGB stars 
with low metallicity ([Fe/H] $\la$ $-$1) and low initial mass \citep{gallino98,travaglio01}.

This paper is focused on the study of the main component.
\\
Compared to the classical analysis, the development of the first AGB stellar models has provided 
a more adequate description of the dynamic environment in which the main component takes place.
The $s$-process nucleosynthesis in AGB stars occurs during the late stages of the stellar
evolution, when the star has a degenerate C-O core, a thin radiative layer (He-intershell) and 
an expanded convective envelope.
During the AGB phase, the star experiences a series of He-shell flashes
called Thermal Pulses (TPs) triggered by the sudden activation of the 3$\alpha$ process at the
base of the He-intershell.
In such a region,  
free neutrons are released by two 
key reactions, \Can and $^{22}$Ne($\alpha$, n)$^{25}$Mg.
\\
The \Can reaction is the main neutron source in low mass AGB stars.
The $^{13}$C forms in a thin zone (the $^{13}$C pocket) via proton captures on the abundant 
$^{12}$C ($^{12}$C(p, \g)$^{13}$N($\beta^+$$\nu$)$^{13}$C).
The most favorable conditions 
for the formation of the $^{13}$C pocket occur during the period immediately following a 
Third Dredge-Up (TDU) episode.
During a TDU, the H-burning shell is switched off and, thus, the convective envelope can penetrate 
inward and carry to the surface the heavy elements previously synthesised in the He-intershell.
The mechanism triggering the formation of the $^{13}$C pocket is far from being understood:
extant models assume that a partial amount of protons
may be diffused from the convective envelope into the He- and C-rich radiative He-intershell
(see discussion by \citealt{straniero06}), otherwise $^{13}$C would be further converted 
to $^{14}$N via proton captures, mainly acting as a neutron poison of the $s$ process via the $^{14}$N(n, 
p)$^{14}$C reaction. Different mechanisms proposed to explain the formation of the $^{13}$C pocket 
are objects of study\footnote{See, e.g., the opacity-induced overshooting at the base of the
convective envelope obtained by \citet{cristallo09} by introducing in the model an exponentially
decaying profile of the convective velocity. Other models investigate the effects of
diffusive overshooting, rotation, magnetic fields or gravity waves 
(\citealt{herwig97,herwig03,langer99,denissenkov03,siess04,piersanti13,busso12}, 
see also \citealt{maiorca12} for results in young open clusters).}.
When the temperature becomes larger than 0.8 $\times$ 10$^8$ K ($kT$ $\sim$ 8 keV, which corresponds to 
$T_8$ $\sim$ 0.9), \Can burns radiatively for 
an extended time scale (some 
10$^4$ yr), releasing a neutron density $N_n$ $\sim$ 10$^7$ cm$^{-3}$ with a large neutron 
exposure\footnote{The neutron exposure $\tau$ is the
time-integrated neutron flux, $\tau$ = $\int$ $N_n$ $v_{th}$ d$t$, where $N_n$ is the 
neutron density and $v_{th}$ the thermal velocity. } \citep{straniero95}. 
\\
The second neutron source, the \Nean reaction, is partially activated at the bottom of 
the convective shells generated by the TPs.
Starting from the large amount of $^{14}$N left in the H-burning shell ashes, $^{22}$Ne is produced 
via the $^{14}$N($\alpha$, $\gamma$)$^{18}$F($\beta$$^{+}$$\nu$)$^{18}$O($\alpha$, 
$\gamma$)$^{22}$Ne nuclear chain.
At a temperature of $T_8$ = 2.5 -- 3 ($kT$ $\sim$ 23 keV corresponds to $T_8$ $\sim$ 2.7),
the \Nean reaction starts releasing neutrons, giving rise to a small neutron exposure with 
a high peak neutron density ($N_{\rm n}$(peak) $\sim$ 10$^{10}$ cm$^{-3}$). 
Although this second burst accounts only for a few percent of the total neutron exposure, it 
regulates the final abundances of the $s$-only isotopes nearby to important branch points
of the $s$ process.

\citet{arlandini99} provided a first interpretation of the solar main component 
by adopting an average between two AGB stellar models with initial masses of 1.5 and 3 $M_\odot$, 
half-solar metallicity, and a specific $^{13}$C-pocket choice (called $case$ $ST$). 
The case ST was 
calibrated by \citet{gallino98} in order to reproduce 
the solar abundances of $s$-only isotopes between 96 $\leq$ $A$ $\leq$ 204 
with half-solar metallicity models. The two 1.5 and 3 $M_\odot$ models were chosen as they represent the
stellar mass range that better reproduces the observations of peculiar $s$-rich disk stars \citep{busso95}.
\\
The complex dependence of the $s$ process on the initial chemical composition of the star was
investigated by \citet{gallino98}. 
They found that the strong component derives from AGB stars of low metallicity. Indeed, 
for any $^{13}$C-pocket strength the number of free neutrons per iron seed increases with
the $^{13}$C/$^{56}$Fe ratio, and the neutron fluence progressively overcomes the first two 
$s$ peaks at neutron magic numbers $N$ = 50 and 82, 
directly feeding $^{208}$Pb (explaining the previously introduced strong component).
\\
The heterogeneity of the $s$ process is also evidenced by spectroscopic observations in different
stellar populations: the discovery of the first three lead-rich low-metallicity stars confirms that 
$^{208}$Pb may be strongly produced in peculiar objects of the Galactic halo \citep{vaneck01}. 
Moreover, the $s$ elements observed in peculiar $s$-rich stars (e.g., MS, S, C, Ba, and Post-AGB stars, 
planetary nebulae, and lead-rich stars, later called CEMP-s stars) show a scatter at a given metallicity. 
This scatter has been 
recognised since the first studies by \citet{smith90}, \citet{busso01}, and \citet{abia02}, 
(for recent analysis see the review by 
\citealt{sneden08,kaeppeler11,karakaslattanzio14}, and references therein). 
\\
At present, stellar models are not able to reproduce the observed scatter without employing a free 
parametrisation in modelling the formation of the $^{13}$C-pocket \citep{herwig97,herwig03,karakas07,cristallo09,piersanti13}.
Most uncertainties of stellar models are indeed related to the treatment of convective/radiative 
interfaces  
\citep{iben83,frost96}, which influence the extension and the $^{13}$C profile of pocket,
as well as the occurrence and deepness of the TDU.
\\
Another key uncertainty of AGB stellar models results from the unknown efficiency of the mass loss rate,
which regulates the number of thermal pulses and the AGB lifetime. This is particularly challenging in 
intermediate-mass AGB \citep{ventura05,ventura10}.
Despite the solar $s$ distribution between 90 $<$ $A$ $\leq$ 204 receives a dominant
contribution by low-mass AGB models of disc metallicity, intermediate-mass or
low-metallicity AGB models are crucial to study the chemical evolution of dwarf galaxies 
and globular clusters showing a clear $s$ process signature (\citealt{tolstoy09,straniero14}, 
and references therein), or the spectroscopic observation of peculiar Galactic stars
(e.g., Rb-rich stars, \citealt{vanraai12,karakas12}; low-metallicity $s$-enhanced stars as 
e.g., CH, CEMP-s, post-AGB stars, \citealt{bisterzo11,lugaro12,DeSmedt14}).
\\
Besides the aforementioned uncertainties, nuclear and solar abundance uncertainties 
affect the $s$ process. Their impact may be substantially reduced by using the unbranched 
$s$-only $^{150}$Sm as a reference isotope of the $s$ distribution (because it has well known solar abundance 
and accurately determined neutron capture cross sections of nearby nuclei; see \citealt{arlandini99}).

 \vspace{2mm}

The $s$-process abundances observed in the Sun are the result of the complex Galactic chemical evolution,
which accounts of the contribution of different stellar generations, with various masses,
metallicities and $s$ process strengths. A Galactic chemical evolution (GCE) model is therefore
needed to interpret the dynamics of the $s$ process over the Galactic history up to the present
epoch (see, e.g., \citealt{travaglio04,romano10,nomoto13}, and references therein).
\\
\citet{travaglio04} have shown that in the GCE context low mass
 AGB stars in the range of 1.5 to 3 $M_\odot$ provide the dominant
 contribution to the solar system main and strong components.
\\
The impact of intermediate AGB stars (IMS; \M $\sim$ 4 to 8 $M_\odot$) on solar abundances
is marginal, with the exception of a few neutron-rich isotopes 
($^{86}$Kr and $^{87}$Rb due to the branches at $^{85}$Kr and $^{86}$Rb; $^{96}$Zr 
affected by the branch at $^{95}$Zr; see also \citealt{bisterzo14ApJ}).   
Indeed, the He-shell of IMS stars reaches  
higher temperatures than low mass AGB stars, and the \Nean reaction is efficiently activated, 
producing higher peak neutron densities ($T_8$ $\sim$ 3.6--3.7; $N_n$ $\sim$ 10$^{11-13}$ cm$^{-3}$).
Under these conditions, the \Nean reaction becomes the major neutron source 
\citep{truran77,karakas06,longland12,vanraai12,karakas12,dOrazi13,straniero14,doherty14}.
Otherwise, in the more massive AGB models the formation of the $^{13}$C pocket and the 
occurrence of efficient TDUs may be inhibited by hot bottom burning and hot third 
dredge-up events \citep{karakas03,herwig04,goriely04}. 
This significantly reduces the overall IMS contribution to the solar $s$ distribution. 
\\
GCE $s$ predictions yield a plausible agreement with the solar abundances of the $s$-only
isotopes between $^{134,136}$Ba and $^{208}$Pb. The solar distribution between 90 $\leq$ $A$
$<$ 140 is instead underestimated by our GCE model: an additional (unknown) 20--30\% contribution 
is required \citep{travaglio04}. 
The light element primary process (LEPP) postulated for this contribution is different from 
the $s$ process in AGB stars and from the weak $s$-process component occurring 
in massive stars. Its origin is largely debated in literature (see, e.g., 
\citealt{frisch12} and \citealt{pignatari13} for a primary $s$ component in massive stars, 
\citealt{arcones13} and \citealt{hansen14} for $\nu p$ process or
$weak$ $r$ process induced by explosive stellar nucleosynthesis).

For an exhaustive discussion about the $s$-process nucleosynthesis
we refer to the reviews by \citet{wallerstein97}, \citet{busso99}, \citet{herwig05}, 
\citet{straniero06}, \citet{kaeppeler11} and \citet{karakaslattanzio14}.

 \vspace{2mm}
 
It is evident from the above considerations that the main component approach does 
not provide a realistic interpretation of the whole solar $s$ abundances, because it does 
not account of all complex aspects of the $s$ process over the Galactic evolution. 
\\
Nevertheless, it is noteworthy that the solar $s$ contributions predicted by main 
component and GCE model are comparable in the atomic mass region between 140 $\leq$ $A$ $\leq$ 
204 (see \citealt{bisterzo11,bisterzo14ApJ}). 
Moreover, despite the remarkable discrepancy estimated by GCE and main component between 
90 $\la$ $A$ $<$ 140, fairly similar $s$ contributions are derived by looking at the 
relative isotopic predictions of a given element. 
The only exceptions are the rarest neutron-rich isotopes ($^{86}$Kr, $^{87}$Rb, $^{96}$Zr), which 
may be significantly produced in intermediate-mass AGB stars.  
\\
Therefore, the study of the main component can be considered a useful 
tool to investigate the nuclear aspects and the sensitivity of the solar $s$ predictions between 
90 $\leq$ $A$ $\leq$ 204, focusing on $s$ isotopes close to the major branches of the $s$ path.

The main goal of this paper is to analyse how the uncertainties of the $\alpha$-induced reactions on 
$^{13}$C and $^{22}$Ne may affect the main component.
  \\  
Both \Can and \Nean reaction rates may be affected by   
contributions of sub-threshold states and resonances,   
making the extrapolation of the laboratory cross section
measurements down to the energy range of stellar burning a complex task \citep{wiescher12}. 
\\
The \Can rate may be influenced by the unknown contribution of the sub-threshold
resonance at 6.356 MeV (owing to the J$^{\pi}$ = 1/2$^+$ state in $^{17}$O).
Up to a factor of three uncertainty was evaluated 
by \citet{angulo99}. Over the years, several analyses have
been dedicated to this neutron source 
(\citealt{drotleff93,denker95,hale97,kubono03,keeley03,johnson06,pellegriti08}). 
Recent investigations have significantly improved the accuracy (down to $\sim$20--30\%;
\citealt{heil0813c,guo12}). Lately, the above resonance has been detected by 
\citet{lacognata13} with the Trojan horse method.
\\
The \Nean rate is dominated by the well studied resonance at 832 keV. The \Nean uncertainty at 
$T_8$ = 3 
is mainly due to the unknown influence of a resonance at 633 keV. The large upper 
limit estimated by \citet{angulo99} (up to a factor of fifty) has been 
reduced by more than a factor of ten with the experimental measurement by \citet{jaeger01}. 
Remarkable experimental and theoretical works were carried out by \citet{kaeppeler94}, \citet{koehler02}, 
\citet{karakas06}, \citet{ugalde07} and \citet{longland12}.

The study of the main component is 
useful to investigate the effects of the major nuclear network uncertainties on 
AGB $s$ predictions, especially close to the branch points. 
\\
A branch in the $s$ path occurs when neutron captures compete with $\beta$ decays
at unstable isotopes with half-lives of a few weeks to a few years, so that the 
$s$ path branches partly by $\beta$ decays following the usual path in 
the stability valley and partly by neutron captures feeding the neutron-rich 
isotopes. The strength of the branching is described by the
branching factor $f_\beta$=$\lambda_\beta$/($\lambda_\beta$+$\lambda_n$), where 
$\lambda_\beta$=1/$\tau$=ln2/$t_{1/2}$ is the $\beta$-decay rate corresponding to the 
half-life $t_{1/2}$ (or to the mean lifetime $\tau$). The decay pattern
may include the $\beta^-$, $\beta^+$ and electron capture channels.
The neutron capture rate $\lambda_n$=$N_n$$<\sigma v>$=$N_n$$v_T$$<\sigma>$ is 
given by the neutron density $N_n$, the mean thermal velocity $v_T$, 
the relative velocity between neutron and target $v$, and the Maxwellian-averaged 
cross section (MACS) defined as $<\sigma>$=$<\sigma v>$/$v_T$.
The neutron capture strength of the branching point is defined as 
$f_{n}$=$\lambda_{n}$/($\lambda_{\beta}$+$\lambda_{n}$). 
\\
The uncertainties in the experimental (n, \g) rates have been highly reduced in
recent years reaching in some cases a precision smaller than one or two percents
(see, e.g., \citealt{kaeppeler11}).
However, the high $s$-process temperatures allow the low-lying excited states to
be populated by the intense and energetic thermal photon bath. Because only the 
ground state is accessible by experiment, the effect of neutron captures in excited
states has to be evaluated theoretically and suffers from large uncertainties
\citep{RT00,rauscher12,reifarth14}.
Even major uncertainties are associated to unstable nuclei for which no experimental measurement 
is available at the present time.
\\
In addition, the $\beta$-decay rates of some radioactive isotopes may be largely affected 
by variations of temperature and electron density. 
Although the laboratory $\beta$-decay rates are accurately known 
\citep{cyburt10}, the contribution of thermally populated excited levels and the effects 
of unknown transitions in a strongly ionised plasma can largely modify the 
$\beta$-decay rates at stellar temperatures. 
\citet{TY87} investigated the $\beta$-decay rates of unstable heavy isotopes 
at temperatures and electron densities typical of stellar interiors 
(5$\times$10$^7$ $\leq$ $T$ $\leq$ 
5$\times$10$^8$ K; 10$^{26}$ $\leq$ $n_e$ $\leq$ 3$\times$10$^{27}$ cm$^{-3}$), 
finding large deviations from the terrestrial values.
The temperature dependence of branchings is even more complex, if they have
isomeric states that are thermalised 
at high temperatures through transitions via mediating states at higher excitation energy.
Consequently, the abundances of the affected $s$-only isotopes  
carry direct information on the physical 
conditions occurring during the $s$ process, i.e., neutron density, temperature, and density
\citep{kaeppeler89}.
\\
In this context, the branch point isotopes $^{151}$Sm, $^{163}$Dy, $^{179}$Hf, $^{176}$Lu
are remarkable examples. The $\beta^-$-decay rate of $^{151}$Sm increases strongly at He-shell
flash temperature and regulates the abundances of the pair $^{152,154}$Gd  
\citep{marrone06,wisshak06}. 
$^{163}$Dy and $^{179}$Hf are terrestrially stable, but start to decay at temperatures
typical of TPs producing a non-negligible $s$-process contribution to $^{164}$Er and 
$^{180}$Ta$^{\rm m}$, both usually bypassed by the $s$ path 
\citep{jaag96,wisshak01,wisshak04}. 
Because of its long half-life, $^{176}$Lu was long considered as a cosmo-chronometer until 
the decay rate was found to exhibit a very strong temperature dependence
\citep{klay91,heil08LuHf,mohr09}. 
\\
The $\beta$-decay rates calculated in stellar environments are subject to nuclear uncertainties 
difficult to estimate. \citet{goriely99} recomputed the calculations of \citet{TY87} by assuming 
a typical uncertainty of $\Delta$log $ft$ = $\pm$0.5 for the decay rates
of unknown transitions and showed that the final stellar rates may vary by up
to a factor of three at typical $s$-process conditions
($T_8$ = 3 and $n_e$ = 10$^{27}$ cm$^{-3}$).
Recently, the nuclear NETwork GENerator (NETGEN; \citealt{xu13}) extrapolated the $\beta$-decay 
rates by \citet{TY87} and their uncertainties \citep{goriely99} to an extended range of temperature 
and electron density.
In a few cases the $\beta$-decay properties of
excited states have been measured in laboratory experiments, e.g. for long-lived isomeric 
states ($^{180}$Ta$^{\rm m}$ determined by \citealt{belic02}; or the key branch for the weak 
$s$ process in massive stars $^{79}$Se$^{\rm m}$ studied by \citealt{klay88}) or for
isotopes with small Q-values, which are sensitive to the bound-state $\beta$-decay
mechanism (examples are $^{163}$Dy and $^{187}$Re studied by \citealt{jung92} and by
 \citealt{bosch96}, respectively). In general, however,
measurements of such data remain a big experimental challenge \citep{reifarth14}.

\vspace{2mm}

\begin{figure*} 
\includegraphics[angle=-90,width=16cm]{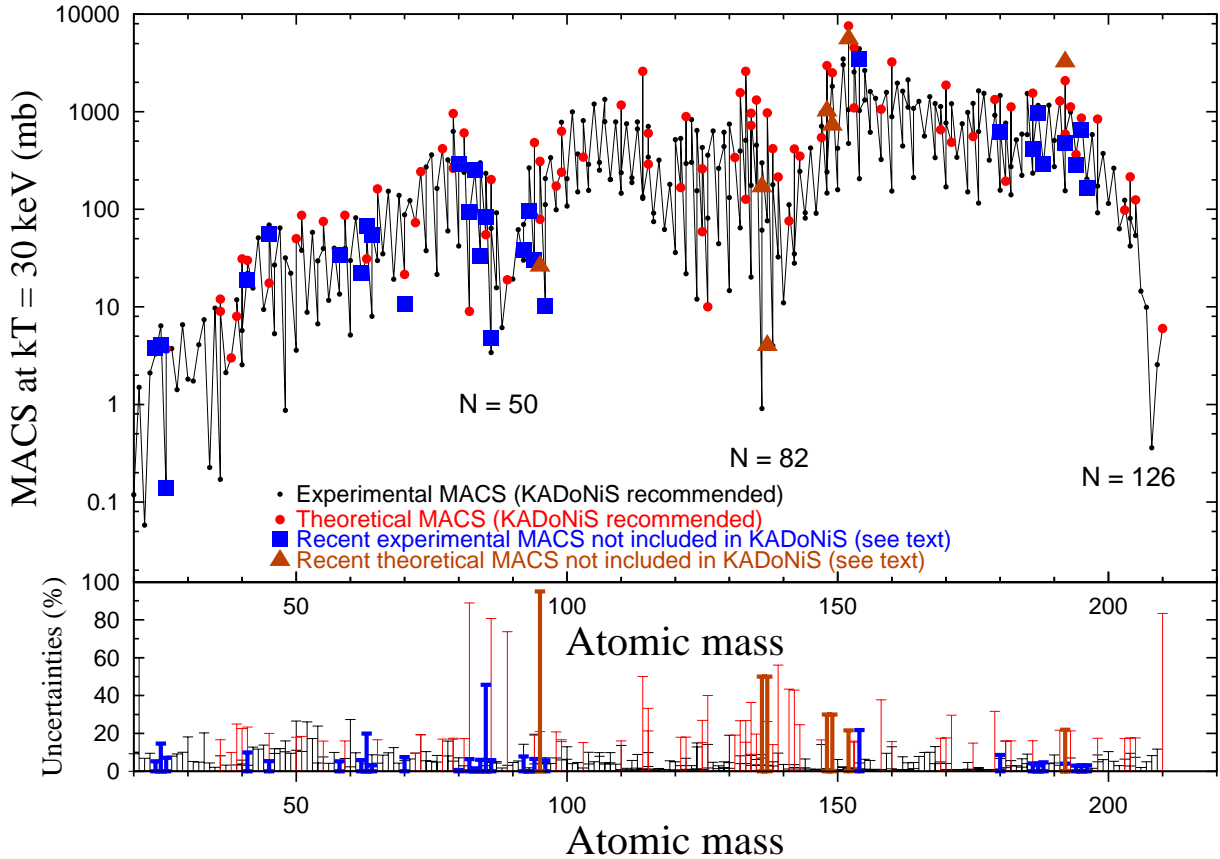}
\vspace{5mm}
\caption{$Top$ $panel$: stellar (n, \g) cross sections at 30 keV for nuclides 
with $A$ $\geq$ 20.
The compilation is taken from the KADoNiS database (v0.3): $small$ $dots$ and $big$ $circles$ refer 
to experimental and theoretical MACS, respectively. 
Recent MACS not included in KADoNiS, but considered in this work are indicated
by $squares$ (experimental measurements) and $triangles$ (theoretical evaluations); 
see text for details.
Neutron-magic numbers at $N$ = 50, 82 and 126 are highlighted.
$Bottom$ $panel$: MACS uncertainties at 30 keV. Colors of error bars are the same as
in the top panel.}
\label{figMACS}
\end{figure*}

 The paper is organised as follows.
We start with a description of the solar main component, updated with a neutron 
capture network that includes the most recent cross section measurements as well as the
solar abundance data by \citet{lodders09}, (Section~\ref{mains}).
The AGB models employed in this work have been outlined by \citet{bisterzo10,bisterzo14ApJ}.
\\
In Sections~\ref{13can} and~\ref{22nean}, we discuss the impact of the \Can and \Nean uncertainties 
on the branches of the main component. 
\\
Relevant branches of the $s$ path and their influence on the $s$ nuclides are analysed in 
Section~\ref{branch}. We distinguish the $s$-only isotopes in different classes,  
according to the characteristics of the related branchings.
First, unbranched $s$-only isotopes (e.g., 
$^{100}$Ru, 
$^{104}$Pd, 
$^{110}$Cd, 
$^{116}$Sn, 
$^{124}$Te, 
$^{150}$Sm, 
$^{160}$Dy, and $^{198}$Hg)  
are useful to constrain the main component (Section~\ref{branchUN}).
Second, we address branchings where the decay of the unstable branch point isotopes 
exhibits no effect of the $s$-process temperature (e.g., $^{96}$Mo, 
which is sensitive to the branch at $^{95}$Zr) 
or where the half-lives are essentially independent of
temperature. These branchings may provide information on the $s$-process neutron
density (e.g., $^{170}$Yb, partially regulated by $^{169}$Er;
$^{142}$Nd, marginally affected by $^{142}$Pr;
$^{186}$Os, which mainly depends on the branch at $^{185}$W;
$^{192}$Pt, influenced by $^{192}$Ir; see Section~\ref{branchNn}).
Third, branchings initiated 
by decays with pronounced dependencies on temperature 
and/or electron density may be
interpreted as $s$-process thermometers (e.g.,  
$^{134}$Ba, owing to the branch at $^{134}$Cs;
$^{152,154}$Gd, strongly sensitive to $^{151}$Sm and $^{154}$Eu;
$^{176}$Lu, which has a short lived isomer; 
$^{204}$Pb, because of the branch at $^{204}$Tl; Section~\ref{branchT}) 
or indicators of the turnover timescale of convective mixing during TP 
(e.g., $^{128,130}$Xe, $^{164}$Er, $^{180}$Ta$^{\rm m}$; Section~\ref{branchMIX}). 
Each of the three classes is illustrated in detail by a few selected
examples: $^{150}$Sm, adopted to normalise the $s$ distribution (Section~\ref{sm150}),
as well as $^{96}$Mo and $^{170}$Yb (Sections~\ref{zr95} and~\ref{yb170}),
$^{134}$Ba (Section~\ref{ba134136}), and $^{180}$Ta$^{\rm m}$ (Section~\ref{ta180}),
which are most sensitive to uncertainties related to the 
cross sections and $\beta$-decay rates of the involved branch point nuclei. Other
branchings are briefly summarised, referring to the complete discussion in Section~B 
(Supporting Information).

In Section~C (Supporting Information), we describe the effect of the $^{13}$C($\alpha$, n)$^{16}$O, 
\Nean and \Neag uncertainties on two AGB models (a 3 $M_\odot$ model at [Fe/H] = $-$1 and a
half-solar metallicity 5 $M_\odot$ model), 
which are selected as representative of a more extended range of stars than usually adopted 
for the main component. 
\\
An overview of the results is provided in Section~\ref{summary}.


\section{Updated solar main component} \label{mains}

Despite the main component does not
provide a realistic description of the solar $s$ distribution between $A$ = 90 and 204, it
still represents a useful approximation to investigate the effect of nuclear uncertainties  
of $s$ isotopes in this atomic mass range.

As shown by \citet{arlandini99}, the ''stellar" main component
is obtained as the averaged yields between two AGB models
with initial masses of 1.5 and 3 $M_\odot$ at half-solar metallicity. 
The $s$-process nucleosynthesis is computed with the post-process
method described by \citet{gallino98}, which is based on input data of full evolutionary FRANEC
models by \citet{straniero97,straniero03}, like  
the temporal history of the temperature and density gradients during the
convective TPs, the number of TPs, the mass of the H shell and He intershell,
the overlapping factor, and the residual mass of the envelope.
\\
In both AGB models, the \Nean reaction operates in a similar range of 
temperatures ($T_8$ $\sim$ 2.5--3). However, the $M$ = 3 \Msun model
achieves $T_8$ $\sim$ 3 for the last 16 TPs, while the $M$ = 1.5 \Msun
model for the last 8 TPs.
The marginal activation of the \Nean neutron burst mainly occurs in the advanced
TPs for a rather short timescale ($\sim$6 yr). 
\\
Most of the $s$-process nucleosynthesis 
takes place radiatively during the interpulse phases via the \Can neutron 
source, which lasts for 3-6$\times$10$^4$ yr (this time decreases with increasing pulse number).
We assume a single $^{13}$C-pocket profile close to case ST described by \citet{gallino98}.
The case ST $^{13}$C-pocket contains 5$\times$10$^{-6}$ \Msun of $^{13}$C and 
2$\times$10$^{-7}$ \Msun of $^{14}$N, and extends in mass for 
$\sim$1$\times$10$^{-3}$ $M_\odot$ ($\sim$1/10 of the typical mass involved in a TP). 
\\
The mass loss is estimated with the Reimers formula \citep{reimers77} with the parameter 
$\eta$ = 0.3 for 1.5 $M_\odot$ and $\eta$ = 1 for 3 $M_\odot$. 
This allows the occurrence of 19 and 25 TPs with TDU,
respectively. Updated opacities and a revised luminosity function of Galactic 
carbon stars \citep{lederer09,guandalini13} suggest a more efficient mass loss than the one adopted 
in our models. This significantly reduces the number of TDUs (e.g., $\sim$15 TDU for 
the above models; \citealt{cristallo11}). To this purpose, note that the overall $s$ 
yields of our AGB models are marginally influenced by the contribution of 
the $s$ abundances predicted in the envelope for TDU numbers higher than 15. 
Moreover, the normalisation of the average $s$ yields to that of $^{150}$Sm allows us 
to overcome some of the major uncertainties of AGB stellar models (e.g., the efficiency of the TDU, 
the mass loss, the number of TPs), which would otherwise dominate the $s$-process 
predictions (see Section~\ref{intro}).
\\
For major details on the AGB models employed we refer to \citet{bisterzo10,bisterzo14ApJ}.

The neutron capture cross section network is based on KADoNiS\footnote{Karlsruhe 
Astrophysical Data Base of Nucleosynthesis in Stars, web site http://www.kadonis.org/, version
KADoNiSv0.3.}.
Different MACS are adopted for 
$^{20,21,22}$Ne by \citet{heil14},
$^{24,25,26}$Mg by \citet{massimi12},
$^{41}$K and $^{45}$Sc by \citet{Heil09pasa},
$^{58,62,63}$Ni by \citet{zugec14} and \citet{lederer13,lederer14},
$^{64,70}$Zn by \citet{reifarth12},
$^{80,82,83,84,86}$Kr by \citet{mutti05},
$^{85}$Kr by \citet{raut13},
$^{92,94,96}$Zr by \citet{tagliente10,tagliente11,tagliente11a}, 
$^{93,95}$Zr by \citet{tagliente13} and \citet{lugaro14},
$^{136,137}$Cs by \citet{patronis04},
$^{148,149}$Pm by \citet{reifarth03},
$^{152,154}$Eu by \citet{best01},
the p-only $^{180}$W by \citet{marganiec10},
$^{186,187,188}$Os by \citet{mosconi10}, 
$^{192,194,195,196}$Pt by \citet{koehler13}, and their theoretical evaluation of the 
$^{192}$Ir MACS.
\\
%
For a few heavy isotopes, the stellar temperatures are high enough for a significant 
population of low-lying nuclear excited states, which modify the neutron capture cross sections 
measured in laboratory. 
These effects are considered in the stellar MACS with the so-called stellar enhancement factor (SEF), 
which accounts of the stellar average over the thermally populated states and the laboratory 
cross section.
The SEF estimated by \citet{bao00} are updated with those recommended by KADoNiS. 
An exception is $^{187}$Os, for which we have adopted \citet{fujii10}, who studied the inelastic 
channel of the neutron capture cross section of $^{187}$Os and employed a Hauser-Feshbach statistical 
model theory (HFSM) to obtain a reliable estimate of the SEF.

The state of the art of all MACS at $kT$ = 30 keV versus atomic mass is shown in Fig.~\ref{figMACS} 
($top$ $panel$). We distinguish between experimental measurements ($small$ $dots$) and theoretical 
evaluations ($big$ $circles$). 
Updated MACS with respect to KADoNiS are indicated by $big$ $squares$ (experimental measurements)
or $triangles$ (theoretical evaluations). 
The $lower$ $panel$ of Fig.~\ref{figMACS} represents the related percentage MACS uncertainties.
\\
As outlined by \citet{kaeppeler11}, most of the MACS involved in the nucleosynthesis of stable isotopes 
heavier than $A$ $\sim$ 90 are known with an accuracy $\la$5\%.
%
Uncertainties of $\sim$10\% 
are evaluated for $^{104}$Pd (10\%), 
$^{139}$La (9.6\%), 
$^{159}$Tb (9.5\%; near the $s$-only isotope $^{160}$Dy), 
$^{166,167,168}$Er (10 to 13\%),  
$^{200}$Hg (10\%),
$^{209}$Bi (11.7\%).
\\
Among the unstable isotopes, the largest MACS uncertainties are quoted for
$^{85}$Kr ($\sim$50\%; recently evaluated from the inverse $^{86}$Kr(\g, n)$^{85}$Kr reaction
and from the $^{86}$Kr(\g, \g') measurement by \citealt{raut13}), 
$^{86}$Rb ($\sim$80\%), 
$^{95}$Zr (up to a factor of two uncertainty, \citealt{lugaro14}), 
$^{141}$Ce and 
$^{142}$Pr ($\sim$40\%). 
Other unstable isotopes with uncertainties from 15\% to 25\%  
are $^{79}$Se (important for the weak $s$-process component), $^{81}$Kr,
$^{99}$Mo, $^{103}$Ru, $^{110}$Ag, $^{115}$Cd, $^{147}$Nd, $^{160}$Tb,
$^{170}$Tm, 
$^{179,182}$Ta, $^{181}$Hf, $^{186}$Re, $^{191}$Os, $^{192}$Ir, $^{193}$Pt,
$^{198}$Au, $^{203}$Hg, $^{204}$Tl,
$^{205}$Pb.

\begin{figure} 
\includegraphics[angle=-90,width=8.5cm]{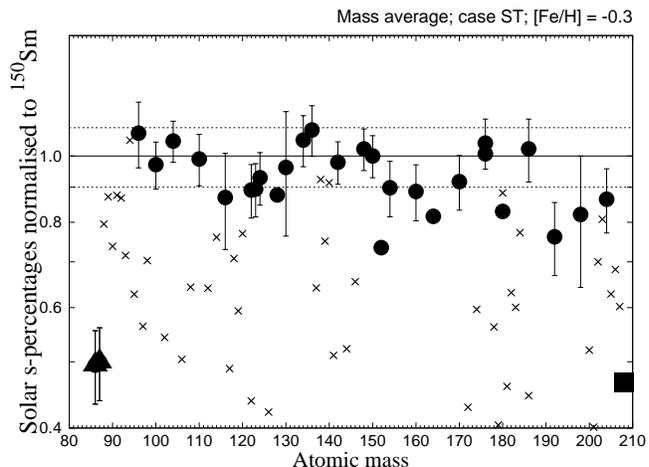}
\caption{The solar main component versus atomic mass is reproduced 
by assuming a ST $^{13}$C-pocket, and by averaging between AGB models with initial masses 
of $M$ = 1.5 and 3 $M_\odot$ at half solar metallicity.
In this Figure, the $s$-production factors are normalised to that of the unbranched 
$s$-only nucleus $^{150}$Sm.
The $s$-production factors in the He-intershell (given in mass fraction '$X_i$' 
over the solar-scaled initial values) correspond to the material cumulatively mixed
with the envelope by the recurrent TDU episodes and eventually dispersed 
in the interstellar medium by efficient stellar winds.
Different symbols are used for $^{86,87}$Sr (which receive additional $s$
contributions, e.g., the weak component in massive stars), and for $^{208}$Pb, $\sim$50\% of 
which is produced by low-mass, low-metallicity AGB stars (strong $s$ process).
The error bars displayed for the $s$-only isotopes account for the uncertainties of the
solar abundances by \citet{lodders09}. 
Note that the solar uncertainties are not shown for $^{128}$Xe, $^{152}$Gd, $^{164}$Er, and 
$^{180}$Ta$^{\rm m}$ because they receive additional non-negligible contributions from the
($\nu$)$p$-process by SNIa and SNII (e.g., \citealt{travaglio14,rauscher02,pignatari13nugrid}).
The solid line corresponds to a 100\% contribution from main component, 
dashed lines represent 10\% of uncertainty.}
\label{figmains}
\end{figure}

The solar main component is shown in Fig.~\ref{figmains}. 
We examine the $s$-production factors in the He-intershell (corresponding to the 
overabundances with respect to the solar-scaled initial values given in mass fractions '$X_i$') 
of isotopes with A $>$ 80 normalised to the unbranched $s$-only nucleus $^{150}$Sm.
\\
A plausible reproduction of $s$-only isotopes heavier than $A$ = 90 (full circles) is obtained 
within the solar abundance uncertainties quoted by \citet{lodders09}. 
Variations with respect to previous
results by \citet{bisterzo11} and \citet{kaeppeler11} partly derive from new solar abundances
by \citet{lodders09} and partly from updates in the nuclear reaction network. 
Among the nuclides with $A$ $>$ 90, a noteworthy variation is the increased solar abundance of Hg
estimated by \citet{lodders09} (+35\%), which reduces the $^{198}$Hg solar $s$-contribution by
\citet{kaeppeler11} from $\sim$100\% to 82\%.
The slightly larger solar Sn uncertainty estimated by \citet{lodders09} (from 10\% by \citealt{AG89}
to 15\%) agrees better with the $s$ prediction for $^{116}$Sn, 
although the solar abundance is still 13\% higher.
Note that the solar Nb and Xe values estimated by \citet{kashiv06} and \citet{reifarth02} were
already considered in \citet{kaeppeler11}.

The updated network produces small variations ($<$5\%) among the $s$-only isotopes. 
Differences larger than $\sim$10\% are obtained for 
$^{164}$Er and $^{180}$Ta$^{\rm m}$, where the 
already ascertained dominant $s$-process contributions 
\citep{best01,wisshak01} have been increased:
\begin{itemize}
\item The $s$ contribution of solar $^{164}$Er is enhanced by 8\% 
(from 74\% to 82\%), because a smaller SEF is evaluated for the MACS of $^{164}$Er  
(the old SEF = 1.24 at 30 keV estimated by \citealt{bao00} is reduced to 1.08 by KADoNiS).
\item $^{180}$Ta$^{\rm m}$ increases from 75\% to 82\% (+7\%), 
mostly due to a decreased SEF of the MACS for $^{180}$Ta$^{\rm m}$ 
(from SEF = 0.96 at 30 keV by \citealt{bao00} to 0.87 by KADoNiS).
\end{itemize}

Starting from these updated results, we analyse in the following sections 
the effect of the two AGB neutron source uncertainties on the $s$ distribution
and discuss the consequences of the MACS and $\beta$-decay rate uncertainties 
of the most relevant branching points of the main component.

Recently, theoretical analyses by \citet{rauscher11} and \citet{rauscher12} showed that the SEF 
correction can cause larger MACS uncertainties than assumed so far. This holds especially for cases, where the 
(experimentally measured) ground state cross section constitutes only a minor fraction of the
MACS. If the ground state contribution, expressed by a new factor $X$, is close
to unity, the uncertainty of the stellar rate is directly connected to the experimental one;
otherwise, the uncertainty of stellar rate may be larger than that of the experimental measurement.
\\
The theoretical uncertainties discussed by \citet{rauscher11} are not included in the present 
work, because they should be considered as upper limits of future more realistic analyses, which 
need to be carried out individually for each nucleus, by accounting of all available nuclear details.
The situation is even more challenging when the experimental rate is inferred from the inverse 
neutron capture reaction, (\g ,n) (e.g., the case of $^{185}$W; \citealt{rauscher14}). 
On the base of the branches analysed in Section~\ref{branch} and in Section~B (Supporting Information),
we highlight outstanding isotopes that would need specific theoretical investigations in 
Section~\ref{summary}.


\section{Uncertainty of the \Can neutron source}\label{13can}

As anticipated in Section~\ref{intro}, the \Can rate is the dominant neutron source
in low mass AGBs and shapes the overall $s$-abundance component.
It burns radiatively during the interpulse periods in the top layers 
of the He-intershell when the temperature reaches $T_8$ $\sim$ 0.8--1.0. 
Under these conditions, the \Can rate is extremely small (10$^{-14}$--10$^{-13}$ 
cm$^3$/mol s) and, therefore, not yet accessible to direct measurements 
in terrestrial laboratories,
owing to the high background signals induced by cosmic $\gamma$-rays. 
Current reaction rates have been determined 
by extrapolation of direct measurements performed at energies higher than 270 keV, 
far from the Gamow window at 140--230 keV where the \Can reaction is effective (see, e.g., 
\citealt{drotleff93}). 
This entails an unknown influence of broad sub-threshold states. A major impact is expected
from the contribution of the sub-threshold resonance due to the J$^{\pi}$ = 1/2$^+$ state in 
$^{17}$O at  
6.356 MeV ($E_\alpha^{\rm lab}$ = $-$3 keV), while minor effects are expected due to
sub-threshold resonances at   
5.939 MeV (J$^{\pi}$ = 1/2$^-$; 
$E_\alpha^{\rm lab}$ = $-$547 keV) and   
5.869 MeV (J$^{\pi}$ = 3/2$^+$; 
$E_\alpha^{\rm lab}$ = $-$641 keV).
\\
Past analyses showed large uncertainties at 8 keV (up to a factor of three) owing to
the unexplored contribution of the 6.356 MeV state (see e.g., \citealt{CF88} and 
\citealt{angulo99}).
More recent evaluations by \citet{heil0813c} and \citet{guo12} are in reasonable agreement and 
provide a significantly reduced uncertainty (down to $\sim$20--30\%, in the AGB energy range of 
interest).
Later, \citet{lacognata13} attained an accuracy of 15\% for the \Can rate 
by means of the Trojan horse method, which allows one to study reactions of
astrophysical interest free of Coulomb suppression and electron screening at astrophysical energies 
with no need for extrapolation. This method clearly displays the presence of the 
 sub-threshold resonance 
corresponding to the 6.356 MeV $^{17}$O state. Their rate is about 35\% higher than that by 
\citet{heil0813c}, still in agreement within uncertainties with most of the results in 
literature at $T_8$ $\sim$ 1. 
\\
In Table~\ref{tabc13ansigma}, we provide a summary of the most relevant
\Can rates found in literature at $T_8$ = 1.

\begin{table}
\caption{The \Can rate at $T$ = 1$\times$10$^8$ K (in units of 10$^{-14}$ cm$^3$/mol s)
determined by  
\citet{CF88},    
\citet{denker95},   
\citet{kubono03},   
\citet{johnson06},     
\citet{heil0813c},	      
\citet{guo12},
NACRE II \citep{xu13NACREII} and \citet{lacognata13}.
Additional studies are not listed explicitly,
but found \Can rates within these ranges.}
 \label{tabc13ansigma}
\centering
\resizebox{8.5cm}{!}{\begin{tabular}{lccc}
\hline
Reference   &   Rate  & Lower & Upper \\ 
                 &  (10$^{-14}$ cm$^3$/mol s)  & Limit  &  Limit \\
\hline
\citet{CF88}        &       2.58   &    --     &   --         \\
\citet{kubono03}    &	   2.02   &   1.49    &  2.54  \\
\citet{johnson06}   &       2.64   &    2.11   & 3.30  \\
NACRE II            &       4.86   &    3.62   &  6.48  \\
\hline
\citet{denker95}    &	   4.32                &    --         &   --         \\
\citet{heil0813c}	   &       4.6    & 3.6  &  5.6   \\
\citet{guo12}	   &       4.19   &  3.46 & 4.90\\
\citet{lacognata13} &       6.20 & 5.32 &  7.13 \\ 
\hline    
\end{tabular}}
\end{table}

\begin{figure}
\includegraphics[width=0.47\textwidth]{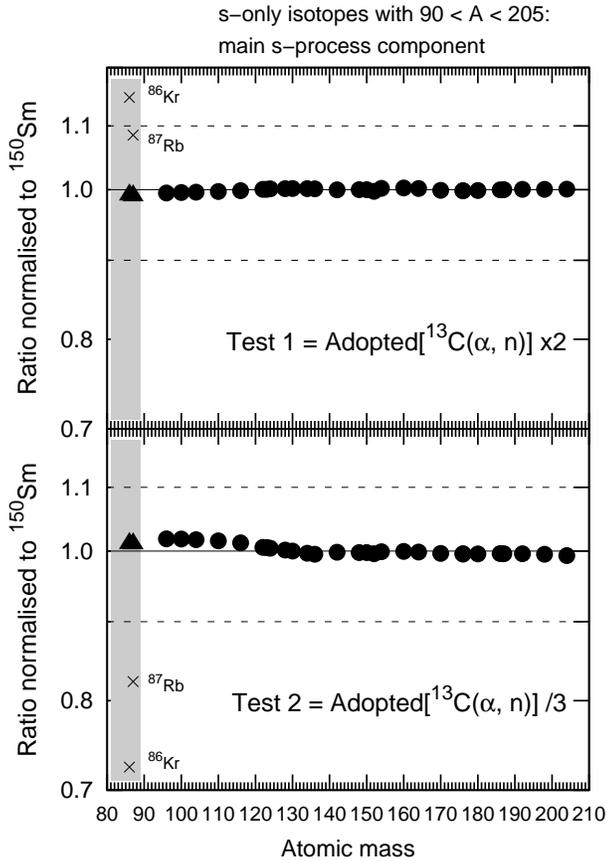}
\caption{Ratios between the main component obtained with a two times higher and a
three times lower \Can rate than our adopted rate shown in Fig~\ref{figmains} ({\bf Test 1} and 
{\bf Test 2} corresponding to the $top$ and $bottom$ $panels$).
We focus on $s$-only isotopes for clarity. 
The shaded area between 80 $<$ $A$ $\la$ 90 indicates the atomic mass region affected by 
additional $s$ contributions (e.g., the weak $s$ process in massive stars; IMS 
AGB stars, see Section B and C, Supporting Information). 
The two neutron-magic nuclei $^{86}$Kr and $^{87}$Rb, which exhibit the largest
variations, are marked by crosses (see text).
A complete version of this Figure (which includes all isotopes from 90 $\leq$ $A$ $\leq$ 210)
is given in Section~A (Supporting Information).}
\label{fig3A}
\end{figure}

In our AGB models we adopt the \Can rate by \citet{denker95}, which is pretty
close to the results by \citet{heil0813c} and \citet{guo12}. 
\\
We carried out two extreme tests in order to analyse the effect 
of the \Can uncertainty on the main component:
\begin{itemize}
\item {\bf Test 1} -- Adopted \Can rate multiplied by a factor of two, close to the upper limit by \citet{lacognata13};
\item {\bf Test 2} -- Adopted \Can rate divided by a factor of three, close to the lower limit
by \citet{kubono03}.
\end{itemize}

The results of the two tests are shown in Fig.~\ref{fig3A} for isotopes with A $>$
80. We focus on $s$-only isotopes for clarity ($circles$).
The ratio of the results obtained in the tests and with the adopted \Can rate are 
plotted in the $upper$ and $lower$ $panels$
together with the 10\% limits indicated by dashed lines.

We see that the main component is marginally affected by both tests:
variations are smaller than 1\% for the $s$-only isotopes with $A$ $>$ 90 and smaller than
3\% between $A$ = 80 and 90.
\\
Significant changes were obtained, however, for the two neutron-magic nuclei $^{86}$Kr 
and $^{87}$Rb (represented by $crosses$), because the branching at $^{85}$Kr is partially
activated during the \Can phase. For a comprehensive
discussion see Section~B (Supporting Information). We note, however, that both
isotopes receive a small contribution from AGB stars: only $\sim$16\% 
and $\sim$18\% of solar $^{86}$Kr and $^{87}$Rb are produced by the main
component, respectively.

\vspace{2mm}

In our AGB models with initial masses 1.5 and 3 
$M_\odot$ and half solar metallicity the amount of $^{13}$C in the pocket assumed for 
case ST
burns radiatively during the interpulse even by reducing the \Can rate
by a factor of three. Only a $^{13}$C mass fraction negligible for the $s$ process 
is ingested in the next convective TP. 
\\
Otherwise, when a substantial amount of $^{13}$C is ingested in the subsequent TP,
it burns at the bottom of the convective shell (at $T_8$ $\sim$ 1.6)
in a very short time scale (of the order of few years), producing 
enhanced neutron densities of up to $\sim$10$^{10-11}$ cm$^{-3}$.
A new generation of AGB models available at the FRUITY database 
by \citet{cristallo09,cristallo11} experiences a 
partial convective burning of $^{13}$C during the first (and second) 
TP(s). 
This mainly occurs in low-mass metal-rich AGB models and influences the abundances of a few 
neutron-rich isotopes during the first TPs (e.g., $^{86}$Kr, $^{87}$Rb, $^{96}$Zr;
and radioactive nuclides as $^{60}$Fe, see \citealt{cristallo06,cristallo11}).
\\
Actually, the recent measurement by \citet{lacognata13} suggests that the \Can
rate adopted in our models must be increased by 40\%, rather than decreased (see 
Table~\ref{tabc13ansigma}). In this case, the ingestion of $^{13}$C into TPs
is strongly reduced.

Despite the main component presented in this work is 
marginally influenced by the uncertainty of the \Can rate, AGB 
predictions in general may show larger sensitivity, when e.g., a 
different amount of $^{13}$C is assumed in the pocket, or a lower initial 
metallicity is adopted. Moreover, the physical characteristics of 
different stellar models employed in the nucleosynthesis calculations 
may influence the effect of the \Can rate. For instance, recent 
low-mass AGB models are characterized by more efficient TDU 
episodes (see the discussion in \citealt{straniero06}\footnote{A more efficient TDU
is required to reproduce the observed luminosity function of the Carbon stars in 
the Milky Way and in the Magellanic Clouds \citep{cristallo11,guandalini13}.}). In particular, 
the first TDU occurs at the very beginning of the TP-AGB phase, when 
the core mass is rather small and, in turn, the temperature developed in 
the He-rich intershell during the interpulse period is lower \citep{cristallo09}. 
In this case, a not negligible amount of $^{13}$C that survives 
after the interpulse is engulfed in the convective zone generated by 
the subsequent TP and burns at relatively high temperature. For example, 
in 1.8 $M_\odot$ models by \citet{lugaro12} and \citet{guo12}, the 
convective burning of the partial $^{13}$C ingested in the TPs 
affects $s$ predictions up to Pb.
This effect may even become dominant for lower mass AGB models (e.g., in \M = 1.25 $M_\odot$ 
models of half-solar metallicity the $^{13}$C neutron source burns convectively 
rather than radiatively; \citealt{raut13}).


\section{uncertainty of the \Nean neutron source and the \Neag rate}\label{22nean}

While the \Can neutron source determines the overall shape of the main
component, the contribution of the \Nean reaction
affects the $s$-only isotopes close to branching points.
\\
For the \Nean reaction, the center-of-mass energy region of interest 
to the $s$ process is $E_{\rm c.m.}$ =
300 to 900 keV. 
At these low energies, direct \Nean measurements 
are limited by cosmic \g-ray background 
because the cross section rate is extremely small.
The lowest well studied J$^{\pi}$ = 2$^+$ resonance at $\sim$832 keV, which 
dominates the \Nean rate \citep{drotleff93}, lies above the relevant 
astrophysical energies.
Theoretical extrapolations of the \Nean reaction measured at higher energies may be 
affected by the unknown influence of low-energy resonances just below the neutron threshold. 
\\
The resonances affecting the ($\alpha$, n) and ($\alpha$, $\gamma$) rates
of $^{22}$Ne correspond to different levels in the compound nucleus.
First investigations derived a large \Nean uncertainty because of the influence of a possible 
resonance at $\sim$635 keV with assigned natural spin-parity J$^{\pi}$ = 1$^-$ (e.g., NACRE,
\citealt{angulo99}; \citealt{kaeppeler94}; \citealt{jaeger01}; and \citealt{koehler02}).
Successively, \citet{longland09} determined the energy and quantum numbers of excited states in 
$^{26}$Mg through a nuclear resonance fluorescence experiment 
and demonstrated that the corresponding level at an excitation energy of
$E_x$ = 11 154 keV has unnatural parity J$^{\pi}$ = 1$^+$, contrary to the previous 
spin-parity assignment.
Because both $^{22}$Ne and $^{4}$He have J$^{\pi}$ = 0$^+$, by angular momentum 
selection rules only natural-parity (0$^+$, 1$^-$, 2$^+$, etc.) states in $^{26}$Mg can participate 
in the \Nean reaction, thus excluding the J$^{\pi}$ = 1$^+$ resonance.
Nevertheless, the \Nean rate may be affected by the possible contribution of other 
unknown low-threshold states of natural parity \citep{ugalde07,deboer10}.
Moreover, if yet unknown low-energy resonances have a strong influence on the \Neag rate, 
the \Neag and \Nean reactions may compete during TPs, further affecting the neutron production 
in AGB stars (see, e.g., IMS stars, \citealt{karakas06}). 
\\
The recent theoretical estimate by \citet{longland12} based on a Monte Carlo evaluation 
suggests a much smaller \Nean uncertainty ($\sim$20\%).

\vspace{2mm}

In previous AGB models (e.g., \citealt{bisterzo10,bisterzo11}), we have adopted the 
lower limit of the \Nean and \Neag rates by \citet{kaeppeler94}. 
These rates were evaluated by neglecting the resonance contribution 
at 633 keV. At $T_8$ = 3, the \Nean rate was
 about 50\% higher than that recommended by 
\citet{jaeger01}, close to the upper limit by \citet{longland12}.

In Table~\ref{tabne22ansigma}, we summarise some of the most significant results achieved
in the estimate of the \Nean and \Neag rates for the maximum temperature  
at the bottom of the advanced TPs in the AGB models adopted here
($T_8$ = 3).

\begin{table}
\caption{The \Nean and \Neag rates at 3$\times$10$^8$ K (in unit of 10$^{-11}$ cm$^3$/mol s)
estimated by \citet{CF88}, NACRE \citep{angulo99},
 \citet{kaeppeler94}, \citet{jaeger01}, 
 \citet{longland12}. The newly evaluated values are labelled with "This work" (see text for 
 explanations).
 We distinguish between experimental measurements ($exp$) and theoretical evaluations 
 ($th$).}\label{tabne22ansigma}
\centering
\resizebox{8.5cm}{!}{\begin{tabular}{lccc}
\hline
Reference          &   Recommended       & Lower & Upper \\ 
                   &   Rate              & Limit & Limit \\
                   &  (10$^{-11}$ cm$^3$/mol s)  &   &   \\
\hline
\multicolumn{4}{l}{$^{22}$Ne($\alpha$, n)$^{25}$Mg rate:} \\
\hline
\citet{CF88}; ($th$) &  1.86  &  -   &  -   \\
NACRE; ($th$)          &  4.06  & 3.37 & 192.0 \\
\citet{kaeppeler94}; ($exp$)& {\bf 9.09 }  & {\bf 4.14 }& {\bf 14.4 }\\
\citet{jaeger01}; ($exp$)  & {\bf 2.69 }  & {\bf 2.63 }& {\bf 3.20 }\\
\citet{longland12}; ($th$)& {\bf 3.36}  & {\bf 2.74 }& {\bf 4.15 }\\
This work; ($th$)        & {\bf 2.24}  & {\bf 1.99 }& {\bf 2.92 }\\ 
\hline
\multicolumn{4}{l}{$^{22}$Ne($\alpha$, $\gamma$)$^{26}$Mg rate:} \\
\hline
\citet{CF88}; ($th$)     &  0.46  &  -   &  -   \\
NACRE; ($th$)            &  2.56   & 0.59 & 20.30 \\
\citet{kaeppeler94}; ($exp$)& {\bf 1.22}   & {\bf 0.81} & {\bf 1.63 }\\
\citet{longland12}; ($th$) & {\bf 1.13}  & {\bf 0.93}  & {\bf 1.38 }\\
This work; ($th$)          & {\bf 0.80}  & {\bf 0.72 } & {\bf 2.62 }\\ 
\hline    
\end{tabular}}
\end{table}

\begin{figure}
\vspace{-10mm}
\includegraphics[width=0.47\textwidth]{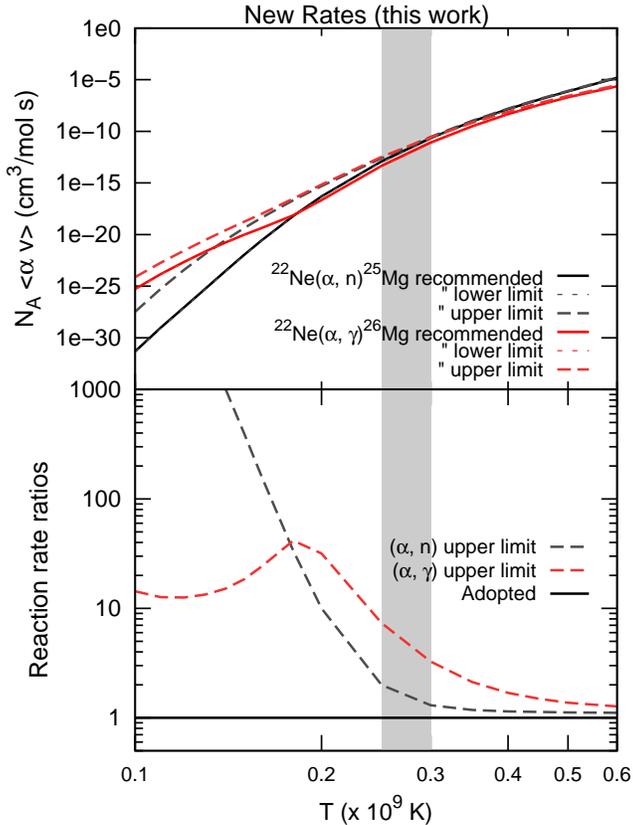}
\vspace{-5mm}
\caption{{\it Top panel}: new evaluated values of the \Nean and \Neag rates
 with black and red lines respectively. 
The difference between lower and recommended rates are almost unnoticeable 
(about 10\% at all temperatures; dotted lines). 
Our upper rates are shown by dashed lines. 
See text for details on the computed uncertainties.
{\it Bottom panel}: The uncertainty bands for the
\Nean and \Neag reactions. The uncertainties are the result of upper
limit resonance contributions and of resonance strength uncertainties.
The shaded areas at $T_8$ = 2.5--3 indicates the temperature range reached during TPs
by low mass AGB stars.
Explicit values of both rates at $T_8$ = 3 are given in Table~\ref{tabne22ansigma}.}
\label{newrates}
\end{figure}

\begin{figure}
\includegraphics[angle=-90,width=0.47\textwidth]{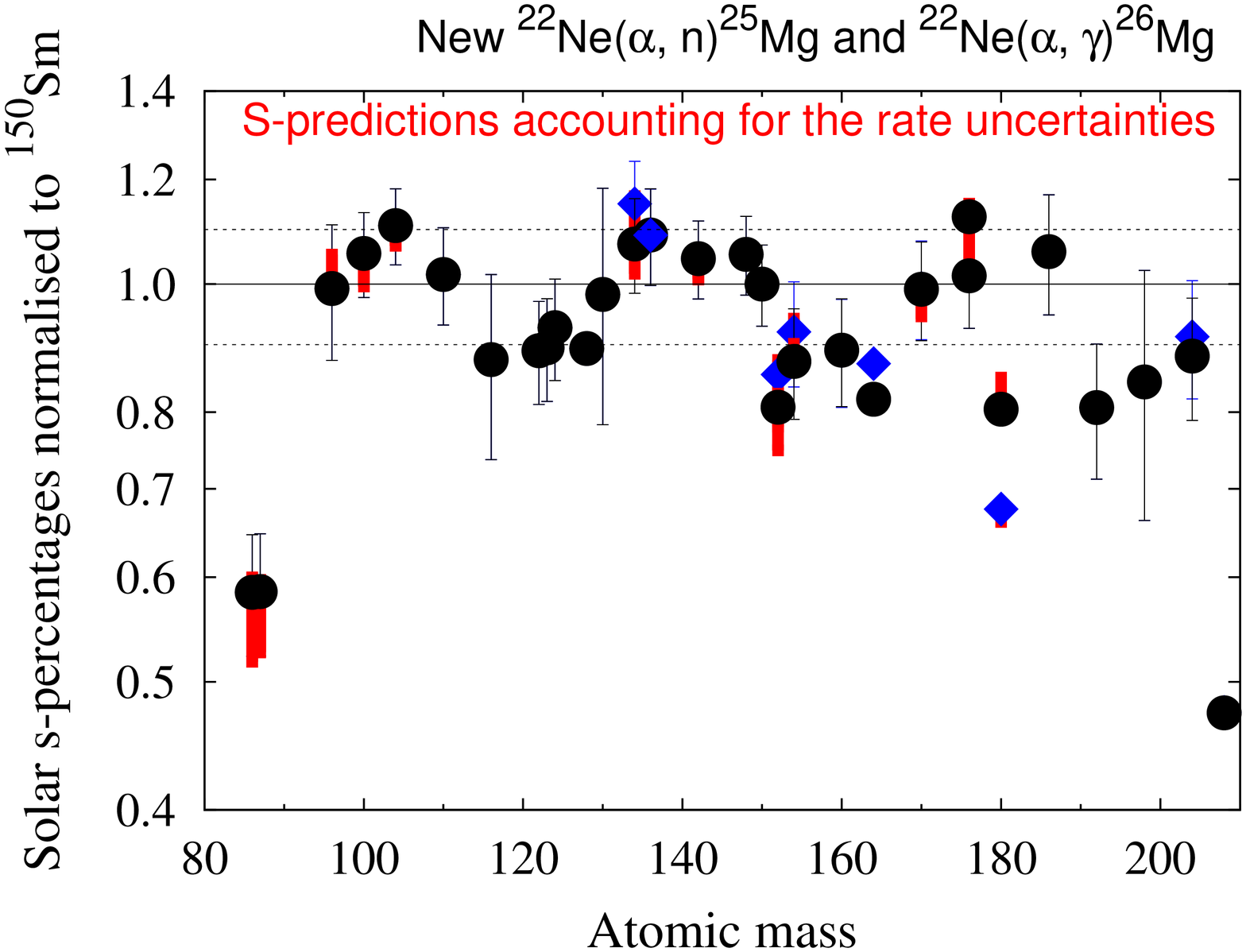}
\caption{The solar main component versus atomic mass (as shown in 
Fig.~\ref{mains}) obtained with the recommended \Nean and \Neag rates
(see Fig.~\ref{newrates}). 
We focus on $s$-only isotopes for clarity ($filled$ $circles$).
Also $^{86,87}$Sr (which receive additional $s$ contributions) and $^{208}$Pb 
(not an $s$-only nuclide) are displayed for completeness.
Thin error-bars account for the uncertainties of the solar abundances by \citet{lodders09}.
Thick bars are obtained by including the uncertainties of the recommended rates discussed
in the text (lower and upper limits) and shown in Fig.~\ref{newrates}.
We indicate the results of an improved treatment of the half-life 
of a few key isotopes strongly sensitive to temperature (and electron density),
which influence the nearby isotopes ($^{134,136}$Ba, $^{152,154}$Gd, $^{164}$Er, 
$^{180}$Ta$^{\rm m}$ and $^{204}$Pb; $filled$ $diamonds$, see text). }
\label{newmains}
\end{figure}

\vspace{2mm}

We have calculated the rates of the \Nean and \Neag reactions (''This work" in 
Table~\ref{tabne22ansigma}; see also Fig.~\ref{newrates}) by accounting for the most recent studies
of all known (directly and/or indirectly detected) resonances as well as nuclear
data available in literature.
\\
 In particular, we adopted the rates calculated on the basis of
experimentally detected levels, i.e. mainly on the basis of the values of \citet{jaeger01}. For
the \Neag rate we added the states which have an indirect determination of
the $\alpha$ width using the $^{22}$Ne($^{6}$Li, d) reaction \citep{giesen93}. In the
temperature range of low mass AGB stars ($T_8$ = 2.5--3), the \Nean recommended
value is about a factor of two lower than that adopted in our models so far, while
the recommended \Neag rate is essentially unchanged.
\\
We have also determined $lower$ and $upper$ $values$ of both rates considering additional
resonances in order to investigate their potential impact (see Fig.~\ref{newrates}, bottom
panel). The $lower$ $limit$ is determined by the lower experimental limit of the 703 keV
resonance (11\%). This reduces the recommended rates by about 10\% at all
temperatures. For the $upper$ $rate$ we introduced all the known (but not directly
measured) states below the 703 keV resonance, adopting a spectroscopical factor of
0.01, which represents a conservative upper limit. Moreover we have estimated a
possible larger effect due to one of the low energy states. We then concentrated on
the lower energy states, which give substantial contributions to the reaction rate
for $T$ $<$ 0.3 GK. This was done to investigate the influence of possible $\alpha$ cluster
states in $^{26}$Mg. These states could have spectroscopic factors as large as 0.1.
Larger values are unlikely because they should have been observed in $\alpha$ transfer
reactions. For the ($\alpha$, $\gamma$) reaction we identified two states, one above (568
keV) and one below (433 keV) the neutron threshold. Also for the ($\alpha$, n) reaction there 
are two states: one at 497 and another at 548 keV.
This corresponds to an increase of the recommended \Nean recommended rate by a factor of
two at $T_8$ = 2.5 and by +30\% at $T_8$ = 3. The recommended \Neag rate increases up to a factor
of seven at $T_8$ = 2.5 and by a factor of three at $T_8$ = 3 (see dashed lines in
Fig.~\ref{newrates},
bottom panel). Note that in network calculation one should not use the upper rate
for both \Nean and \Neag reactions at the same time because the
change of spectroscopic factors has influence on both rates.

\vspace{2mm}

Based on the above results, we have updated the solar main component
discussed in Section~\ref{mains} with the recommended \Nean and \Neag 
rates (see Fig.~\ref{newmains}; $filled$ $circles$).
By comparing these revised results 
with the previous $s$-distribution displayed in Fig.~\ref{mains}, major variations are shown 
close to the branches of the $s$ path.
\\
The impact of the uncertainties evaluated for the recommended \Nean and \Neag rates (Fig.~\ref{newrates})
has a rather small effect on the $s$-distribution (see thick bars in Fig.~\ref{newmains}).
The abundances of $s$-only isotopes are reproduced within the solar uncertainties.
\\
However, it should be noted that the several ambiguities remaining in the nuclear data for the 
$^{22}$Ne + $\alpha$ reaction rates need to be resolved. 
 Indeed, the theoretical \Nean and \Neag rates presented in this work (and all estimations given 
in literature as well) are based on old experimental measurements (e.g., \citealt{jaeger01}) and 
the uncertainties may be underestimated. The presence 
of low energy unknown states, which have been identified in several indirect experiments, makes 
the evaluation at relevant energy still uncertain. New direct experimental investigations 
are needed to measure the resonances with higher accuracy. 
\\
Future investigations are in progress to shed light on this issue
(e.g., via indirect measurements at n$\_$TOF-EAR2, \citealt{chiaveri12}, \citealt{massimi14}; 
or at accelerator 
facilities deep underground where the cosmic-ray background into detectors is reduced by several 
orders of magnitude,
LUNA\footnote{See \citet{costantini09} and references therein; web http://luna.lngs.infn.it/.} 
and DIANA\footnote{See \citet{lemut11}.}).

Our post-process AGB models generally adopt constant $\beta$-decay rates, 
which are based on a geometric average of the rates given by \citet{TY87} at different 
temperatures (and electron densities) over the convective pulse.
Note that this approximation does not affect the prediction of most of the $s$-only branched
isotopes, because the $\beta$-decay rates of close unstable nuclei are almost 
constant in the temperature range of the TP, or they do not compete with neutron capture 
rates.
Relevant exceptions are $^{134}$Ba, $^{152,154}$Gd, $^{164}$Er, $^{176}$Lu/$^{176}$Hf, 
$^{180}$Ta$^{\rm m}$, $^{204}$Pb: in these cases, the $\beta$-decay rates of close unstable 
isotopes (e.g., $^{134}$Cs, 
$^{151}$Sm and $^{154}$Eu, $^{164}$Ho, $^{176}$Lu, $^{180}$Ta$^{\rm m}$, $^{204}$Tl, 
respectively) vary by order(s) of magnitude over the large temperature and density 
gradients that characterise the convective zone (0.2 $\la$ $T_8$ $\la$ 3 and 10 $\la$ $\rho$ 
$\la$ 10$^4$ g/cm$^3$), competing with neutron captures. 
The above branches require an improved treatment of the $\beta$-decay rates over the TP.
\\
The treatment of the branches close to $^{176}$Lu/$^{176}$Hf and $^{180}$Ta$^{\rm m}$ 
was already refined in recent AGB models (see \citealt{heil08LuHf,wisshak06Lu}).
We extend the improvement to the branches close to 
$^{134}$Ba, $^{152,154}$Gd, $^{164}$Er and $^{204}$Pb.
We further implement the $s$ prediction of $^{180}$Ta$^{\rm m}$, by including the same 
treatment to nearby branched isotopes ($^{179}$Ta and $^{179}$Hf).
\\
We provide detailed calculations by dividing each TP in 30 convective meshes of constant mass. 
In each mesh temperature (and density) can be considered constant during each time step.
We follow the production and destruction of the unstable isotopes close to the above $s$-only
nuclides in each mesh with the $\beta$-decay rates computed at each mesh temperature. 
The neutron density resulting from the \Nean reaction 
is computed in the various meshes at each time step. Neutron densities for an efficient 
$s$ process are only reached in the bottom region of the convective zone
(see Fig.~A1, Section~A, Supporting Information).
The abundances resulting in each zone after neutron irradiations are periodically mixed to account 
for the turnover time of the convective zone (of the order of a few hours).
\\
As shown in Fig.~\ref{newmains}, $^{134}$Ba, $^{152,154}$Gd, $^{164}$Er and $^{180}$Ta$^{\rm m}$ 
are mainly affected by the improved treatment of the $\beta$-decay rates (see $filled$ $diamonds$); 
a variation smaller than 5\% is displayed by 
$^{204}$Pb. We will provide a more detailed discussion on these branches 
in Section~\ref{branch} and Section~B (Supporting Information).

\subsection{Tests of the \Nean and \Neag rates}\label{sub1}

As discussed in Section~\ref{22nean}, 
future direct experimental measurements of the \Nean and \Neag reaction rates
may evidence the presence of low energy unknown states. Also the 
contribution of known resonances needs to be determined with high precision.
\\
These considerations encourage us to work in a more conservative range of uncertainty than that estimated in 
 Section~\ref{22nean}:
starting from our recommended \Nean and \Neag rates (hereafter adopted to compute the reference 
$s$ distribution; see Fig.~\ref{newmains}),
we investigate the effects of the following tests on the solar main component:
\vspace*{-2mm}
\begin{itemize}
\item {\bf Test A} -- Recommended \Nean rate multiplied by a factor of 
four, close to the recommended value by \citet{kaeppeler94}, while the \Neag rate is 
almost unchanged. 
\item {\bf Test B} -- Recommended lower limits of the \Nean and \Neag rates.
\item {\bf Test C} -- Recommended \Nean rate and upper limit for the \Neag rate.
This test evaluates the competition between ($\alpha$,
$\gamma$) and ($\alpha$, n) rates in AGB models of low initial mass.
 \end{itemize}

\begin{figure}
\includegraphics[width=0.47\textwidth]{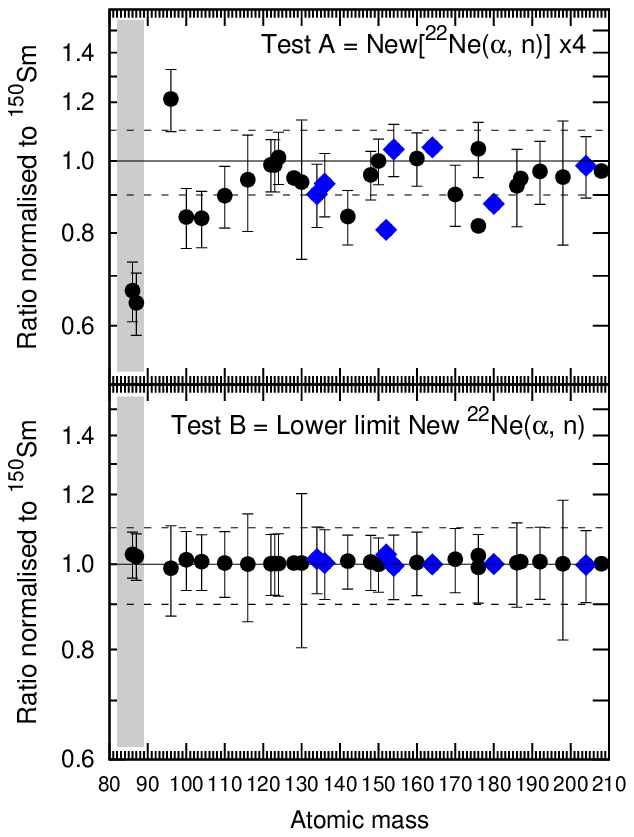}
\caption{Ratios between the main component obtained with 
{\bf Tests A} and {\bf B} and our results obtained with the recommended \Nean and \Neag rates 
(Fig~\ref{newrates}).
We focus on $s$-only isotopes for clarity ($filled$ $circles$).
An improved treatment of the branches close to $^{134}$Ba, $^{152,154}$Gd, $^{164}$Er, 
$^{180}$Ta$^{\rm m}$ and $^{204}$Pb is included ($filled$ $diamonds$).
 A complete version of this Figure (which includes all isotopes from 90 $\leq$ $A$ $\leq$ 210)
is given in Section~A (Supporting Information).
As in Fig.~\ref{fig3A}, the shaded area between 80 $<$ $A$ $<$ 90 indicates the atomic mass region affected by 
additional $s$ contributions. }
\label{fig3}
\end{figure}

In Fig.~\ref{fig3}, we show the effect of {\bf Test A} ($upper$ $panel$) and {\bf Test B} 
($lower$ $panel$) on the solar main component of $s$-only isotopes with $A$ $>$ 80. 
In both panels, a 10\% uncertainty is indicated by dashed lines. 
A detailed list of the results of both tests 
that includes all isotopes with $A$ $\geq$ 70 is given in Table~A1,  
(Supporting Information).
\\
Values are obtained by normalising the $s$-production factors to that of $^{150}$Sm 
in both tests. 
Note that for $^{150}$Sm we obtain an $s$-production factor of X($^{150}$Sm)/X$_{ini}$($^{150}$Sm) 
= 1133.9 with the recommended \Nean and \Neag rates.
{\bf Test A} provides X($^{150}$Sm)/X$_{ini}$($^{150}$Sm) = {1275.5 (+12.5\%), and {\bf Test B} yields 
X($^{150}$Sm)/X$_{ini}$($^{150}$Sm) = 1128.4 ($-$1.0\%).
The $s$ production of $^{150}$Sm increases by +12.5\% with increasing the \Nean by a factor of four.
This difference is independent of any branch (see Section~\ref{sm150}) and affects 
the entire $s$-process distribution in the same way.

As displayed by Test A, the \Nean rate mainly affects $^{86,87}$Sr, which decrease 
by $\sim$30\% being regulated by the branches at $^{85}$Kr and $^{86}$Rb.
We remind that the main component synthesises about half of solar $^{86,87}$Sr (see 
discussion in Section~B and~C, Supporting Information).
\\ 
Noteworthy variations ($\sim$10--20\%) are obtained for
$^{96}$Mo, 
$^{134}$Ba, $^{142}$Nd, $^{152}$Gd, $^{170}$Yb, and $^{176}$Lu.
The $s$ prediction to $^{96}$Mo is affected by the branch at $^{95}$Zr (Sections~\ref{zr95});
the large sensitivity of the branch at $^{134}$Cs to neutron density
modifies the $^{134}$Ba/$^{136}$Ba ratio (up to 10\%; Section~\ref{ba134136}); 
$^{142}$Nd 
is regulated by the branch at $^{141}$Ce 
(Section~B, Supporting Information);
the branches at $^{151}$Sm and $^{154}$Eu influence the $s$ production of 
$^{152,154}$Gd (e.g., variations of 20\% of $^{152}$Gd; Section~B, Supporting Information);
$^{170}$Yb is affected by the branch at $^{170}$Tm (10\%; Section~\ref{yb170}); 
finally, the branch at $^{176}$Lu modifies the $^{176}$Lu/$^{176}$Hf ratio (see 
discussion in Section~B, Supporting Information). 
\\
Note that the $s$ contribution to $^{100}$Ru and $^{104}$Pd also shows important variation, 
although the branches at $^{99}$Mo and $^{103}$Ru (with strongly reduced half-lives at stellar
temperatures) are only marginally open. In both cases, the $\sim$16\% variation is 
mainly explained by the cumulative effect of a different $s$ contribution to $^{150}$Sm, 
adopted to normalise the $s$ distribution.

Other $s$-only isotopes show differences of less than 5\%. 
\\
A comprehensive description of the most important branchings of the main component
is provided in Section~\ref{branch}.
\\
 Note that the variations shown by {\bf Test B} are marginal because our recommended 
\Nean rate lies close to a plausible lower limit ($\sim$ $-$10\%; Section~\ref{22nean}).
Indeed, the dominant effect of the resonance at 800 keV indicates that smaller values are 
unlikely.

\begin{figure}
\includegraphics[width=0.47\textwidth]{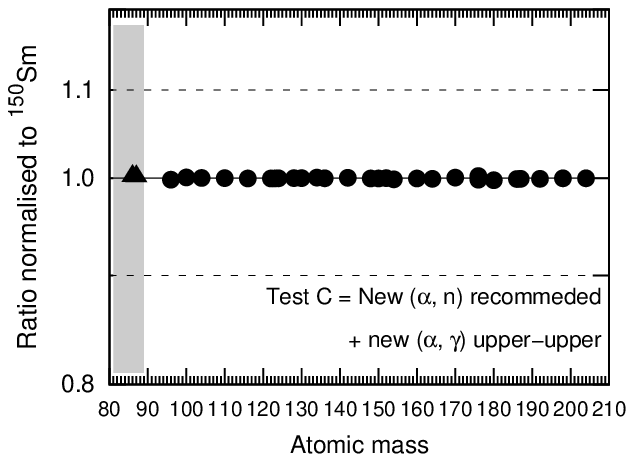}
\vspace{-50mm}
\caption{Ratios between the main component obtained with 
the recommended \Nean and \Neag rates compared to 
{\bf Test C} in which the upper limit for the \Neag rate is adopted,
while the \Nean reaction is unchanged. 
Symbols are the same as Fig.~\ref{fig3}.
A complete version of this Figure (which includes all isotopes from 90 $\leq$
A $\leq$ 210) is given in Section A (Supporting Information).}
\label{fig4}
\end{figure}

\vspace{2mm}

In our AGB models with initial masses 1.5 and 3 \Msun and half-solar metallicity,
the competition between the \Neag and \Nean reactions is marginal:
in spite of the rather generous margins considered for the \Neag rate, {\bf Test C}
confirms that the effect of the \Neag rate for the $s$-abundance distribution is negligible
(see Fig.~\ref{fig4}). 
\\
Appreciable variations are only seen for $^{26}$Mg, directly involved in the ($\alpha$,
$\gamma$) reaction.
Due to its very small MACS (0.126 $\pm$ 0.009 mb at 30 keV, 
\citealt{bao00}), $^{26}$Mg is accumulated in the $s$ process.

\vspace{2mm}

In stellar models with higher initial mass (e.g., IMS stars), where the temperature 
at the bottom of the TP increases sufficiently to efficiently activate both
\Nean and \Neag reactions, their competition becomes important for the nucleosynthesis
of Mg isotopes (see e.g., \citealt{karakas06}, \citealt{longland12}).
Above all, the \Nean rate plays a key role in the production of $s$ isotopes, as
the \Can neutron source is expected to have small or negligible effects in IMS stars
(see Section~\ref{intro}). 
\\
In Section~C (Supporting Information), we discuss the impact of the \Nean and \Neag rates 
on two AGB models: a 5 $M_\odot$ model at half-solar metallicity taken as example of IMS stars
and a 3 $M_\odot$ model at [Fe/H] = $-$1 chosen as representative of low-metallicity models.
\\
In a 5 $M_\odot$ model at half-solar metallicity, the maximum temperature at the bottom 
of the convective TPs is $T_8$ $\sim$ 3.6. In this condition, the \Nean neutron burst efficiently
operates, reaching a neutron density of $N_n$ $\sim$ 10$^{11}$ cm$^{-3}$.
The whole $s$ distribution increases by up to a factor of $\sim$2 by including the upper 
limit of the recommended \Nean rate, and by one order of magnitude with {\bf Test A} (recommended
[\Nean]$\times$4).
Key neutron-rich isotopes ($^{86}$Kr, $^{87}$Rb, $^{96}$Zr) are largely produced being  
the branches at $^{85}$Kr and $^{95}$Zr easily open.
The impact of the \Neag rate on $s$ isotopes remains rather small ($\la$6\%).

\vspace{2mm}

In low-mass AGB stars with [Fe/H] $<$ $-$0.3 the \Nean uncertainty also produces a larger impact
on $s$ predictions than that observed on the solar main component. 
For instance, in a 3 $M_\odot$ model at [Fe/H] = $-$1 the maximum temperature at the bottom of 
the advanced TPs reaches $T_8$ = 3.5. In this case, both \Nean and \Can neutron sources operate 
efficiently. In addition, the $s$ distribution is largely modified by the lower initial metallicity: 
as discussed in Section~\ref{intro}, by decreasing the metallicity 
the abundances of isotopes with neutron magic numbers $N$ = 50 and 82 are progressively overcome 
(thus reducing the whole distribution between 90 $\la$ $A$ $\la$ 130 and 140 $\la$ $A$ $\la$ 204, 
respectively), while $^{208}$Pb is mainly 
produced.
\\
We refer to Section~C (Supporting Information) for major details.


\section{Uncertainties of major branches of the main component}\label{branch}

We distinguish the $s$-only isotopes in three classes,
according to their related unstable branch point isotopes:
\begin{itemize}
\item unbranched $s$-only isotopes (Section~\ref{branchUN}), with unstable isobars having 
half-lives shorter than a couple of days (thus forbidding neutron captures during TPs);
\item $s$-only isotopes sensitive to neutron density only (Section~\ref{branchNn}), with 
unstable isobars having half-lives (almost) constant in stellar environments; 
\item $s$-only isotopes affected by branch point with unstable isobars having $\beta$-decay 
rates quickly changing under stellar conditions: we distinguish between nuclides sensitive 
to both neutron density and stellar temperature (and/or electron density; Section~\ref{branchT}) 
and nuclides less affected by neutron density, but dominated by stellar temperature and/or 
electron density gradients during TP (Section~\ref{branchMIX}).
\end{itemize}


\subsection{Unbranched $s$-only isotopes}\label{branchUN}

As mentioned in Section~\ref{mains}, a few $s$-only isotopes of the $s$ path ($^{100}$Ru, %
$^{110}$Cd, $^{116}$Sn, 
$^{124}$Te, $^{150}$Sm, $^{160}$Dy, see \citealt{arlandini99}; as well as $^{104}$Pd and 
$^{198}$Hg) are marginally affected by nearby branch points with short half-lives.
Thus, only a few percent of the $s$ path ($<$3\%) bypasses the above $s$-only nuclides, which 
become useful constraints for the main component.


\subsubsection{The $s$-only isotope $^{150}$Sm}\label{sm150}

\begin{figure}
\includegraphics[angle=0,width=8cm]{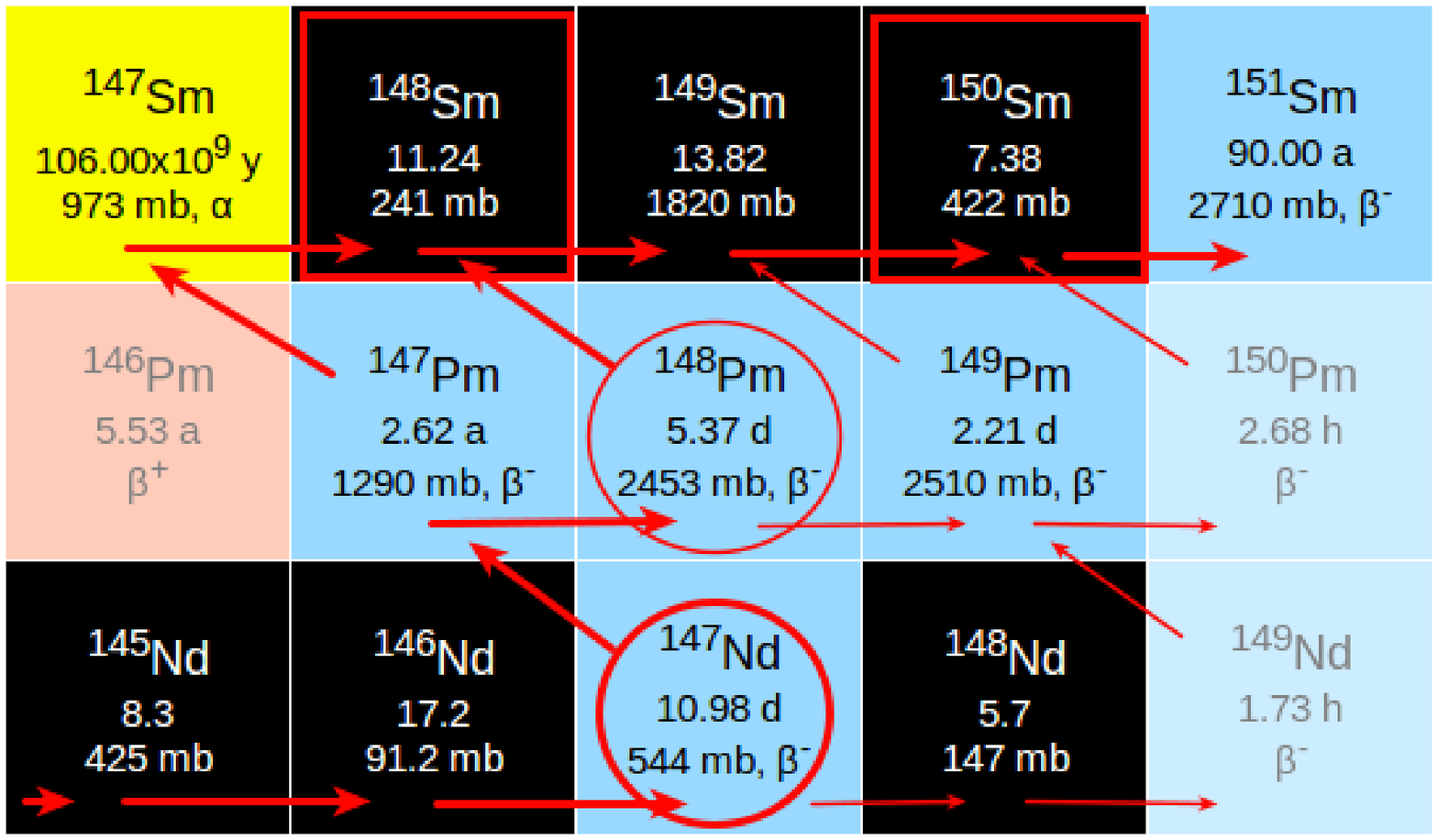}
\caption{Schematic representation of the branches close to the
$s$-only isotopes $^{148,150}$Sm (red squares). 
Thick lines represent the $s$-process
nucleosynthesis during the \Can neutron irradiation, while thin lines correspond
to the neutron capture channels open by the marginal
activation of the \Nean reaction during TPs. Major branches at $^{147}$Nd and $^{148}$Pm,
which regulate the $s$ contribution to $^{148}$Sm, are highlighted by circles. 
All branches of the $s$ path join at $^{150}$Sm, which is adopted to normalise the 
overall $s$ distribution.
(\textit{This and the following Figures are adapted from http://www.kadonis.org/. See the 
electronic edition of the Journal for a colour version of the Figures.})}
\label{branchpm147}
\end{figure}

\begin{table*}
\caption{The $s$-production factors of $^{148}$Sm and $^{150}$Sm and their relative ratio (which
corresponds to the $s$ contribution to solar $^{148}$Sm) for different choices of the \Nean rate.}
\label{sm148150nean}
\centering
\resizebox{18cm}{!}{\begin{tabular}{l|cccc}
 \hline
$s$-production factors & Recomm. \Nean rate $(^a)$ &  {\bf TEST A }      &   {\bf TEST  B}      & {[\Nean]}=0 $(^b)$ \\
                      &  &Recomm. $^{22}$Ne($\alpha$, n)$\times$4&Recomm. $^{22}$Ne($\alpha$, n)$\times$0.9 &  \\
\hline
$^{148}$Sm             &  1194.1        &   1286.9 (+7.8\%)   &   1195.4 ($<$1\%) &  1258.5 (+5.4\%)   \\
$^{150}$Sm             &  1133.9        &   1275.5 (+12.5\%)  &   1128.4 ($<$1\%) &  1088.4 ($-$4\%) \\
\hline
$^{148}$Sm/$^{150}$Sm (in \%)  &   1.05     &  1.01 ($-$4\%)  &  1.06 (+1\%)  &  1.156  (+10.1\%) \\
\hline
\multicolumn{5}{l}{$(^a)$ Results obtained with our adopted \Nean rate (the same as Fig.~\ref{newmains}).}\\
\multicolumn{5}{l}{$(^b)$ Results with the contribution of the \Can neutron source alone.}\\
\hline
 \end{tabular}}
\end{table*}

$^{150}$Sm is one of the few $s$-only isotopes that is exposed to the full $s$ flow, 
because all branches in the Nd-Pm-Sm region join at $^{149,150}$Sm (see Fig.~\ref{branchpm147}).
Indeed, the short half-lives of $^{149}$Nd ($t_{1/2}$ = 2.21 d) and $^{150}$Pm ($t_{1/2}$ 
= 2.68 h) leave $^{150}$Sm virtually unbranched.
\\
As anticipated in Section~\ref{mains}, $^{150}$Sm is particularly suited to normalise the
overall $s$-process predictions.
We select $^{150}$Sm because, besides being an unbranched $s$-only isotope, $^{150}$Sm 
has very well known MACS 
($\sim$1\% uncertainty, $\sigma$[$^{150}$Sm(n, \g)] = 422 $\pm$ 4 mbarn; KADoNiS) and 
solar abundance (5\% uncertainty, 0.265 $\pm$ 0.013 number of atoms per 10$^6$ Si atoms; 
\citealt{lodders09}).

Unlike $^{150}$Sm, 
the $^{148}$Sm abundance and, thus, the $^{148}$Sm/$^{150}$Sm ratio are otherwise 
regulated by the branches at $^{147}$Nd, $^{147}$Pm, and $^{148}$Pm. 
\\
During the $^{13}$C-pocket phase, the neutron density is not sufficient 
to bypass $^{148}$Sm at the branching points $^{147}$Nd ($t_{1/2}$ = 11 d) and $^{148}$Pm ($t_{1/2}$ = 
5.37 d) efficiently, whereas the branch at $^{147}$Pm ($t_{1/2}$ = 2.62 yr) is partially activated
(at $N_n$ = 1$\times$10$^7$ cm$^{-3}$, $\sim$20\% of 
the $s$ path feeds $^{148}$Pm, bypassing the long-lived $^{147}$Sm). 
\\
On the other hand, the $^{147}$Nd and $^{148}$Pm neutron capture channels are activated during the 
\Nean irradiation.
At the beginning of the TP, the neutron density strongly increases and $^{148}$Sm 
is largely bypassed by the $s$ path. 
Under these conditions, the amount of $^{148}$Sm produced during the previous $^{13}$C-pocket
phase is progressively depleted but starts to recover as the neutron density falls during the TP and is
almost completely restored at the end of the TP ($\sim$94\% of the value produced at the end of
the previous interpulse is reestablished).
In Fig.~\ref{stampesm150}, we plot the temporal evolution of the neutron density and of the isotopic
abundances of $^{147,148,149,150}$Sm, $^{147,148}$Nd and $^{147,148,149}$Pm  
during the 15$^{\rm th}$ He shell flash for an AGB model with 1.5 $M_\odot$ and half-solar metallicity. 
The final $^{148}$Sm abundance is predominantly determined by the freeze-out of 
the neutron supply, intuitively defined by the time when 
an isotopic abundance reaches 90\% of its final value \citep{arlandini99,cosner80}. 
The abundance of a branched nucleus is frozen when the 
probability of further neutron captures (for the nearby unstable isotope) is marginal. 
Thus, the starting moment of the freeze-out depends on the MACS of the unstable isotope: 
the larger the MACS is, the later the freeze-out occurs.
\\
The branch at $^{147}$Pm ($t_{1/2}$ = 2.62 yr, which following \citealt{TY87} is reduced 
to 1.2 yr at $T_8$ = 3) mainly affects the production of $^{147}$Sm.
The remaining $^{147}$Pm abundance decays eventually into $^{147}$Sm at the end of the 
TP.

\begin{figure}
\vspace{5 mm}
\includegraphics[angle=-90,width=9cm]{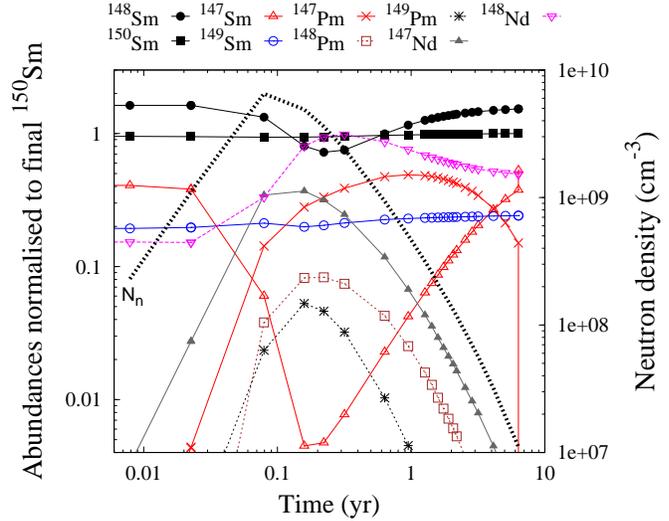}
\caption{Temporal evolution of the neutron density and of the isotopic abundances of 
$^{147,148,149,150}$Sm, $^{147,148}$Nd and $^{147,148,149}$Pm during the 15$^{\rm th}$ He shell
flash in an AGB star with 1.5 $M_\odot$ and half-solar metallicity. The abundance values are given as number
fractions normalised to $^{150}$Sm at the end of the He 
shell flash. The time-scale starts when the temperature at the bottom of the convective TP reaches 
$T_8$ = 2.5, which corresponds to the onset of the \Nean reaction. The initial abundances are
those left from the previous $^{13}$C-pocket phase.}
\label{stampesm150}
\end{figure}

\vspace{2mm}

The main component reproduces the solar abundance of $^{148}$Sm (+5\% with respect
to solar), in  
agreement within the solar uncertainty.

As for $^{150}$Sm, also the MACS of $^{148}$Sm is accurately known (241 $\pm$2 mbarn at 30 keV, KADoNiS; 
$<$1\%).
\\
While the short half-lives of $^{147}$Nd and $^{148}$Pm are prohibitive for measuring their 
(n, \g) cross sections with present techniques, the MACS of $^{147}$Pm could be successfully 
determined via the activation method (709 $\pm$ 100 mbarn, 14\%; \citealt{reifarth03}; KADoNiS).
Uncertainties of $\sim$17\% have been estimated for the calculated MACS values of $^{147}$Nd
and $^{148}$Pm (at 30 keV, $\sigma$[$^{147}$Nd(n, \g)] = 544 $\pm$ 90 mb, KADoNiS; 
$\sigma$[$^{148}$Pm(n, \g)] = 1014 $\pm$ 175 mbarn, \citealt{reifarth03}).
The MACS uncertainties of the branch point isotopes $^{147}$Nd, $^{147}$Pm and $^{148}$Pm 
have small effects on $^{148}$Sm ($<$4\% in total).

Although \Can is the major source of neutrons in low-mass AGB stars, the $^{148,150}$Sm 
$s$ abundances are slightly influenced by the efficiency of the \Nean reaction: the $s$ 
production of $^{150}$Sm increases by 4\% by including the recommended \Nean rate, and rises up to 
+12.5\% with Test A, while the relative contribution $^{148}$Sm/$^{150}$Sm shows up to $\sim$4\% 
variations (see Table~\ref{sm148150nean}). 
This variation reflects a more (or less) efficient $s$-process contribution during the TPs
neutron burst independently of any branch, and affects the entire $s$-process 
distribution. On the other hand, the \Nean neutron source is crucial for regulating the 
$^{148,150}$Sm abundances: the solar $^{148}$Sm/$^{150}$Sm ratio would be overestimated by 
$\sim$16\% by excluding its additional neutron irradiation (see column~5 of Table~\ref{sm148150nean}).


\subsubsection{Additional unbranched isotopes}

Besides $^{150}$Sm (Section~\ref{sm150}), $^{100}$Ru, $^{104}$Pd, $^{110}$Cd, 
$^{116}$Sn, $^{124}$Te,  
$^{160}$Dy, and $^{198}$Hg are noteworthy $s$-only isotopes essentially unaffected 
by branches, and therefore also important for characterising
the entire $s$-process distribution.
\\
At TP temperatures, the $\beta$-decay half-lives of their potential branch point
isotopes are of the order of a couple of days, so that their decay rates clearly dominate over the
respective neutron capture rates (compare e.g., the stellar half-lives of 
$^{99}$Mo, 
$^{103}$Ru, $^{110}$Ag, $^{115}$Cd, 
$^{122}$Sb, $^{160}$Tb, 
and $^{198}$Au; \citealt{TY87}). 
\\
Accordingly, the variations of the $s$ contributions to $^{100}$Ru, $^{104}$Pd, and $^{110}$Cd 
in Fig.~\ref{fig3} include the cumulative effect of the normalisation to $^{150}$Sm 
(Section~\ref{sm150}). 
The most well known isotopes are 
$^{110}$Cd, 
$^{124}$Te, 
and $^{150}$Sm, 
which have 5--7\% solar uncertainty and MACS values are known at $\sim$1\%.
Less accurate MACS data with uncertainties of 6 and 10\% are affecting the
$s$ prediction of $^{100}$Ru and $^{104}$Pd, 
although their solar abundances are well determined.
\\
\citet{utsunomiya13} have indirectly determined the $^{99}$Mo MACS via the inverse (\g, n) reaction.
Although their estimated value at 30 keV is significantly larger than the theoretical MACS recommended
by KADoNiS (410 instead of 240 mb), the effect on the $s$ prediction for $^{100}$Ru is only $\sim$1\%.
\\
The $s$ abundances of $^{116}$Sn and $^{124}$Te are reproduced in agreement within the solar uncertainties
(15 and 7\%, respectively; \citealt{lodders09}).

The $s$ abundances of $^{160}$Dy and $^{198}$Hg are difficult to
determine as discussed in detail in Section~B (Supporting Information). 
\\
About 10\% of solar $^{160}$Dy
is missing. This dearth may be reconciled within the uncertainties: 
although
the (n, \g) cross section of $^{160}$Dy is very well determined in laboratory 
($\sigma$[$^{160}$Dy(n, \g)] = 890 $\pm$ 12 mbarn at 30 keV, 1.4\%; KADoNiS), 
the MACS remains rather uncertain in this case due to a significant stellar 
enhancement factor (SEF = 1.12 at 30 keV; KADoNiS). This correction may
provide the 10\% missing $s$-process contribution to solar
$^{160}$Dy. 
The solar abundance of Dy is instead well determined (5\% by \citealt{lodders09}).
\\
The $s$ contribution to solar $^{198}$Hg is very uncertain, mainly 
because Hg is a volatile element, and its solar abundance is affected by 20\% 
uncertainty \citep{lodders09}.
An additional 8.7\% uncertainty derives from the $^{198}$Hg MACS (173 $\pm$ 15 mbarn; 
dated back to \citealt{beer85}; KADoNiS).
\\
Note that the large uncertainties associated to the $^{116}$Sn  
and $^{198}$Hg $s$-contributions derive from the poorly known Sn and Hg
solar abundances.


\subsection{The $s$-only isotopes sensitive to the neutron density}\label{branchNn}

In this Section, we analyse the $s$ contribution to $^{96}$Mo,  
which is influenced by the branch at $^{95}$Zr. This branching depends only on the 
neutron density because the $\beta$-decay rate of $^{95}$Zr remains constant at the 
relevant $s$-process temperatures.
\\
The $s$-only isotopes $^{170}$Yb, $^{142}$Nd, $^{186}$Os and $^{192}$Pt belong to 
this class of branchings as well. 
Although the half-lives of their related branch point nuclides exhibit a marginal 
sensitivity to stellar temperature, the respective MACS uncertainties still dominate
over the temperature effects.


\subsubsection{The $s$-only isotope $^{96}$Mo (the branch at $^{95}$Zr)}\label{zr95}

In this atomic mass region, the long-lived $^{93}$Zr behaves as a stable isotope during the 
main $s$ process ($t_{1/2}$ = 1.5$\times$10$^6$ yr; Fig.~\ref{branchzrnbmo}). 
The $s$ path directly feeds $^{94}$Zr, bypassing $^{92,94}$Mo (two $p$-only isotopes, 
mainly destroyed during AGB nucleosynthesis).
The radioactive decay of $^{93}$Zr into $^{93}$Nb 
occurs after the end of the
TP-AGB phase, when the $s$-process nucleosynthesis stops (e.g., \citealt{wallerstein88}).
This decay produces a decrease of the relative [Zr/Nb] ratio. Similarly to the discovery of Tc in 
the spectrum of an AGB star of spectral type S \citep{merrill52,lambert95}, the [Zr/Nb] ratio supplies 
spectroscopic information about the synthesis of heavy elements in evolved stars (see, e.g., 
\citealt{ivans05,kashiv10}).

The abundance of $^{96}$Mo is dominated by the branch point at $^{95}$Zr ($t_{1/2}$ = 64.03 d). 
At the low neutron density reached during the $^{13}$C-pocket phase ($\sim$10$^{7}$ cm$^{-3}$) 
most of the $s$ flow proceeds towards $^{95,96}$Mo, but moderate neutron captures on $^{95}$Zr 
are allowed at the peak neutron density achieved during TPs ($N_n$ $\ga$ 1$\times$10$^{9}$ 
cm$^{-3}$; $f_n$ $\ga$ 0.1). Under these conditions the $s$ path feeds the neutron-rich $^{96}$Zr
(about 3\% of solar Zr), and $^{95,96}$Mo are partially bypassed. 
\\
Fig.~\ref{stampezr95} shows the temporal evolution of the neutron density and of the isotopic 
abundances of $^{95,96}$Mo, $^{95,96}$Zr and $^{95}$Nb during 
the 15$^{\rm th}$ He shell flash in an AGB star with 1.5 $M_\odot$ and half-solar metallicity.
At the peak neutron density $^{96}$Mo is depleted, but as soon as the neutron density decreases, 
$^{96}$Mo starts rising again to exceed the values at the end of the previous interpulse by 18\%.
\\ 
While $^{96,97}$Mo and $^{96}$Zr are only influenced by the branch at $^{95}$Zr, $^{95}$Mo is 
mostly bypassed by the $s$ path owing to the branch at $^{95}$Nb ($t_{1/2}$ = 34.99 d;
$N_n$ $>$ 4$\times$10$^9$ cm$^{-3}$, $f_n$ $>$ 0.5).

\begin{figure}
\includegraphics[angle=0,width=8.35cm]{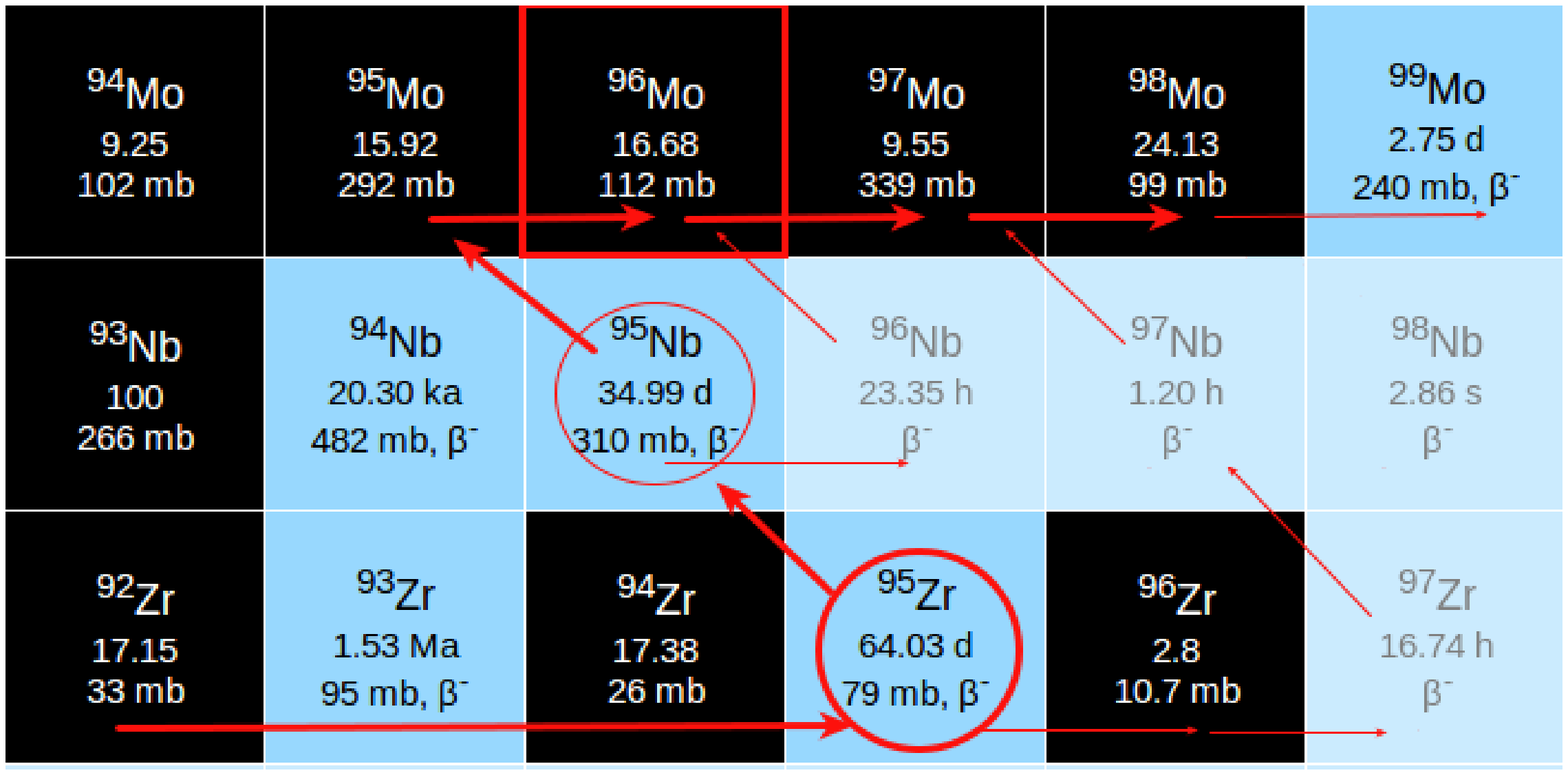}
\caption{The same as Fig.~\ref{branchpm147}, but for the $s$-path region close to the $s$-only 
isotope $^{96}$Mo (red rectangle). While $^{93}$Zr is practically
stable on the time scale of the $s$ process, $^{95}$Zr acts as the main branching
point.}
\label{branchzrnbmo}
\end{figure}

\begin{figure}
\vspace{5 mm}
\includegraphics[angle=-90,width=9cm]{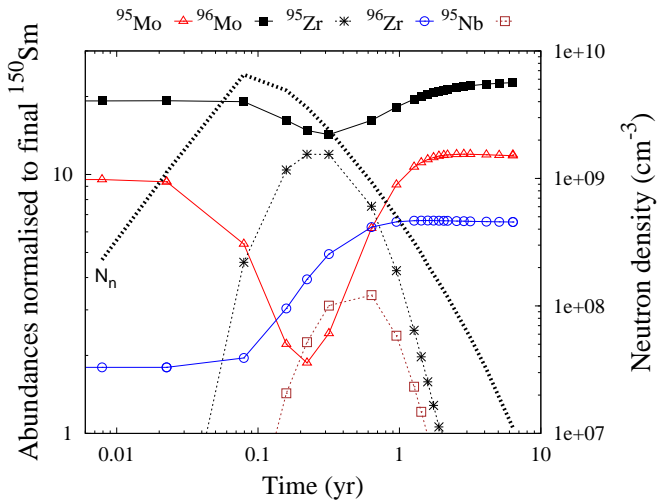}
\caption{The same as Fig.~\ref{stampesm150}, but for the isotopic abundances of 
 $^{95,96}$Mo, $^{95,96}$Zr and $^{95}$Nb. }
\label{stampezr95}
\end{figure}

The $^{95}$Zr MACS is largely uncertain, and discrepant values are found in literature:
KADoNiS recommends the value estimated by \citet{bao00} (79 $\pm$ 12 mb at 30 keV), with a 
rather small uncertainty (15\%), whereas \citet{touk90} estimated a value of 50 mb at
$kT$ = 30 keV.
Recently,
\citet{lugaro14} provided a value
that was 50\% lower than that of \citet{touk90} and about three times lower
than in KADoNiS. Accordingly, the MACS of $^{95}$Zr is still affected by an
uncertainty of about a factor of two.
\\
In our AGB models, a value close to that suggested by \citet{touk90} was adopted so far 
for the $^{95}$Zr MACS.

In Table~\ref{tab8}, we list the $s$ contributions to isotopes from $^{94}$Zr 
to $^{98}$Mo. By including the old \Nean rate, the $s$ contribution to solar $^{96}$Mo
was overestimated by +4\% (column~2; which include our guess on the $^{95}$Zr MACS); this
value slightly increases to +8\% by adopting the new theoretical 
$^{95}$Zr MACS evaluated by \citet{lugaro14}, (column~3). 
\\
Updated $s$ predictions (which include the recommended \Nean rate and the $^{95}$Zr MACS 
by \citealt{lugaro14}) reproduce 99\% of solar $^{96}$Mo.
Note that the solar $^{96}$Mo is largely overestimated (+20\%) with increasing the recommended \Nean rate by 
factor of four (Test A). This exceeds the solar Mo uncertainty (10\%; \citealt{lodders09})
and the $^{96}$Mo MACS uncertainty (7\%; 112 $\pm$ 8 mbarn at 30 keV; KADoNiS),
which dated back to an earlier measurement by \citet{winters87}.
\\
Major changes are shown by $^{96}$Zr.
Although $^{96}$Zr is a neutron-rich isotope, it receives an important $s$ contribution 
during the \Nean neutron burst in AGB stars: once built-up by the activation of the branch 
at $^{95}$Zr, $^{96}$Zr is only marginally destroyed by neutron captures owing to its small 
MACS (10.3 $\pm$ 0.5 mb at 30 keV; \citealt{tagliente11a}). 
Starting from \citet{arlandini99}, the main component was known to produce about half 
of the solar $^{96}$Zr. Present calculations obtained with the recommended \Nean rate 
reduce this contribution from 50\% to 14\%.
This value is extremely sensible to the \Nean rate: {\bf Test A} yields 
an increase by a factor 6 (87\%).
Moreover, a factor of two uncertainty is still associated to the theoretical $^{95}$Zr MACS.

We remind that $^{96}$Zr is a well known indicator of the initial AGB mass. 
An additional $s$ contribution to $^{96}$Zr
comes from IMS AGB stars \citep{travaglio04}.
Indeed, although IMS stars marginally produce heavy $s$ isotopes, the strong neutron
density reached during TPs ($N_n$ $\sim$ 10$^{11}$ cm$^{-3}$) may significantly
increase the abundance of $^{96}$Zr (see Section~C, Supporting Information).

\begin{table*}
\caption{ Old solar main component isotopic percentage contributions from
$^{94}$Zr to $^{98}$Mo (column~2; computed with our estimated $^{95}$Zr MACS, close 
to that recommended by \citet{touk90}; "OLD"), compared with our previous calculations
(column~3; which include the $^{95}$Zr MACS by \citealt{lugaro14}; "L14") and with 
$s$ predictions obtained by assuming the $^{95}$Zr MACS by KADoNiS (column~4; "KAD").
In column~5 we list the updated main component obtained with the recommended
\Nean rate (Fig.~\ref{newmains}). In column~6 we provide the results of Test A in 
Section~\ref{sub1}, in which we multiplied by a factor of four the recommended \Nean rate.
Only variations larger than 5\% are given in brackets.}
 \label{tab8}
\resizebox{13cm}{!}{\begin{tabular}{l|cccccc}
\hline    
    & & \multicolumn{5}{l}{Tests of the $^{95}$Zr(n, \g)$^{86}$Kr MACS}  \\
          \hline
Isotope    &     OLD                 &   L14         &  KAD        & & L14  &  L14    \\  
           &\multicolumn{3}{c}{ {\bf [Old $^{22}$Ne($\alpha$,n)]}} & &{\bf [Recomm. $^{22}$Ne($\alpha$,n)]}&{\bf [Recomm. $^{22}$Ne($\alpha$,n)$\times$4]}\\
\hline
$^{94}$Zr  & 105.4        &105.5   &   105.4          & & 107.4        &  104.2        \\
$^{96}$Zr  &  50.0 (1.40) & 35.6   &    84.3 (2.37)   & & 14.3 (0.40)  &   87.4 (2.46) \\
$^{93}$Nb  & 71.5         & 71.5   &    71.6          & & 69.2         &   73.5        \\
$^{95}$Mo  & 60.8         & 62.8   &    56.2 (0.90)   & & 60.4         &   61.7        \\
$^{96}$Mo  & 104.0        &108.0   &    94.8 (0.88)   & & 99.2 (0.92)  &  120.5 (1.12) \\
$^{97}$Mo  & 55.0         & 56.3   &    51.9 (0.92)   & & 53.4         &   65.3 (1.16) \\
$^{98}$Mo  & 69.6         & 70.3   &    67.8          & & 74.7 (1.06)  &   71.0        \\
\hline
\end{tabular}}
\end{table*}


\subsubsection{The $s$-only isotope $^{170}$Yb (the branches at $^{169}$Er and $^{170}$Tm)}\label{yb170}

\begin{figure}
\includegraphics[angle=0,width=8cm]{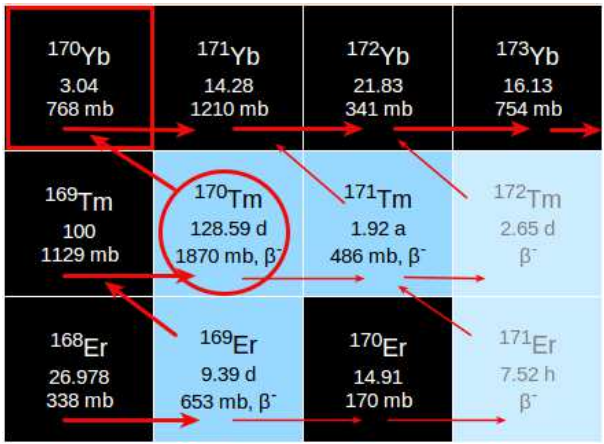}
\caption{The same as Fig.~\ref{branchpm147}, but for the $s$ path close to the $s$-only isotope 
$^{170}$Yb (red rectangle).
The branch at $^{170}$Tm (marked by the circle) is activated mainly during TPs.}
\label{branchertm}
\end{figure}

\begin{figure}
\vspace{5 mm}
\includegraphics[angle=-90,width=9cm]{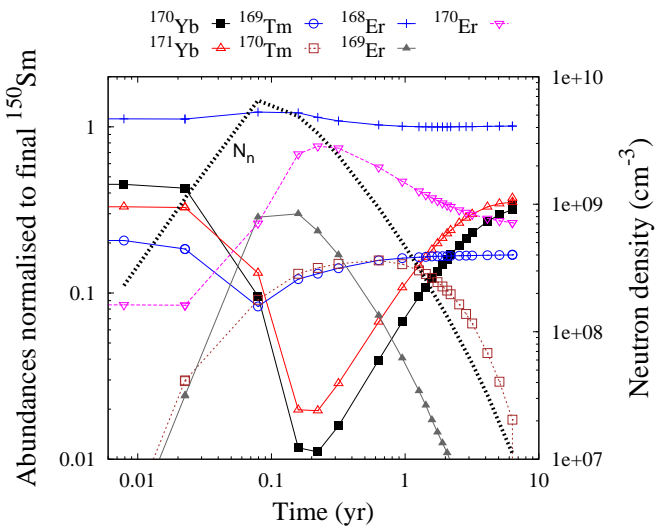}
\caption{The same as Fig.~\ref{stampesm150}, but for the isotopic abundances of $^{170,171}$Yb,
$^{169,170}$Tm and $^{168,169,170}$Er. }
\label{stampeertmyb}
\end{figure}

The abundance of $^{170}$Yb is regulated by the branches at $^{170}$Tm 
and to a minor extent at $^{169}$Er (whose terrestrial half-life $t_{1/2}$ = 9.4 d decreases to $t_{1/2}$ 
$\sim$ 7.5 d at $T_8$ = 3; \citealt{TY87}). Neutron captures on $^{169}$Er are only relevant
at high neutron densities ($N_n$ $\ga$ 3$\times$10$^9$ cm$^{-3}$; $f_n$ $\ga$ 0.3).
\\
The $s$ contribution to $^{170}$Yb is essentially determined by the branch at $^{170}$Tm
due to its comparably long half-life ($t_{1/2}(\beta^-)$ = 128.59 d,
almost independent of temperature). $^{170}$Yb is efficiently bypassed ($f_n$ $\ga$ 0.5) by
the $s$-process flow for neutron densities larger than about 1.3$\times$10$^8$ cm$^{-3}$ when
neutron captures prevail over $\beta$ decays.
Fig.~\ref{stampeertmyb} shows the temporal evolution of the neutron density and of the 
isotopic abundances of $^{170,171}$Yb, $^{169,170}$Tm and $^{168,169,170}$Er during the 
15$^{\rm th}$ He shell flash in a 1.5 $M_\odot$ model and half-solar metallicity. 
$^{170}$Yb is strongly depleted during the peak neutron density, when both $^{169}$Er and 
$^{170}$Tm branches are activated and the $s$ path proceeds mainly via $^{171}$Tm ($t_{1/2}$ 
= 1.92 yr), thus bypassing $^{171}$Yb as well.
Most of $^{170}$Yb is restored as the neutron density decreases ($\sim$74\% of 
the $^{170}$Yb abundance produced during the previous $^{13}$C pocket is reestablished).
Note that the $^{170}$Tm $\beta^+$-decay channel to $^{170}$Er is negligible \citep{TY87}.

The main component reproduces 99\% of solar $^{170}$Yb. About 10\% of solar 
$^{170}$Yb is missing with increasing the recommended \Nean rate by a factor of four.
The MACS of $^{170}$Yb is very well determined with less than 1\% uncertainty 
(768 $\pm$ 7 at 30 keV; KADoNiS).
The major uncertainty that affects the abundance of $^{170}$Yb is related to its SEF 
($\sim$1.08 at $kT$ = 23 keV; KADoNiS).
\\
Other uncertainties derive from the theoretical MACS of $^{170}$Tm (1870 $\pm$ 330 at 30 keV; 
17.6\%), which produces up to 6\% variation on the $^{170}$Yb $s$-predictions, 
and from the $^{170}$Tm $\beta^-$-decay rate, for which \citet{goriely99} estimated 
up to a factor of 1.7 uncertainty at $T_8$ = 3, affecting the $s$ contribution to $^{170}$Yb
by $\sim$4\%.


\subsubsection{Additional branches sensitive to the neutron density}

 Besides $^{96}$Mo and $^{170}$Yb, we underline $^{142}$Nd, $^{186}$Os and $^{192}$Pt
among the isotopes mainly sensitive to the \Nean neutron burst.

Similar to the branch at $^{95}$Zr, the $\beta$-decay rates of $^{85}$Kr and $^{86}$Rb
are constant at stellar temperature, and the $s$ contributions to $^{86,87}$Sr are mainly
sensitive to the neutron density reached during the \Nean irradiation.
The effect of the above branchings on the abundances of $^{86,87}$Sr as well as on the 
neutron-magic nuclei $^{86}$Kr and $^{87}$Rb during the $^{13}$C-pocket and TP phases 
are discussed in Section~B (Supporting Information).

The solar abundance of the neutron-magic isotope $^{142}$Nd is affected by the branches at 
$^{141}$Ce and $^{142}$Pr (partially open during TP), and decreases by $\sim$15\% with 
increasing the recommended \Nean rate by a factor of four.
\\
While the solar abundances of $^{96}$Mo, $^{170}$Yb and $^{142}$Nd are plausibly 
represented by the main component, the interpretation of solar $^{186}$Os and $^{192}$Pt 
is more problematic.
\\
The studies carried out by \citet{mosconi10} and \citet{fujii10} have substantially
increased the accuracy of the $^{186}$Os MACS (4.1\%). 
The $s$ abundance of $^{186}$Os is mainly influenced by the branch at $^{185}$W, while smaller 
variations are produced by the branch at $^{186}$Re ($\beta$ decay rather constant at stellar 
temperatures).
Although the half-life of $^{185}$W is slightly sensitive to temperature and electron
density ($t_{1/2}$ = 75.10 d decreases to 50 d at $T_8$ = 3; \citealt{TY87}) the $s$ abundance 
of $^{186}$Os is dominated by the uncertainty of the theoretical MACS of $^{185}$W.
 The present $^{185}$W(n, \g) cross section overestimates the solar $^{186}$Os abundance 
by 20--30\%, hardly to be reconciled with the 8\% uncertainty of the solar abundance 
\citep{lodders09}. The accuracy of the $^{185}$W MACS recommended by KADoNiS (9\%) could be 
largely underestimated (see \citealt{rauscher14}), being based on an average among inverse 
(\g, n) reactions \citep{sonnabend03,mohr04,shizuma05}. 
In order to reproduce the solar abundance with our AGB models, we adopt a MACS for $^{185}$W 
which is about 80\% higher than recommended in KADoNiS.
\\
Present calculations underestimate the solar $^{192}$Pt abundance by $\sim$20\%.
The abundance of $^{192}$Pt is mainly determined by the branch at $^{192}$Ir.
Similar to $^{186}$Os, the $s$ contribution of $^{192}$Pt is also dominated by the uncertainty 
of the neutron capture channel rather than by the $\beta$-decay branch (slight reduction 
of the half-life of $^{192}$Ir from 77.54 d to 55.33 d at $T_8$ = 3; \citealt{TY87}).
The MACS measurements on the Pt isotopes by \citet{koehler13} included also a new calculation 
of the values for $^{192}$Ir. The discrepancy is reduced by accounting of a 20\% 
uncertainty of the theoretical $^{192}$Ir MACS, which increases the $^{192}$Pt $s$-prediction 
to $\sim$85\%. 
However, a 15\% missing contribution scarcely agrees with the much improved accuracy of the 
$^{192}$Pt MACS measured by \citet{koehler13} (4\%), and with an 8\% solar uncertainty 
\citep{lodders09}. A more detailed theoretical analysis of the branch 
at $^{192}$Ir would help to improve the $^{192}$Pt $s$-prediction.
\\
In Section~B (Supporting Information), the uncertainties affecting the $s$ predictions of 
$^{142}$Nd, $^{186}$Os and $^{192}$Pt are analysed in detail.

While the $s$ contribution of $^{192}$Pt is marginally influenced by the \Nean 
uncertainty, the present $s$ contributions to solar $^{96}$Mo and $^{170}$Yb better agree with our newly 
evaluated \Nean rate. However, the MACS of $^{170}$Yb is affected by a non-negligible 
SEF correction (1.15 at 30 keV; KADoNiS) although its neutron capture cross section is well 
determined experimentally (see also \citealt{rauscher11}).


\subsection{The $s$-only isotopes strongly sensitive to stellar temperature 
(and/or electron density) and neutron density}\label{branchT}
 
The $\beta$-decay rates of several unstable isotopes exhibit a strong dependence
on temperature and electron density \citep{TY87}. 
As anticipated in Section~\ref{newmains}, the uncertainty affecting the stellar $\beta$-decay rate of 
a few branches may produce wide variations of the $s$ predictions.

The branches at issue are $^{134}$Cs (affecting the $s$ contribution to solar $^{134}$Ba), 
$^{151}$Sm and $^{154}$Eu (which influence the $^{152,154}$Gd abundances),
$^{176}$Lu (making it an $s$-process thermometer rather than a cosmic clock),
and $^{204}$Tl (with consequences for the $s$ contribution of $^{204}$Pb).


\subsubsection{The $s$-only pair $^{134,136}$Ba (the branch at $^{134}$Cs)}\label{ba134136}

\begin{figure}
\includegraphics[angle=0,width=8.35cm]{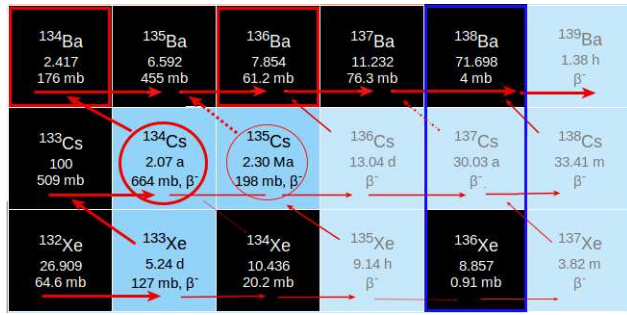}
\caption{The same as Fig.~\ref{branchpm147}, but for the $s$ path close to the $s$-only isotopes
$^{134}$Ba and $^{136}$Ba (red squares). 
The main branch point occurs at $^{134}$Cs. 
The electron-capture (EC) channel of $^{134}$Cs is negligible. 
The long-lived $^{135}$Cs can be considered almost stable during the $s$ process,
and its radiogenic contribution to $^{135}$Ba occurs after the TP-AGB phase
(thick dashed arrow); the abundance of $^{137}$Cs stored during TPs $\beta^-$ decays to 
 $^{137}$Ba during interpulse periods (thin dashed arrow). 
 The neutron-magic nuclei at N = 82 are indicated by a blue box.}
\label{branchxexecs}
\end{figure}

The production of the two $s$-only isotopes $^{134,136}$Ba occurs via both
neutron irradiations, in the $^{13}$C pocket and during TPs.
\\
Most of solar $^{134}$Ba is produced in the $^{13}$C pocket, when the $s$ path 
proceeds via $^{132}$Xe and $^{133}$Cs directly to $^{134}$Ba, because the low 
neutron density does not allow neutron captures on $^{133}$Xe and $^{134}$Cs 
(Fig.~\ref{branchxexecs}). 
In this way the solar $^{134}$Ba/$^{136}$Ba ratio would be overestimated by about 
a factor of two.
The additional \Nean neutron burst partially activated during TP 
is essential to regulate the $^{134,136}$Ba predictions in order to reproduce the solar
abundances. 
The main component is expected to reproduce the $s$ contribution of the $s$-only pair 
$^{134}$Ba and $^{136}$Ba so that $^{134}$Ba:$^{136}$Ba = 1:1 within solar and nuclear
uncertainties.

 The dominant uncertainty affecting the $^{134}$Ba/$^{136}$Ba ratio derives 
from the $\beta^-$-decay rate of $^{134}$Cs:
the terrestrial half-life of $^{134}$Cs ($t_{1/2}$ = 2.07 yr) is strongly reduced under stellar 
conditions \citep{TY87}: it decreases by a factor of three at $T_8$ = 1 (0.67 yr) 
and by two orders of magnitude at $T_8$ = 3 (3.8 d). 
Note that $\lambda^-$($^{134}$Cs) does not change with electron (or mass) density.
\\
In previous AGB models (see filled circles in Fig.~\ref{newmains}), we have adopted 
a constant $\beta^-$-decay rate for $^{134}$Cs, which was based on a
geometric average of the rates given by \citet{TY87} at different temperatures 
over the convective pulse ($\lambda^-$($^{134}$Cs) = 1.6$\times$10$^{-7}$ s$^{-1}$, 
$t_{1/2}$ = 50 d; see \citealt{bisterzo11,liu14}).
This value reproduces the solar $^{134}$Ba/$^{136}$Ba ratio, and the predicted $s$
contributions exceed the solar abundances of $^{134}$Ba and $^{136}$Ba by only 6 
and 9\%, respectively, well within the uncertainties quoted for the solar abundances 
(6\% for Ba, 5\% for Sm; \citealt{lodders09}) and for the MACS values of barium (3.2\% 
at 30 keV; KADoNiS).

The strong temperature dependence of the half-life of $^{134}$Cs 
is now considered by an improved treatment of the $\beta$-decay rate as a function of 
the temperature gradient within the TPs.
The maximum temperature reached by a given AGB model at the bottom of the convective 
intershell rises from pulse to pulse. 
The convective TP itself is characterised by a very large gradient in temperature and density, 
which decreases from $T_8$ $\sim$ 3 and $\rho$ = 2$\times$10$^4$ g/cm$^3$
at the bottom of the advanced TP to $T_8$ $\sim$ 0.1 and $\rho$ = 10 g/cm$^3$ in the top layers.
Under these conditions, the $^{134}$Cs half-life varies from 2 yr in the top layers of the 
convective TP (where no neutrons are available) down to 3.8 d in the region of highest 
neutron density at the bottom. Accordingly, the production of $^{134}$Cs depends strongly 
on the locus inside the TP: for $T_8$ $\sim$ 3 one finds $t_{1/2}$ = 3.8 d, and at 
$N_n$ $\sim$ 4$\times$10$^{9}$ cm$^{-3}$ only $\sim$25\% of the $s$ flow bypasses $^{134}$Ba. 
Slightly further outward, at $T_8$ $\sim$ 2.7, neutrons are still produced via $^{22}$Ne($\alpha$, 
n)$^{25}$Mg, 
but the half-life of $^{134}$Cs has increased to 50 d: under these conditions, neutron captures 
dominate over $\beta$ decays starting from $N_n$ $\sim$ 10$^{9}$ cm$^{-3}$ ($f_n$ $\sim$ 0.5), 
and $^{134}$Ba is more efficiently bypassed.
\\
We have interpolated (with a cubic-spline method) the $\beta^-$-decay rates given by \citet{TY87}
as a function of the stellar temperature in the range from $T_8$ = 0.5 to 5. 
The adopted rate results very close to that recommended by NETGEN\footnote{web site  
http://www.astro.ulb.ac.be/Netgen.}.
$^{134}$Cs is freshly produced in the thin bottom layers of the TP, but even in this hot environment
 $^{134}$Cs has half-life of a few days, 
much larger than the turnover time, so that it is quickly brought to the outer and cooler layers of the TPs. 
Here, the $^{134}$Cs half-life increases up to 2 yr, but the lack of neutrons does not allow further
neutron captures and it decays into $^{134}$Ba.
Because a small amount of $^{134}$Cs is converted to $^{135}$Cs during the peak neutron density 
at the bottom of the TPs, $^{134}$Ba is temporarily reduced (by $\sim$30\%). 
The initial amount of $^{134}$Ba 
is almost fully reestablished as soon as the neutron density decreases.
Consequently, almost the entire $s$ flow is directed towards $^{134}$Ba, resulting in a 
large overestimation of solar $^{134}$Ba (+30\%).
\\
This problem with the branching at $^{134}$Cs can be somewhat relaxed if one 
considers the uncertainty in the temperature-dependent half-life. \citet{goriely99}
estimated a possible decrease of the decay rate by a factor of $\sim$3 at $T_8$ = 2 -- 3. 
Assuming twice that change as 2$\sigma$-uncertainty would reduce the overestimation of 
the solar $^{134}$Ba (see $filled$ $diamond$ in Fig.~\ref{newmains}).
In this case, $^{134}$Ba is largely depleted at the beginning of the advaced TPs
(see, e.g., Fig.~\ref{stampecs134}, which shows the temporal evolution of the isotopic
abundances during the 15$^{\rm th}$ TP in a 1.5 $M_\odot$ model at half-solar metallicity).
As the neutron density decreases, the $\beta^-$-decay channel is favoured
and the $^{134}$Ba abundance is increasing again ($N_n$ $\la$ 2$\times$10$^{8}$ cm$^{-3}$; 
$f_n$ $\leq$ 0.1). 
About 76\% of $^{134}$Ba produced during the previous $^{13}$C pocket is restored 
at the tail of the neutron density.
The resulting $s$ contributions to $^{134}$Ba and $^{136}$Ba exceed the solar abundances 
by +12\% and +9\%, respectively.
This may suggest that the stellar $^{134}$Cs $\beta^-$decay rate by \citet{TY87} has 
been overestimated. Further investigations on this topic are advised.
\\
An additional important effect concerning the $s$ predictions for the Ba isotopes derives 
from the \Nean rate. The above $s$ contributions to $^{134}$Ba and $^{136}$Ba may be reduced 
to +4\% and +2\% with increasing the recommended \Nean rate by a factor of four (Test A).

\begin{figure}
\vspace{5 mm}
\includegraphics[angle=-90,width=9cm]{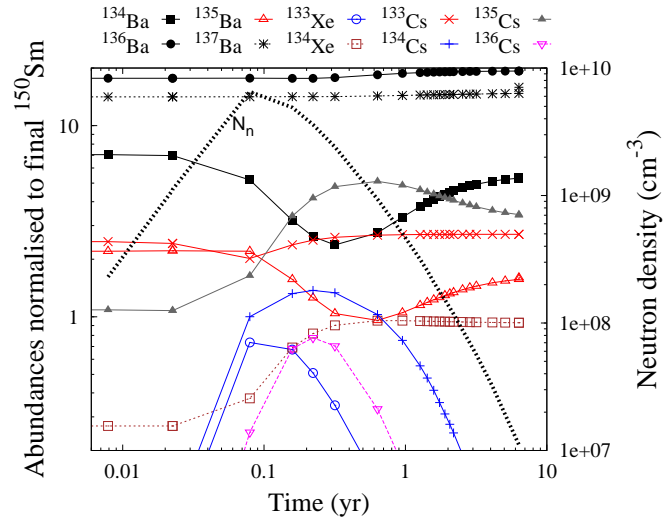}
\caption{The same as Fig.~\ref{stampesm150}, but for $^{134,135,136,137}$Ba, $^{133,134}$Xe 
and $^{133,134,135,136}$Cs isotopic abundances. }
\label{stampecs134}
\end{figure}

\vspace{2mm}

A marginal impact on the $^{134,136}$Ba $s$-predictions derives from the branches 
at $^{133}$Xe and $^{135,136}$Cs.
The short half-life of $^{133}$Xe ($t_{1/2}$ = 5.24 d) does not allow efficient neutron 
captures (less than 2\% of the $s$ flow feeds $^{134}$Xe). 
\\
Although the terrestrial half-life of the long-lived $^{135}$Cs ($t_{1/2}$ = 2.3 Myr) 
is reduced by four orders of magnitude at $T_8$ = 3 
(e.g., $t_{1/2}$ = 267 yr at $\rho$ = 3000 g/cm$^3$; \citealt{TY87}), it is practically 
stable compared to the 6 year timescale of the \Nean irradiation with 
a TP.
Therefore, once the $s$ flow feeds $^{135}$Cs, it proceeds towards $^{136}$Cs, and $^{135}$Ba 
may be easily bypassed (see Fig.~\ref{stampecs134}). The abundance of the long-lived $^{135}$Cs 
stored during TPs $\beta^-$ decays into $^{135}$Ba at the end of the TP-AGB phase. 
\\
The $s$ flow at $^{136}$Cs ($t_{1/2}$ = 13.04 d) continues
mainly via $\beta^-$ decay to $^{136}$Ba, where the two $s$ branches formed
at $^{134}$Cs join again.
Accordingly, the $s$ contribution to $^{136}$Ba keeps increasing during the entire \Nean 
irradiation, thus enhancing the abundance from the previous $^{13}$C pocket by +10\%.

The $^{134}$Cs MACS has been estimated semi-empirically by \citet{patronis04}, who provide 
a rather small uncertainty of 9\% (KADoNiS), 
 resulting in a $\sim$2.5\% variation of the $^{134}$Ba abundance.
The uncertainties of the theoretical MACS values of $^{133}$Xe 
(KADoNiS) and $^{136,137}$Cs 
\citep{patronis04} are presently estimated between 30 and
50\%, but have almost no impact ($<$2\%) on $^{134,136}$Ba.


\subsubsection{Additional branches strongly sensitive to stellar
temperature (and/or electron density) and neutron density}

A few branches behave as thermometers of the $s$ process in AGB stars because their $\beta$-decay
rates deeply rise (or drop) with the large temperature and electron density gradients that characterise
a convective TP \citep{TY87}. In a few cases, this $\beta$-decay behaviour is responsible for crucial
uncertainties.

Similar to the way $^{134}$Cs affects the $s$ contribution to $^{134}$Ba,
the decay rates of  $^{151}$Sm and $^{154}$Eu depend strongly on temperature
and electron density during TPs with corresponding consequences for the $s$
abundances of $^{152}$Gd and $^{154}$Gd, respectively.
\\
The $s$ predictions to $^{152,154}$Gd have been improved by including in AGB models an 
appropriate treatment of the nearby $\beta$-decay rates over the full convective TPs (see
$filled$ $diamonds$ in Fig.~\ref{newmains}): about 85\% and 92\% of solar $^{152}$Gd and $^{154}$Gd
are produced by AGB stars.
According to \citet{goriely99} both rates may vary up to a factor of three at $T_8$ = 3.
This would produce an extreme impact on the $^{152}$Gd $s$-prediction (up to a factor of 2),
and up to 10-15\% variations of the $^{154}$Gd $s$-contribution. 
The dominant effect of the $^{151}$Sm half-life prevents the $s$ contribution to solar 
$^{152}$Gd from being accurately assessed.
Additional uncertainties may derive from the $^{151}$Sm and $^{154}$Eu MACS (see \citealt{rauscher11}).
\\
 Certainly, the $s$ predictions of the above isotopes are largely influenced by the uncertainties 
of the \Nean rate.
A detailed analysis of these aspects is given in Section~B (Supporting Information).

Two other isotopes are worth mentioning, where
the AGB contributions depend strongly on temperature:
the long-lived $^{176}$Lu (more properly classified as an $s$-process thermometer 
rather than a cosmic clock),
and $^{204}$Tl (which regulates the $s$ contribution of solar $^{204}$Pb).
\\
 At present, the discrepancy between experimental data and the astrophysical treatment of 
the branch at $^{176}$Lu remains to be solved \citep{heil08LuHf,mohr09,cristallo10LuHf}.
The solution may be found in the nuclear coupling scheme between thermally populated levels 
in $^{176}$Lu \citep{gintautas09,dracoulis10,gosselin10}.
Besides the branch at $^{176}$Lu, the \Nean rate and the SEF estimate of the $^{176}$Hf MACS 
are directly influencing the predicted $^{176}$Lu/$^{176}$Hf ratio. 
\\
 Although the MACS of $^{204}$Pb is accurately known (3\%), the 
present $^{204}$Pb $s$ prediction is jeopardised by the uncertainty of the theoretical
$^{204}$Tl neutron capture cross section and by the evaluation of its stellar $\beta^-$-decay 
rate ($\sim$10\% variations). 
An additional 10\% uncertainty from the solar Pb abundance \citep{lodders09} has to be 
considered as well.

The impact of the present uncertainties is discussed in Section~B (Supporting Information).


\subsection{The $s$-only isotopes mainly sensitive to stellar temperature 
(and electron density)}\label{branchMIX}

The branches at issue are $^{179}$Hf (responsible for the $s$ contribution to $^{180}$Ta$^{\rm m}$; see, e.g.,
\citealt{kaeppeler04,wisshak04}), $^{164}$Dy (which produced about half of solar amount of the 
proton-rich isotope $^{164}$Er; \citealt{jaag96}), and $^{128}$I (which regulates the $s$ production to 
 $^{128}$Xe, \citealt{reifarth04}).
\\
Because of their low sensitivity to the neutron density, the branches at $^{180}$Ta$^{\rm m}$ 
and $^{128}$I have provided information on the convective mixing time scale.
However, the large uncertainties of the $\beta$-decay rates of the respective branch
point nuclei have led to problems in the interpretation of the branchings
towards $^{164}$Er and $^{180}$Ta$^{\rm m}$.


\subsubsection{The $s$ contribution to $^{180}$Ta$^{\rm m}$ (the branches at $^{179}$Hf, 
$^{179}$Ta and $^{180}$Hf)}\label{ta180}

\begin{figure}
\includegraphics[angle=0,width=8cm]{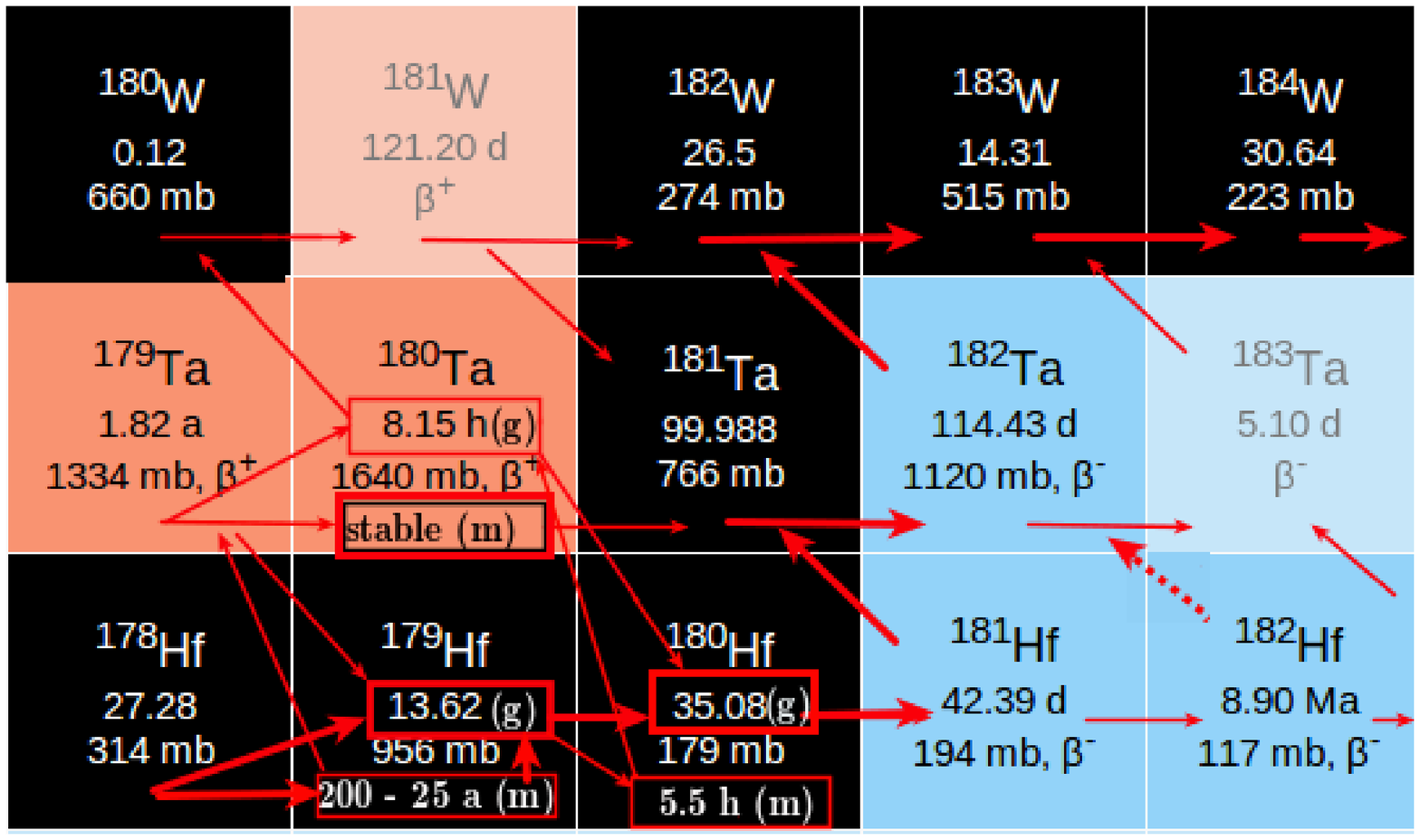}
\caption{The same as Fig.~\ref{branchpm147}, but for the $s$ path close to $^{180}$Ta$^{\rm m}$. 
During the \Can neutron irradiation $^{179}$Hf is stable and the $s$ path follows
the thick red lines; during \Nean neutron burst (thin red lines) the isomeric state 
of $^{179}$Hf becomes unstable (with half-life strongly temperature dependent; 
\citep{TY87}), allowing the branch of the $s$ flow towards $^{179}$Ta, and the stable 
$^{180}$Ta$^{\rm m}$.
The long-lived $^{182}$Hf ($t_{1/2}$ = 8.9 Myr) can be considered almost stable during 
the $s$ process, and its radiogenic contribution to $^{182}$Ta occurs 
after the TP-AGB phase (red dashed-thick line).}
\label{branchhftaw}
\end{figure}

\begin{figure}
\vspace{5 mm}
\includegraphics[angle=-90,width=9cm]{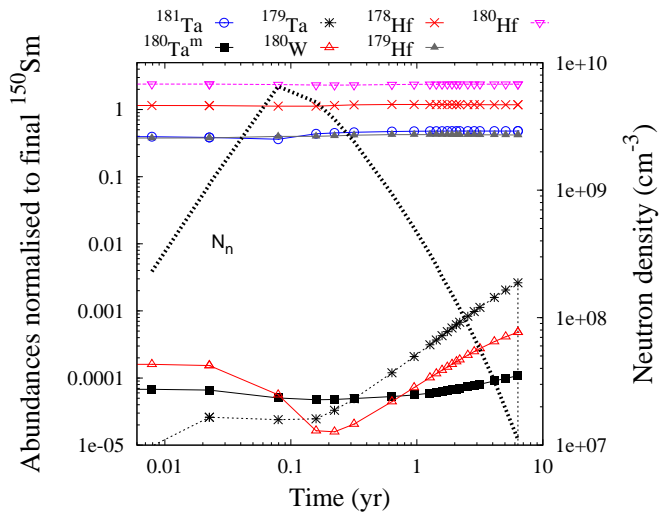}
\caption{The same as Fig.~\ref{stampesm150}, but for the isotopic abundances of 
$^{180}$Ta$^{\rm m}$, $^{179,181}$Ta,$^{180}$W and $^{178,179,180}$Hf. }
\label{stampehftaw}
\end{figure}

In the solar system, $^{180}$Ta occurs only in its stable isomer ($t_{1/2}$ $>$ 10$^{15}$ 
yr). $^{180}$Ta$^{\rm m}$ is the rarest stable isomer in Nature and constitutes 0.012\% 
of solar Ta. The ground state of $^{180}$Ta is unstable and decays with a half-life of
8.15 h by $\beta^-$ and $\beta^+$ emission to $^{180}$W and $^{180}$Hf, respectively. 
\\ 
The origin of $^{180}$Ta$^{\rm m}$ has been a challenge for years. 
Being proton-rich, the expected contributions from p(\g)- and $\nu$p-processes have been 
largely investigated (e.g, \citealt{rayet95,rauscher02,heger05}).  
As illustrated in Fig.~\ref{branchhftaw}, the $s$ flow is bypassing $^{180}$Ta$^{\rm m}$
because $^{179,180}$Hf are stable during the \Can irradiation. At the higher temperatures during
TPs, however, higher lying nuclear states in $^{179}$Hf and $^{180}$Hf are thermally
populated, opening two branches towards $^{180}$Ta$^{\rm m}$: 
(i) via $\beta^-$ decay of the thermally populated $^{179}$Hf state at 214 keV, which 
$\beta^-$ decays to $^{179}$Ta and produces $^{180}$Ta$^{\rm m}$ via neutron captures, and 
(ii) via the $^{179}$Hf partial neutron capture cross section towards the weakly populated 
$^{180}$Hf isomer, which $\beta^-$ decays quickly to $^{180}$Ta$^{\rm m}$.

Several studies have been dedicated to the $s$-process nucleosynthesis of $^{180}$Ta$^{\rm m}$
(e.g., \citealt{nemeth92,kaeppeler04,wisshak04,wisshak06Hf,mohr07}).
\\
Briefly, about half of the $s$ contribution to $^{180}$Ta$^{\rm m}$ is produced 
via the branch at $^{179}$Hf$^{\rm m}$. At $T_8$ = 3, the $^{179}$Hf half-life 
becomes sensitive to temperature and electron density (e.g., $t_{1/2}$ from 85 to 25 yr; \citealt{TY87}),
feeding $^{179}$Ta in small amounts. 
Neutron captures on $^{179}$Ta are dominating over the small $\beta^+$ decay back to $^{179}$Hf 
(the terrestrial $^{179}$Ta half-life 
increases by a factor of two at TP temperature), producing mainly $^{180}$Ta$^{\rm g}$, which
$\beta^+$- and $\beta^-$ decays into $^{180}$Hf and $^{180}$W, respectively.
This channel yields about 5\% of the solar $p$-rich $^{180}$W.
A minor amount of $^{179}$Ta (IR = 0.04) feeds the stable isomer $^{180}$Ta$^{\rm m}$,
contributing 
44\% of its solar abundance. 
At the end of the TP, the residual $^{179}$Ta abundance $\beta^+$ decays to its stable Hf
isobar (Fig.~\ref{stampehftaw}).
\\
An additional 39\% $s$ contribution to $^{180}$Ta$^{\rm m}$ derives from
the $\beta^-$ decay of $^{180}$Hf$^{\rm m}$. The production of the $^{180}$Hf$^{\rm m}$ 
isomeric state is weak:
only $\sim$1.3\% of the total $^{179}$Hf MACS feeds $^{180}$Hf$^{\rm m}$ 
(at 30 keV $\sigma$[$^{179}$Hf(n, \g)$^{180}$Hf] = 922 $\pm$ 8 mbarn; 
$\sigma$[$^{179}$Hf(n, \g)$^{180}$Hf$^{\rm m}$] = 11.4 $\pm$ 0.6 mbarn; KADoNiS).
$^{180}$Hf$^{\rm m}$ has short half-life ($t_{1/2}$ = 5.5 h), and decays directly to
$^{180}$Ta$^{\rm m}$.
This contribution is independent of temperature and is completely determined by the partial 
 cross section to $^{180}$Hf$^{\rm m}$.

The final $s$ abundance of $^{180}$Ta$^{\rm m}$ is largely sensitive to temperature 
and electron density \citep{TY87}, and needs to be properly evaluated by accounting 
of the temperature and density gradients over the convective He flashes.
\citet{belic99,belic02} 
carried out a photoactivation experiment of $^{180}$Ta$^{\rm m}$ to study
the probability for connecting isomer and ground state in $^{180}$Ta via thermally
induced transitions to higher lying mediating states. Whereas the direct
internal decay of the isomer to the ground state is highly forbidden by selection
rules, thermal excitations of such mediating states are allowed under stellar
conditions. They found that the two states are fully thermalised 
at $T_8$ = 3: thus, $^{180}$Ta$^{\rm m}$ should be destroyed in the 
bottom layers of the advanced TPs, where such temperatures are reached. 
Instead, the fast convective mixing (of the order of a few hours) occurring during TP prevents 
the destruction of $^{180}$Ta$^{\rm m}$ (see Fig.~\ref{stampehftaw}). 
\\
Starting from \citet{wisshak01,wisshak04} and \citet{kaeppeler04}, 
our AGB models account for the thermally induced destruction of $^{180}$Ta$^{\rm m}$ obtained 
by the photoactivation measurement by \citet{belic99,belic02}, 
by following the strong half-life variation with temperature and density gradients together with
convective mixing at each TP. This method firstly pointed out that $^{180}$Ta$^{\rm m}$ receives 
a dominant contribution from the main component, increasing the previous estimate from $\sim$49\% 
\citep{arlandini99} to $\sim$80\% \citep{kaeppeler04,wisshak04}. In addition, it provides 
information on the convective turnover time during He-shell flashes, and underlines that temperature
gradient and neutron freeze-out effects are not sufficient to analyse the abundances of such peculiar 
isotopes correctly.

The main component reproduces about 83\% of solar $^{180}$Ta$^{\rm m}$ (see $filled$ $circle$ 
in Fig.~\ref{newmains}). 
This prediction is affected by a number of uncertainties,
however. Therefore, the fraction of $\sim$20\%, which have to be contributed by other sources, 
could well be significantly larger. 
\\
In order to improve the estimated $s$ contribution to $^{180}$Ta$^{\rm m}$, we have 
included the same treatment to the $^{179}$Ta $\beta^+$-decay and $^{179}$Hf $\beta^-$-decay rates 
over the full convective zone.
The $s$ contribution to $^{180}$Ta$^{\rm m}$ is reduced  
from $\sim$83\% to $\sim$68\% (see $filled$ $diamond$ in Fig.~\ref{newmains}).  
This value remains strongly sensitive to the uncertain $\beta$-decay rate of
$^{179}$Hf estimated by \citet{goriely99} who suggested that the $^{179}$Hf $\beta^-$-decay 
rate may decrease by a factor of three, resulting in a strong  
impact on the contribution to solar $^{180}$Ta$^{\rm m}$ ($-$30\%).
The smaller $^{179}$Ta $\beta^+$-decay uncertainty estimated by \citet{goriely99} 
does not affect $^{180}$Ta$^{\rm m}$.
\\
As a consequence, AGB stars remain to be
the major nucleosynthesis sites for $^{180}$Ta$^{\rm m}$, but leave a larger fraction for
abundance contributions by 
other processes.
\\
In light of this result, we suggest that an investigation of the $^{179}$Hf half-life
under stellar conditions would be important in order to assess the contribution
to $^{180}$Ta$^{\rm m}$ from AGB stars more reliably.

We examine additional uncertainties associated to the $^{180}$Ta$^{\rm m}$ $s$ prediction.
Although the $^{180}$Ta$^{\rm m}$ MACS is rather well known with 6.8\% uncertainty 
(1465 $\pm$ 100 mbarn at
30 keV; \citealt{wisshak04}), it is affected by a non negligible stellar enhancement
factor (SEF = 0.87 at 30 keV; KADoNiS). 
Moreover, the $^{179}$Ta MACS is only estimated theoretically with $\sim$30\%
uncertainty (1334 $\pm$ 422 at 30 keV; KADoNiS), producing $\sim$10\% variations of the
$^{180}$Ta$^{\rm m}$ $s$ abundance. 
\\
Finally, the $^{180}$Ta$^{\rm m}$ $s$-prediction decreases by $\sim$12\% with increasing
the \Nean rate by a factor of four (see Table~A1, Supporting Information).

\subsubsection{Updated $s$ contributions to $^{182}$W and the short-lived $^{182}$Hf}\label{hf182}

The $s$ and $r$ process contributions to solar $^{182}$W and $^{182}$Hf are mainly regulated by the 
branch at $^{181}$Hf (see Fig.~\ref{branchhftaw}).
Present AGB predictions marginally produce the short-lived $^{182}$Hf ($t_{1/2}$ = 8.9 Myr), because 
the terrestrial half-life of $^{181}$Hf is strongly reduced during TPs 
(from $t_{1/2}$ = 42.39 d to 1.26 d at $T_8$ = 3; \citealt{TY87}) and the $s$ path mainly
feeds $^{182}$W.
We estimate that 65\% of solar $^{182}$W is synthesised by the main component (see Table~A.1, 
Appendix~A, Supporting Information), which corresponds to a residual 35\% $r$ contribution, in agreement 
with previous calculations by \citet{wisshak06Hf} and \citet{vock07}.
This value includes the radiogenic contribution of $^{182}$Hf at the end of the TP-AGB phase.
Indeed, no $\beta$-decay occurs during TPs because the decrease of the $^{182}$Hf half-life
is not sufficient to compete with neutron captures ($t_{1/2}$ = 22 yr at $T_8$ = 3; \citealt{TY87}).
\\
\citet{lugaro14hf181} have recently demonstrated that the present AGB contributions to $^{182}$Hf
and $^{182}$W have so far been underestimated. Starting from the experimental work by 
\citet{bondarenko02}, who did not find any evidence for the existence of the $^{181}$Hf levels
responsible for the strong enhancement of the $\beta$-decay rate (at 68 keV, 170 keV, and 298 keV),
\citet{lugaro14hf181} argued that the terrestrial $\beta$-decay rate of $^{181}$Hf remains rather 
unchanged in stellar condition. This favours the neutron capture channel towards $^{182}$Hf, 
which increases the abundance of $^{182}$W after the TP-AGB phase.
By assuming a constant $^{181}$Hf $\beta$-decay rate, we find a negligible $r$ contribution to 
solar $^{182}$W, which would be fully produced by AGBs. 
On the other hand, \citet{lugaro14hf181} estimated a new $^{181}$Hf $\beta$-decay rate
by removing the 68 keV level (see their Fig.~S2; top panel, blue line).
By including this value in our calculations, the $s$ contribution of 
$^{182}$W increases from 65\% to 80\%, reducing the $r$ component by a factor of two.
The lower limit given by the authors (obtained by removing the three levels; top panel, red line)
further increases the solar $s$ abundance of $^{182}$W to 86\%. 
The $^{182}$W $s$ prediction shows $-$3\% variations within the \Nean rate uncertainties.
\\
This result is of great relevance to the chronometry of the early solar system (see  
\citealt{wasserburg06} for a review on short-lived isotopes), or in presolar stardust grains
\citep{avila12}.
 
In addition, it provides another key to explain the apparent anomaly in the $r$-process residuals
at A = 182, together with the uncertainty associated to the neutron capture cross sections advanced
by \citet{vock07}.


\subsubsection{Additional branches mainly sensitive to stellar
temperature (and/or electron density)}

Similarly to $^{180}$Ta$^{\rm m}$, also the $p$-rich isotope $^{164}$Er receives a dominant
contribution from the $s$ process owing to the activation of the branch at $^{163}$Dy 
(which becomes unstable in stellar environment).
\\
About 80\% of solar $^{164}$Er is produced during TPs ($filled$ $circle$ in Fig.~\ref{newmains}).
This value rises to 87\% by including an improved treatment of the $\beta$-decay rates 
of nearby unstable isotopes over the convective TPs (see $filled$ $diamond$ in Fig.~\ref{newmains}).
Unfortunately, the $s$ contribution to $^{164}$Er can not be accurately assessed,
being largely sensitive to the competition between the $\beta^-$ and 
$\beta^+$-decay rates of $^{164}$Ho, and in particular to the uncertainty of the $^{164}$Ho 
$\beta^-$-decay rate estimated under stellar conditions (10--20\%). 
We discuss in Section~B (Supporting Information) the main uncertainties affecting the $s$ contribution to 
$^{164}$Er.

\citet{reifarth04} provided a test for the convective mixing time scale 
during TPs, by examining the branch at $^{128}$I,
which regulates the production of the $s$-only pair $^{128,130}$Xe. 
The $s$ predictions of $^{128,130}$Xe exhibit a weak dependence on neutron density owing to the short half-lives
of $^{127}$Te ($t_{1/2}$ = 9 h) and $^{128}$I ($t_{1/2}$ = 25 min), 
and the competition between $^{128}$I $\beta$-decay and electron captures may constrain 
the convective mixing timescale during the He shell flash. 
 While the $\beta^-$-decay channel of $^{128}$I shows only a weak 
dependence on temperature, the electron capture rate is strongly
temperature-dependent: from $t_{1/2}(EC)$ $\sim$ 6 h at $T_8$
= 0.5 up to $\sim$8 d at $T_8$ = 3. By assuming a sufficiently short turnover mixing timescale, 
the $^{128}$I produced via neutron captures in the hot bottom layers of the TP is promptly 
brought to the cooler external layers of the convective zone, allowing a partial activation
of the EC channel with the result that $\sim$5 to 6\% of the $s$-process flow is bypassing
$^{128}$Xe. The corresponding $s$ predictions for $^{128}$Xe and $^{128}$Xe are in 
agreement with the $^{128,130}$Xe ratio observed in SiC grains and confirm the results 
of \citet{reifarth04}, (see Section~B, Supporting Information).
The $^{128,130}$Xe $s$-predictions are rather well determined, because the branching is 
completely regulated by the decay of $^{128}$I, so that the influence of uncertain theoretical 
MACS values of nearby unstable isotopes is negligible at stellar temperatures.



\section{Summary and Conclusions}\label{summary}

 Despite a Galactic chemical evolution model would provide a more realistic description
of the abundances observed in the solar system, the main component is still a useful 
tool to investigate the $s$-process nucleosynthesis of nuclei with atomic mass between 90 and 204.
\\
We have studied the major uncertainties affecting the nuclear network.
The analysis has been carried out with the most recent neutron capture cross sections and 
with updated solar abundance data.

\vspace{2mm}

We have examined the impact of the present uncertainties of the two neutron sources operating
in AGB stars, the \Can and \Nean reactions, focusing on the $s$-only isotopes sensitive
to the most important branch points of the main component.

The overall $s$ distribution of isotopes heavier than $A$ $\sim$ 90 shows negligible variations
(up to $\sim$1\%) by changing the \Can rate by about factor of two.
Only the two neutron-magic nuclei $^{86}$Kr and $^{87}$Rb are influenced by the \Can neutron 
irradiation, because of the marginal activation of the branch at $^{85}$Kr.
Note, however, that in the AGB models we adopt to reproduce the solar main component ($M$ = 1.5 and 3 
$M_\odot$ at half solar metallicity), the $^{13}$C abundance in the pocket is exhausted radiatively
during the interpulse period (only a negligible amount of $^{13}$C is engulfed in the subsequent TP).
Thus, our AGB models do not experience a partial convective burning of $^{13}$C during
the first TPs, which could affect the production of a few neutron-rich isotopes (e.g., 
$^{86}$Kr and $^{87}$Rb, or $^{96}$Zr as well, see \citealt{cristallo06}).
The new measurement by \citet{lacognata13} suggests that the \Can rate adopted in our
models should be increased by 40\%, rather than decreased. 
In this case, the convective $^{13}$C burning during TPs is significantly
reduced.
\\
However, beside the undeniable progress made by means of 
indirect measurements \citep{lacognata13}, the existence of a sub-threshold state makes the
evaluation of the rate at astrophysical energies still uncertain. A further increase of this
rate would likely have marginal consequences on the main component, a substantial 
reduction may increase the amount of $^{13}$C engulfed in the convective pulse and burned at 
relatively high temperature.

The present uncertainty of \Nean (e.g., \citealt{kaeppeler94,jaeger01,longland12}) mainly 
influences the isotopes close to and within the branchings of the $s$ path.
\\
We have provided new evaluated values of the \Nean and \Neag rates that account for 
all known and potential resonances as well as all nuclear data available. 
In the temperature range of AGB stars ($T_8$ = 2.5--3), the recommended \Nean value is about 
a factor of two lower than that adopted in our models so far, while the recommended \Neag rate 
is essentially unchanged.
The recommended rates are in agreement within the errors with the recent values presented in literature
\citep{jaeger01,longland12}. 
\\
However, the \Nean and \Neag rates are mainly based on the knowledge of the 832 
keV resonance. As discussed for the \Can neutron source, the presence of unknown low-energy states, 
which have been identified in several indirect experiments, makes the \Nean and \Neag recommended
values uncertain at stellar energy. A direct determination of these reaction rates at temperatures  
below the present experimental limits will shed light on the actual efficiency of the AGB neutron sources. 
Because the available low-energy measurements were mainly limited by a 
significant neutron background, deep underground laboratories are the most promising places 
to plan future experimental investigations on the two AGB neutron sources (LUNA, DIANA).
\\
For this reason we have analysed the solar main component within a more cautious range of 
uncertainty, corresponding to variations by a factor of four starting from our recommended \Nean rate.
\\
Major variations are shown by the $s$-only nuclides close to the branchings, which
are most sensitive to neutron density as $^{96}$Mo, $^{142}$Nd, and $^{170}$Yb (due to the 
branch points at $^{95}$Zr,
$^{141}$Ce, $^{170}$Tm), $^{134}$Ba and $^{152}$Gd (which also depend on the $^{134}$Cs and 
$^{151}$Sm half-lives, respectively, both strongly reduced in stellar environments), and 
$^{176}$Lu (affected by the contribution of isomeric mediating states, see \citealt{mohr09}).
\\
In low mass AGB stars (which are the major contributors to the solar $s$ abundances; 
\citealt{travaglio04,bisterzo14ApJ}), the \Neag reaction is not efficiently activated during TPs owing
to the rather small temperature reached at the bottom of the advanced convective He flashes.
Thus, the competition between \Neag and \Nean reactions is marginal and does not influence
the solar $s$ distribution.
\\
A larger impact of the \Neag and \Nean rates is expected from intermediate AGB stars, 
where the \Nean is the major neutron source, or from low-metallicity AGB models.
The study of these models provides information about the $s$-process nucleosynthesis in, e.g., 
globular clusters, dwarf galaxies, intrinsic or extrinsic peculiar stars showing $s$ enhancement.
To this purpose, we have analysed the rate uncertainties in a 3 $M_\odot$ model at [Fe/H] = $-$1 and a 
half-solar metallicity 5 $M_\odot$ model chosen as representative of an extended range 
of AGB stars.

\vspace{2mm}

\vspace{2mm}

The status of the (n, \g) stellar cross sections has been significantly improved in the 
last decade, with accuracies of less than 5\% for a number of isotopes in the 
mass region 120 $<$ $A$ $<$ 180.
On the other hand, the accuracy of neutron capture and $\beta$-decay rates of isotopes
that act as important branching points of the $s$ path plays a crucial role in the 
production of a few $s$-only nuclides. First, because the MACS of unstable isotopes
are barely accessible to direct measurements, only theoretical estimates
(with large uncertainty) are available in most cases. 
Second, because the stellar 
$\beta$-decay rates are poorly known for branch point nuclei, in particular if they 
are extremely sensitive to temperature and electron density.
 Moreover, \citet{rauscher11} and \citet{rauscher12} highlighted that the 
(experimentally measured) ground state cross section may constitute only a minor fraction of the
MACS. Thus, the theoretical uncertainties associated to the MACS may be in a few cases 
underestimated (note that \citealt{rauscher11} provide upper limits of MACS uncertainties; 
specific theoretical investigations carried out individually for each nucleus are strongly 
needed for a few isotopes\footnote{Specific cases are $^{134}$Cs, $^{151}$Sm and $^{154}$Eu, $^{164}$Ho, $^{179}$Hf,
$^{185}$W, $^{192}$Ir, in order to improve the $s$ contributions of $^{134}$Ba, $^{152,154}$Gd, 
$^{164}$Er, $^{180}$Ta$^{\rm m}$, $^{186}$Os, $^{192}$Pt, as well as $^{160}$Dy, $^{164}$Er, 
$^{170}$Yb, $^{176}$Hf, $^{180}$Ta$^{\rm m}$, which are directly influenced.}).
\\
In this study, we have discussed the present major nuclear uncertainties that affect 
the $s$-only isotopes.
We have distinguished the $s$-only nuclei in different classes, according to the type 
of information that can be deduced from their abundances with respect to the physical 
conditions in AGB stars: unbranched 
nuclides (useful to constrain the $s$ distribution), isotopes mainly affected by neutron
density, and isotopes strongly sensitive 
to temperature and electron density (which help to address the characteristics of stellar
models).
For each class, the specific problems and suggestions for a possible improvement are 
given.
We suggest that an investigation of the $\beta$-decay rates as a function
of the stellar environment would be important for $^{134}$Cs, $^{151}$Sm, $^{179}$Hf, 
and $^{164}$Ho in order to improve the $s$ contributions of $^{134}$Ba, $^{152}$Gd,
$^{180}$Ta$^{\rm m}$, and $^{164}$Er.

\vspace{2mm}

In conclusion, we find that the solar main component may still reproduce 
the $s$-only isotopes within the present uncertainties.


\section*{Acknowledgments}

We are indebted to I. Dillmann and T. Rauscher for sharing information concerning
the nuclear network and the theoretical analysis that significantly improved
the present discussion in many ways.\\
This work has been supported by JINA (Joint Institute for Nuclear Astrophysics,
University of Notre Dame, IN) and by KIT (Karlsruhe Institute of Technology, Karlsruhe, Germany).



\section*{SUPPORTING INFORMATION}

Additional Supporting Information may be found in the online version of this article:

{\bf Appendix A.} The results for the updated main component obtained in 
the analysis described in Section~2 are listed in Table A1 for the isotopes with A $>$ 70. 
We also report the effects of the tests discussed in Section~4.1.

{\bf Appendix B.} This Appendix completes the information about the most important branches 
of the main component discussed in the paper. Following the classification given in 
Section~5, we analyse here the additional branchings of each category, which are only briefly 
outlined in the text.

{\bf Appendix C.} In this Appendix we have analysed the uncertainties of the neutron
sources in a 3 $M_\odot$ model at [Fe/H] = $-$1 and a half-solar metallicity 5 $M_\odot$
model chosen as representative of an extended range of AGB stars.

\onecolumn

\appendix

\newpage

%
%
%
%
%
%
%
%
%
%
%
%
%
%
%
%

\clearpage

\section{The main component of the solar system}

The results for the updated main component shown in Fig.~5 are listed in 
Table~A1 for the isotopes with A $>$ 70.
The recommended \Nean and \Neag rates described in Section~4 have been adopted.
\\ 
We also report the effects of a conservative range of uncertainty associated to the \Nean 
and \Neag rates (see discussion in Section~4.1).
\\
The main component is listed in column~2. All values are normalised to $^{150}$Sm (see Section~5.1.1).
In columns~3 to~6 we list the results of two tests on the \Nean rate discussed in Section~4.1:
{\bf Test A} corresponds to column~3 (recommended $^{22}$Ne($\alpha$, n)$\times$4, top panel of Fig.~6),
{\bf Test B} corresponds to column~5 (lower limit estimated in Section~4, bottom panel of Fig.~6).
The $s$-only isotopes are highlighted in bold, while $s$-only nuclides ($^{80}$Kr, $^{86}$Sr, $^{128}$Xe, 
$^{152}$Gd) and $s$ isotopes ($^{90}$Zr, $^{96}$Zr, $^{164}$Er, $^{180}$Ta$^{\rm m}$) with significant $p$-process
contributions are written in italic. 
Isotopes with negligible $s$-process contribution ($<$5\%) are excluded from this Table.
 Uncertainties (in \%) listed for $s$-only isotopes in column~7 refer to solar abundances by Lodders et al. (2009).

 A complete version of Figs.~3, ~6 and ~7 is displayed in Figs.~A1, ~A2, and~A3, respectively.



\begin{footnotesize}

\begin{longtable}{|l||r||r|r|r|r|c|}
\caption*{ \textbf{Table~A1}.} \\
\hline
 Isotope   & Main component (in \%) & {\bf Test A (in \%)}  &  Ratio & {\bf Test B (in \%)}  &  Ratio &Solar  \\
           & Recomm.  $^{22}$Ne($\alpha$, n) &Recomm. $^{22}$Ne($\alpha$, n)$\times$4& (3)/(2) &Recomm. $^{22}$Ne($\alpha$, n)$\times$0.9 & (5)/(2)&uncertainty (in \%) \\
\hline
\endfirsthead
\multicolumn{7}{c}%
{\tablename\ A1 
$-$$-$ \textit{Continued from previous page}} \\
\hline
 Isotope   & Main component (in \%) & {\bf Test A (in \%)}  &  Ratio & {\bf Test B (in \%)}  &  Ratio& Solar \\
            & Recomm. $^{22}$Ne($\alpha$, n) &Recomm. $^{22}$Ne($\alpha$, n)$\times$4& (3)/(2) &Recomm. $^{22}$Ne($\alpha$, n)$\times$0.9 & (5)/(2)&uncertainty (in \%)  \\
\hline
\endhead
\hline \multicolumn{7}{r}{\textit{Continued on next page}} \\
\endfoot
\hline
\endlastfoot
 {\bf $^{{\bf 70}}$Ge}        &  {\bf   6.0} &  {\bf   4.9}  &   {\bf  .81} &   {\bf   6.1}  &   {\bf 1.02} & {\bf 10} \\ 
      $^{     72 }$Ge         &         7.5  &         6.2   &         .82  &          7.6   &        1.01  &  \\ 
      $^{     73 }$Ge         &         7.3  &         6.2   &         .84  &          7.4   &        1.00  &  \\ 
      $^{     74 }$Ge         &         8.4  &         7.4   &         .89  &          8.4   &        1.00  &  \\ 
      $^{     75 }$As         &         5.9  &         5.3   &         .90  &          5.9   &        1.00  &  \\    
 {\bf $^{{\bf 76}}$Se}        &  {\bf  13.2} &  {\bf  11.7}  &   {\bf  .88} &   {\bf  13.3}  &   {\bf 1.01} & {\bf 7} \\
      $^{     77 }$Se         &         6.5  &         5.6   &         .87  &          6.5   &        1.01  &  \\ 
      $^{     78 }$Se         &        14.3  &        12.7   &         .88  &         14.4   &        1.00  &  \\ 
      $^{     80 }$Se         &         8.0  &         7.6   &         .95  &          8.0   &        1.00  &  \\ 
      $^{     79 }$Br$^*$     &         7.9  &         7.5   &         .96  &          7.8   &         .99  &  \\  
      $^{     81 }$Br$^*$     &        10.3  &         9.1   &         .88  &         10.5   &        1.01  &  \\ 
 {\it $^{{\it 80}}$Kr}        &  {\it   9.9} &  {\it   5.7}  &   {\it  .58} &   {\it  10.5}  &   {\it 1.06} & $-$ \\ 
 {\bf $^{{\bf 82}}$Kr}        &  {\bf  23.3} &  {\bf  19.5}  &   {\bf  .84} &   {\bf  23.5}  &   {\bf 1.01} & $-$ \\
      $^{     83 }$Kr         &         8.2  &         7.0   &         .85  &          8.3   &        1.00  &  \\ 
      $^{     84 }$Kr         &        10.5  &        10.4   &         .99  &         10.4   &         .99  &  \\ 
      $^{     86 }$Kr         &        16.4  &        25.8   &        1.57  &         16.1   &         .98  &  \\ 
      $^{     85 }$Rb         &        12.0  &        15.2   &        1.27  &         11.7   &         .97  & \\ 
      $^{     87 }$Rb         &        18.3  &        41.1   &        2.25  &         17.2   &         .94  &  \\ 
 {\it $^{{\it 86}}$Sr}        &  {\it  58.5} &  {\it  39.1}  &   {\it  .67} &   {\it  60.0}  &   {\it 1.03} & {\bf 7} \\ 
 {\bf $^{{\bf 87}}$Sr}        &  {\bf  58.5} &  {\bf  37.7}  &   {\bf  .64} &   {\bf  59.7}  &   {\bf 1.02} & {\bf 7} \\ 
      $^{     88 }$Sr         &        82.6  &        72.3   &         .87  &         83.0   &        1.00  &  \\ 
      $^{     89 }$Y          &        85.8  &        85.6   &        1.00  &         85.5   &        1.00  & \\ 
 {\it $^{{\it 90}}$Zr}        &  {\it  76.0} &  {\it  69.1}  &   {\it  .91} &   {\it  76.3}  &   {\it 1.00} &  \\ 
      $^{     91 }$Zr         &        85.4  &        86.0   &        1.01  &         84.9   &         .99  &  \\ 
      $^{     92 }$Zr         &        83.3  &        89.5   &        1.07  &         82.8   &         .99  &  \\ 
      $^{     94 }$Zr         &       107.4  &       103.5   &         .96  &        107.8   &        1.00  &  \\ 
 {\it $^{{\it 96}}$Zr}        &        14.3  &        86.8   &        6.06  &         12.0   &         .84  &  \\ 
      $^{     93 }$Nb$^*$     &  {\it  69.2} &  {\it  74.3}  &   {\it 1.07} &   {\it  69.0}  &   {\it 1.00} &  \\ 
      $^{     95 }$Mo         &        60.4  &        61.7   &        1.02  &         59.7   &         .99  &  \\ 
 {\bf $^{{\bf 96}}$Mo}        &  {\bf  99.2} &  {\bf 120.3}  &   {\bf 1.21} &   {\bf  98.2}  &   {\bf  .99} & {\bf 10} \\ 
      $^{     97 }$Mo$^*$     &        53.4  &        65.1   &        1.22  &         53.4   &        1.00  &  \\ 
      $^{     98 }$Mo         &        74.1  &        70.5   &         .95  &         74.8   &        1.01  &  \\ 
      $^{     99 }$Ru$^*$     &        27.2  &        25.0   &         .92  &         27.5   &        1.01  &  \\ 
 {\bf $^{{\bf 100}}$Ru}       &  {\bf 105.5} &  {\bf  88.6}  &   {\bf  .84} &   {\bf 106.7}  &   {\bf 1.01} & {\bf 6} \\ 
      $^{     101}$Ru         &        16.8  &        14.1   &         .84  &         17.0   &        1.01  &  \\ 
      $^{     102}$Ru         &        57.2  &        48.1   &         .84  &         57.6   &        1.01  &  \\ 
      $^{     103}$Rh         &        15.6  &        13.0   &         .84  &         15.7   &        1.01  & \\ 
 {\bf $^{{\bf 104}}$Pd}       &  {\bf 110.7} &  {\bf  92.7}  &   {\bf  .84} &   {\bf 111.5}  &   {\bf 1.01} & {\bf 5} \\ 
      $^{     105 }$Pd        &        14.2  &        12.0   &         .84  &         14.3   &        1.01  &  \\ 
      $^{     106 }$Pd        &        52.4  &        45.0   &         .86  &         52.7   &        1.00  &  \\ 
      $^{     108 }$Pd        &        66.4  &        58.4   &         .88  &         66.7   &        1.00  &  \\ 
      $^{     107 }$Ag$^*$    &        15.1  &        13.0   &         .86  &         15.2   &        1.01  & \\ 
      $^{     109 }$Ag        &        29.7  &        26.2   &         .88  &         29.8   &        1.00  &  \\ 
 {\bf $^{{\bf 110}}$Cd}       &  {\bf 101.7} &  {\bf  91.3}  &   {\bf  .90} &   {\bf 102.0}  &   {\bf 1.00} & {\bf 7} \\ 
      $^{     111 }$Cd        &        32.7  &        29.5   &         .90  &         32.7   &        1.00  &  \\ 
      $^{     112 }$Cd        &        65.1  &        60.0   &         .92  &         65.1   &        1.00  &  \\ 
      $^{     113 }$Cd        &        37.1  &        34.3   &         .92  &         37.2   &        1.00  &  \\ 
      $^{     114 }$Cd        &        76.6  &        72.2   &         .94  &         76.5   &        1.00  &  \\ 
      $^{     116 }$Cd        &         8.6  &        20.0   &        2.33  &          7.7   &         .89  &  \\ 
      $^{     115 }$In        &        38.6  &        36.4   &         .94  &         38.6   &        1.00  &  \\ 
 {\bf $^{{\bf 116}}$Sn}       &  {\bf  87.7} &  {\bf  82.8}  &   {\bf  .94} &   {\bf  87.7}  &   {\bf 1.00} & {\bf 15} \\ 
      $^{     117 }$Sn        &        48.9  &        47.2   &         .97  &         48.9   &        1.00  &  \\ 
      $^{     118 }$Sn        &        69.3  &        70.0   &        1.01  &         69.1   &        1.00  &  \\ 
      $^{     119 }$Sn        &        58.5  &        58.7   &        1.00  &         58.4   &        1.00  &  \\ 
      $^{     120 }$Sn        &        76.2  &        76.3   &        1.00  &         76.1   &        1.00  &  \\ 
      $^{     122 }$Sn        &        32.6  &        64.3   &        1.97  &         31.2   &         .96  &  \\ 
      $^{     121 }$Sb        &        37.6  &        37.2   &         .99  &         37.7   &        1.00  &  \\ 
 {\bf $^{{\bf 122}}$Te}       &  {\bf  89.1} &  {\bf  88.1}  &   {\bf  .99} &   {\bf  89.1}  &   {\bf 1.00} & {\bf 7} \\ 
 {\bf $^{{\bf 123}}$Te}       &  {\bf  89.5} &  {\bf  88.4}  &   {\bf  .99} &   {\bf  89.6}  &   {\bf 1.00} & {\bf 7} \\ 
 {\bf $^{{\bf 124}}$Te}       &  {\bf  92.7} &  {\bf  93.7}  &   {\bf 1.01} &   {\bf  92.8}  &   {\bf 1.00} & {\bf 7} \\
      $^{     125 }$Te        &        20.6  &        20.6   &        1.00  &         20.7   &        1.00  &  \\
      $^{     126 }$Te        &        42.7  &        41.4   &         .97  &         42.7   &        1.00  &  \\
 {\it $^{{\it 128}}$Xe}       &  {\it  89.4} &  {\it  84.8}  &   {\it  .95} &   {\it  89.6}  &   {\it 1.00} & $-$ \\
 {\bf $^{{\bf 130}}$Xe}       &  {\bf  98.2} &  {\bf  92.0}  &   {\bf  .94} &   {\bf  98.5}  &   {\bf 1.00} & $-$ \\
      $^{     131 }$Xe        &         7.4  &         7.0   &         .94  &          7.4   &        1.00  &  \\
      $^{     132 }$Xe        &        27.5  &        27.9   &        1.01  &         27.4   &         .99  &   \\
      $^{     133 }$Cs        &        13.9  &        14.0   &        1.01  &         13.8   &        1.00  &  \\  
 {\bf $^{{\bf 134}}$Ba}       &  {\bf 115.0} &  {\bf 103.8}  &   {\bf  .90} &   {\bf 116.6}  &   {\bf 1.01} & {\bf 6} \\
      $^{     135 }$Ba$^*$    &        26.3  &        29.6   &        1.13  &         26.0   &         .99  &  \\ 
 {\bf $^{{\bf 136}}$Ba}       &  {\bf 108.9} &  {\bf 101.5}  &   {\bf  .93} &   {\bf 109.2}  &   {\bf 1.00} & {\bf 6} \\ 
      $^{     137 }$Ba        &        60.3  &        70.0   &        1.16  &         60.1   &        1.00  &  \\ 
      $^{     138 }$Ba        &        93.5  &        89.9   &         .96  &         93.5   &        1.00  &  \\ 
      $^{     139 }$La        &        76.2  &        73.3   &         .96  &         76.3   &        1.00  &  \\ 
      $^{     140 }$Ce        &        95.3  &        85.2   &         .89  &         95.8   &        1.00  &  \\ 
      $^{     142 }$Ce        &         9.5  &        37.5   &        3.97  &          8.1   &         .86  & \\   
      $^{     141 }$Pr        &        50.4  &        49.6   &         .98  &         50.0   &         .99  &  \\ 
 {\bf $^{{\bf 142}}$Nd}$^{**}$&  {\bf 104.5} &  {\bf  88.0}  &   {\bf  .84} &   {\bf 105.4}  &   {\bf 1.01} & {\bf 5} \\ 
      $^{     143 }$Nd        &        33.4  &        32.1   &         .96  &         33.4   &        1.00  &  \\ 
      $^{     144 }$Nd        &        52.8  &        51.1   &         .97  &         52.8   &        1.00  &  \\ 
      $^{     145 }$Nd        &        26.4  &        25.8   &         .98  &         26.4   &        1.00  &  \\ 
      $^{     146 }$Nd        &        67.3  &        63.5   &         .94  &         67.6   &        1.00  &  \\ 
      $^{     148 }$Nd        &        15.6  &        20.1   &        1.29  &         14.7   &         .94  &  \\ 
      $^{     147 }$Sm        &        27.1  &        25.3   &         .93  &         27.2   &        1.00  &  \\ 
 {\bf $^{{\bf 148}}$Sm}       &  {\bf 105.3} &  {\bf 100.9}  &   {\bf  .96} &   {\bf 105.9}  &   {\bf 1.01} & {\bf 5} \\ 
      $^{     149 }$Sm        &        12.9  &        12.9   &        1.01  &         12.8   &        1.00  &  \\ 
{\bf $^{{\bf 150}}$Sm}$^{***}$&  {\bf 100.0} &  {\bf 100.0}  &   {\bf 1.00} &   {\bf 100.0}  &   {\bf 1.00} & {\bf 5} \\ 
      $^{     152 }$Sm        &        22.5  &        23.1   &        1.02  &         22.4   &        1.00  &  \\ 
      $^{     151 }$Eu        &         6.0  &         6.0   &        1.00  &          6.0   &        1.00  &  \\ 
      $^{     153 }$Eu        &         6.1  &         6.2   &        1.02  &          6.0   &        1.00  &  \\ 
 {\it $^{{\it 152}}$Gd}       &  {\it  85.4} &  {\it  69.0}  &   {\it  .81} &   {\it  87.6}  &   {\it 1.03} & {\bf 5} \\ 
 {\bf $^{{\bf 154}}$Gd}       &  {\bf  92.0} &  {\bf  95.4}  &   {\bf 1.04} &   {\bf  91.6}  &   {\bf 1.00} & {\bf 5} \\ 
      $^{     155 }$Gd        &         5.8  &         6.1   &        1.04  &          5.8   &        1.00  &  \\ 
      $^{     156 }$Gd        &        17.6  &        17.9   &        1.02  &         17.6   &        1.00  &  \\ 
      $^{     157 }$Gd        &        10.9  &        11.0   &        1.02  &         10.9   &        1.00  &  \\ 
      $^{     158 }$Gd        &        27.1  &        27.4   &        1.01  &         27.2   &        1.00  &  \\ 
      $^{     159 }$Tb        &         7.9  &         7.9   &        1.01  &          7.9   &        1.00  &  \\ 
 {\bf $^{{\bf 160}}$Dy}       &  {\bf  89.0} &  {\bf  89.7}  &   {\bf 1.01} &   {\bf  89.4}  &   {\bf 1.00} & {\bf 5} \\ 
      $^{     161 }$Dy        &         5.2  &         5.3   &        1.01  &          5.2   &        1.00  &  \\ 
      $^{     162 }$Dy        &        15.9  &        15.9   &        1.00  &         16.0   &        1.01  &  \\ 
      $^{     164 }$Dy        &        24.0  &        20.9   &         .87  &         24.5   &        1.02  &  \\ 
      $^{     165 }$Ho        &         8.5  &         8.2   &         .96  &          8.6   &        1.01  &  \\ 
 {\it $^{{\it 164}}$Er}       &  {\it  87.1} &  {\it  90.8}  &   {\it 1.04} &   {\it  87.0}  &   {\it 1.00} & {\bf 5} \\ 
      $^{     166 }$Er        &        15.9  &        15.8   &         .99  &         16.0   &        1.00  &  \\ 
      $^{     167 }$Er        &         9.7  &         9.6   &        1.00  &          9.7   &        1.00  &  \\ 
      $^{     168 }$Er        &        32.7  &        30.2   &         .92  &         33.2   &        1.01  &  \\ 
      $^{     170 }$Er        &        12.6  &        12.1   &         .96  &         12.2   &         .97  &  \\ 
      $^{     169 }$Tm        &        14.8  &        13.3   &         .90  &         15.0   &        1.02  &  \\ 
 {\bf $^{{\bf 170}}$Yb}       &  {\bf  99.3} &  {\bf  89.5}  &   {\bf  .90} &   {\bf 100.6}  &   {\bf 1.01} & {\bf 5} \\ 
      $^{     171 }$Yb        &        23.9  &        20.3   &         .85  &         24.2   &        1.01  &  \\ 
      $^{     172 }$Yb        &        44.7  &        40.7   &         .91  &         44.7   &        1.00  &  \\ 
      $^{     173 }$Yb        &        28.0  &        26.3   &         .94  &         28.0   &        1.00  &  \\ 
      $^{     174 }$Yb        &        61.9  &        56.9   &         .92  &         62.2   &        1.00  &  \\ 
      $^{     176 }$Yb        &         6.0  &         8.4   &        1.40  &          5.6   &         .93  &  \\ 
      $^{     175 }$Lu        &        18.4  &        16.8   &         .91  &         18.5   &        1.01  &  \\ 
 {\bf $^{{\bf 176}}$Lu}       &  {\bf 112.4} &  {\bf  92.0}  &   {\bf  .82} &   {\bf 115.0}  &   {\bf 1.02} & {\bf 5} \\ 
 {\bf $^{{\bf 176}}$Hf}       &  {\bf 101.5} &  {\bf 105.4}  &   {\bf 1.04} &   {\bf 100.6}  &   {\bf  .99} & {\bf 5} \\ 
      $^{     177 }$Hf        &        17.4  &        16.4   &         .95  &         17.5   &        1.01  &  \\ 
      $^{     178 }$Hf        &        58.7  &        53.3   &         .91  &         59.1   &        1.01  &  \\ 
      $^{     179 }$Hf        &        42.1  &        38.2   &         .91  &         42.3   &        1.01  &  \\ 
      $^{     180 }$Hf        &        91.6  &        83.6   &         .91  &         92.0   &        1.00  &  \\ 
 {\it $^{{\it 180}}$Ta$^{m}$} &  {\it  67.6} &  {\it  59.2}  &   {\it  .88} &   {\it  67.6}  &   {\it 1.00} & {\bf 10} \\ 
      $^{     181 }$Ta        &        47.7  &        43.6   &         .91  &         47.9   &        1.00  &  \\ 
      $^{     182 }$W$^*$     &        64.9  &        61.0   &         .94  &         65.2   &        1.00  &  \\ 
      $^{     183 }$W         &        65.0  &        56.2   &         .87  &         65.9   &        1.01  &  \\ 
      $^{     184 }$W         &        82.4  &        71.2   &         .86  &         83.0   &        1.01  &  \\ 
      $^{     186 }$W         &        42.3  &        42.5   &        1.00  &         41.2   &         .98  &  \\ 
      $^{     185 }$Re        &        28.3  &        24.2   &         .85  &         28.5   &        1.01  &  \\ 
      $^{     187 }$Re        &         9.3  &         9.6   &        1.04  &          9.0   &         .97  &  \\ 
 {\bf $^{{\bf 186}}$Os}       &  {\bf 105.8} &  {\bf  98.0}  &   {\bf  .93} &   {\bf 106.2}  &   {\bf 1.00} & {\bf 8} \\ 
      $^{     187 }$Os$^{\dag}$&       38.2  &        36.2   &         .95  &         38.5   &        1.01  & {\bf 8} \\ 
      $^{     188 }$Os        &        28.2  &        28.4   &        1.00  &         28.1   &         .99  &  \\ 
      $^{     190 }$Os        &        14.7  &        14.1   &         .96  &         14.8   &        1.00  &  \\ 
 {\bf $^{{\bf 192}}$Pt}       &  {\bf  80.6} &  {\bf  78.1}  &   {\bf  .97} &   {\bf  81.1}  &   {\bf 1.01} & {\bf 8} \\ 
      $^{     194 }$Pt        &         6.6  &         6.2   &         .94  &          6.6   &        1.00  &  \\
      $^{     196 }$Pt        &        14.0  &        13.2   &         .94  &         14.1   &        1.00  &  \\ 
      $^{     197 }$Au        &         6.1  &         5.8   &         .94  &          6.2   &        1.00  &  \\ 
 {\bf $^{{\bf 198}}$Hg}       &  {\bf  84.3} &  {\bf  80.2}  &   {\bf  .95} &   {\bf  84.4}  &   {\bf 1.00} & {\bf 20} \\ 
      $^{     199 }$Hg        &        22.1  &        21.0   &         .95  &         22.1   &        1.00  &  \\ 
      $^{     200 }$Hg        &        53.1  &        49.6   &         .94  &         53.1   &        1.00  &  \\ 
      $^{     201 }$Hg        &        40.5  &        38.4   &         .95  &         40.5   &        1.00  &  \\ 
      $^{     202 }$Hg        &        71.8  &        67.1   &         .94  &         72.0   &        1.00  &  \\ 
      $^{     204 }$Hg        &         8.5  &         5.8   &         .69  &          8.8   &        1.03  &  \\ 
      $^{     203 }$Tl        &        80.9  &        79.0   &         .98  &         80.7   &        1.00  &  \\ 
      $^{     205 }$Tl$^*$    &        70.8  &        50.9   &         .72  &         72.4   &        1.02  &  \\ 
 {\bf $^{{\bf 204}}$Pb}       &  {\bf  91.3} &  {\bf  89.9}  &   {\bf  .99} &   {\bf  91.1}  &   {\bf 1.00} & {\bf 7} \\ 
      $^{     206 }$Pb$^{**}$ &        70.2  &        64.2   &         .92  &         70.4   &        1.00  &  \\ 
      $^{     207 }$Pb        &        61.3  &        58.5   &         .96  &         61.3   &        1.00  &  \\ 
      $^{     208 }$Pb        &        47.4  &        45.9   &         .97  &         47.5   &        1.00  & \\  
      $^{     209 }$Bi        &         6.3  &         9.5   &        1.52  &          6.1   &         .97  &  \\ 
\hline                                                                                                                               
\multicolumn{7}{l}{($^*$) The overabundances of $^{79}$Br, $^{81}$Br, $^{93}$Nb, $^{97}$Mo, $^{99}$Ru, $^{107}$Ag, $^{135}$Ba, $^{182}$W, $^{205}$Tl include the decay of their short-lived isobars $^{79}$Se,}\\
\multicolumn{7}{l}{ $^{81}$Kr, $^{93}$Mo and $^{93}$Zr, $^{97}$Tc, $^{99}$Tc, $^{107}$Pd, $^{135}$Cs, $^{182}$Hf, $^{205}$Pb.  }\\
\multicolumn{7}{l}{($^{**}$) $^{142}$Nd and $^{206}$Pb account for the $\alpha$-decay of $^{146}$Sm and $^{210}$Bi.} \\
\multicolumn{7}{l}{($^{***}$) Values are normalised to $^{150}$Sm: we obtain X($^{150}$Sm)/X$_{ini}$($^{150}$Sm) = {\bf 1133.9} with the recommended \Nean } \\
\multicolumn{7}{l}{ and \Neag rates (column~2). {\bf Test A} provides X($^{150}$Sm)/X$_{ini}$($^{150}$Sm) = {\bf 1275.5 (+12.5\%),} and {\bf Test B} yields } \\ 
\multicolumn{7}{l}{ X($^{150}$Sm)/X$_{ini}$($^{150}$Sm) = 1128.4 ($-$1\%). See Table~3 of the manuscript. } \\ 
\multicolumn{7}{l}{($^{\dag}$) $^{187}$Os is a nominal $s$-only isotope, despite it receives a cosmogenic contribution by the decay of the very long-lived $^{187}$Re.}\\
\end{longtable}

\end{footnotesize}
          

Noteworthy isotopes are $^{96}$Mo and $^{176}$Lu: with TEST A the $s$ contribution to $^{96}$Mo 
is about 20\% higher than solar, while $^{176}$Lu is clearly underestimated (about 9\% of solar $^{176}$Lu 
is missing). This may suggest that the adopted conservative upper limit of the recommended \Nean rate 
($\times$4) is overestimated. However, a large uncertainty is associated to these isotopes (e.g., 
the branches at $^{95}$Zr and the unknown contribution of isomeric $^{176}$Lu mediating states). 
\\ 
Apart from some $s$ isotopes, several nuclides show variations larger than 10\%.
Among them, we highlight $^{96}$Zr (whose $s$ prediction increases from 14\% to 87\%),
$^{137}$Ba and $^{205}$Tl, which receive a dominant $s$-process contribution, and
$^{86}$Kr, $^{87}$Rb, $^{116}$Cd, $^{122}$Sn, $^{135}$Ba, $^{142}$Ce, $^{148}$Nd, $^{185}$Re.
We remind that $^{86}$Kr, $^{87}$Rb, $^{96}$Zr (and $^{137}$Ba) are also produced in IMS stars (see Section~C).
Note that the neutron-rich isotopes $^{176}$Yb and $^{208}$Hg, as well as $^{209}$Bi are marginally 
produced by the main component ($\la$10\%).
In Fig~\ref{FigA3}, we display the results of both tests for isotopes that receive an $s$-process contribution
higher than 50\%.

\twocolumn

\clearpage
\newpage

\begin{figure}
\includegraphics[angle=-90,width=8cm]{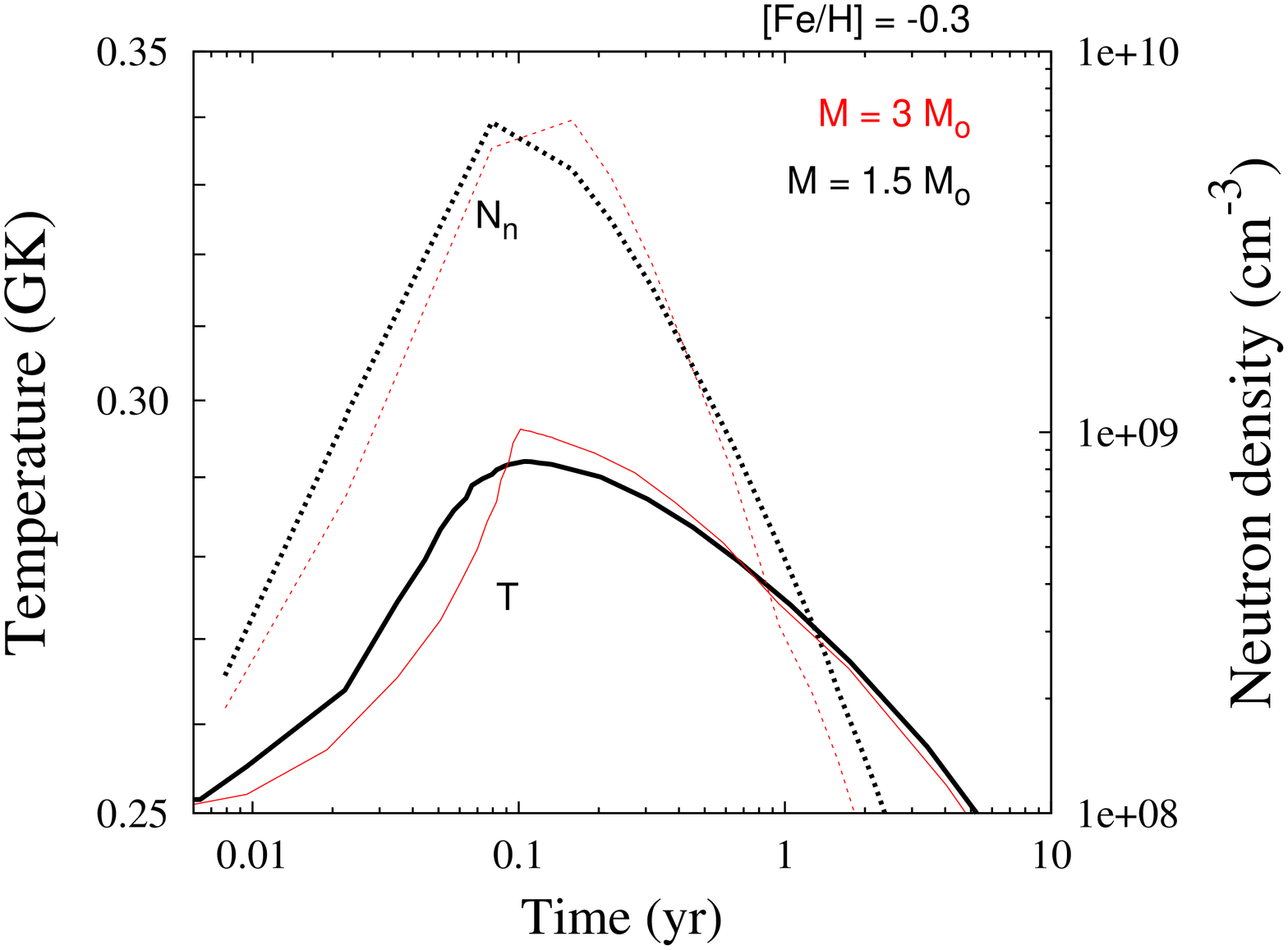} 
\caption{Temporal evolution of the neutron density ($N_n$) and the maximum temperature at the bottom
of the convective zone ($T$) during the 15th He 
shell flash in AGB stars with 1.5 and 3 \Msun (thick and thin lines, respectively) and 
half-solar metallicity. The time-scale starts when the temperature at the bottom
of the convective TP reaches $T_8$ = 2.5, which corresponds to the
onset of the \Nean reaction. }
\label{FigA0}
\end{figure}

\begin{figure}
\includegraphics[angle=0,width=8cm]{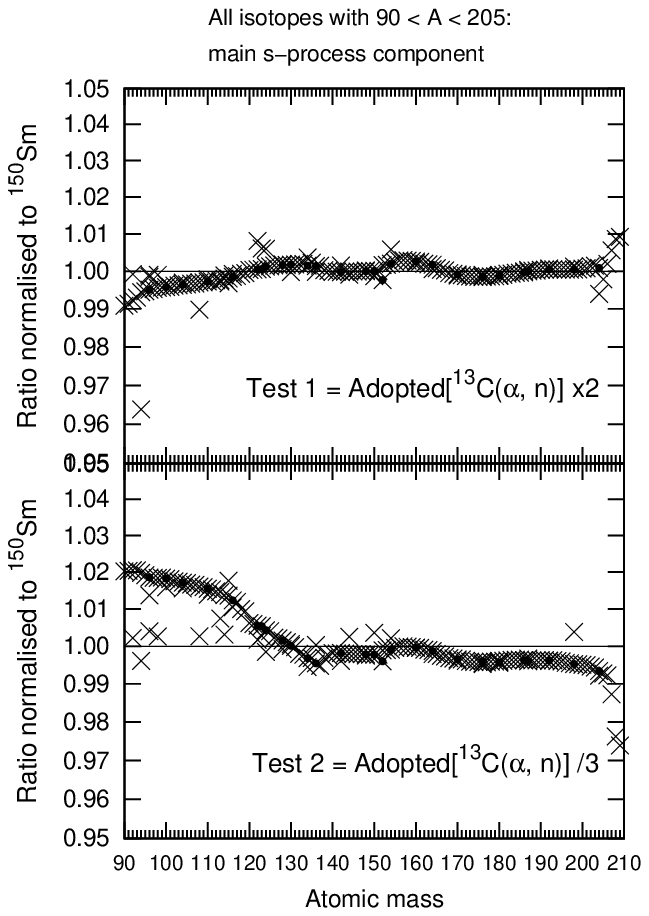}
\vspace{-0.7cm}
\caption{Ratios between the main component obtained with a two times higher and a
three times lower \Can rate than our adopted rate shown in Fig~2 ({\bf Test 1} and 
{\bf Test 2} corresponding to the $top$ and $bottom$ $panels$).
This is a complete version of Fig.~3 showing all isotopes from 90 $\leq$ $A$ $\leq$ 210.}
\label{FigA1}
\end{figure}

\begin{figure}
\includegraphics[angle=0,width=8cm]{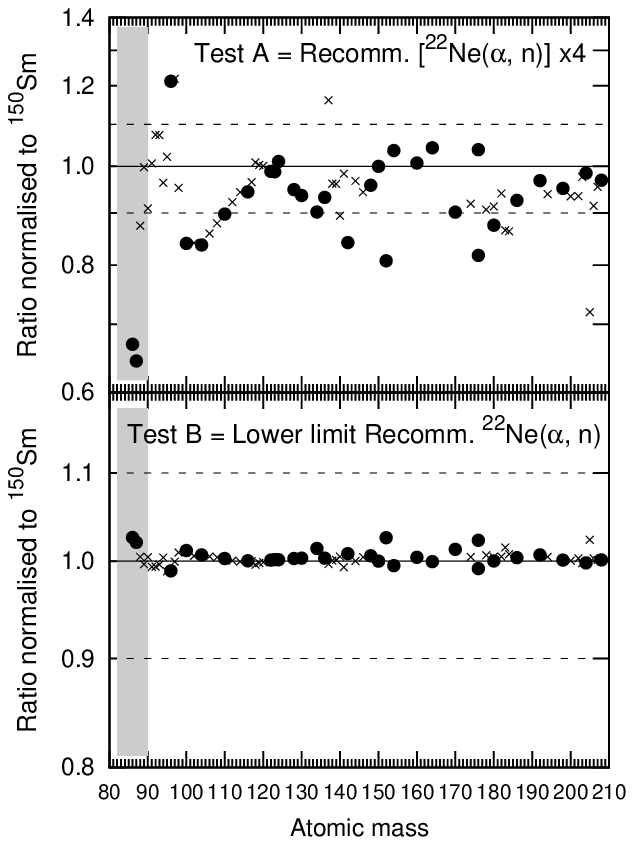}
\vspace{-0.7cm}
\caption{Ratios between the main component obtained within a conservative range 
of uncertainty associated to the \Nean and \Neag rates, as discussed in Section~4.1.
This is a complete version of Fig.~6 showing isotopes from 90 $\leq$ $A$ $\leq$ 210 that receive
an $s$-process contribution higher than 50\%.
As in Fig.~6, an improved treatment for the branches close to $^{134}$Ba, $^{152,154}$Gd, $^{164}$Er, 
$^{180}$Ta$^{\rm m}$ and $^{204}$Pb is included.}
\label{FigA3}
\end{figure}

\begin{figure}
\includegraphics[angle=0,width=8cm]{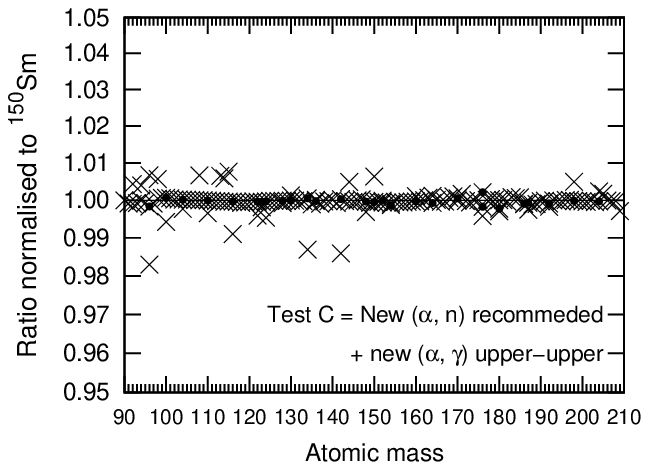}
\vspace{-5.5cm}
\caption{Ratios between the main component obtained with 
the recommended \Nean and \Neag rates compared to {\bf Test C} in which 
the upper limit for the \Neag rate is adopted, while the \Nean reaction is unchanged. 
This is a complete version of Fig.~7 showing all isotopes from 90 $\leq$ $A$ $\leq$ 210.}
\label{FigA3}
\end{figure}

\clearpage
\newpage


\section{Additional branches}

This Appendix completes the information about the most important branches 
of the main component discussed in the paper.
Following the classification given in Section~5, we 
analyse here the additional
branchings of each category, which are only briefly outlined in the
text.


\subsection{Additional unbranched isotopes}


\subsubsection{The $s$-only isotope $^{160}$Dy}\label{dy160}

\begin{figure}
\includegraphics[angle=0,width=5cm]{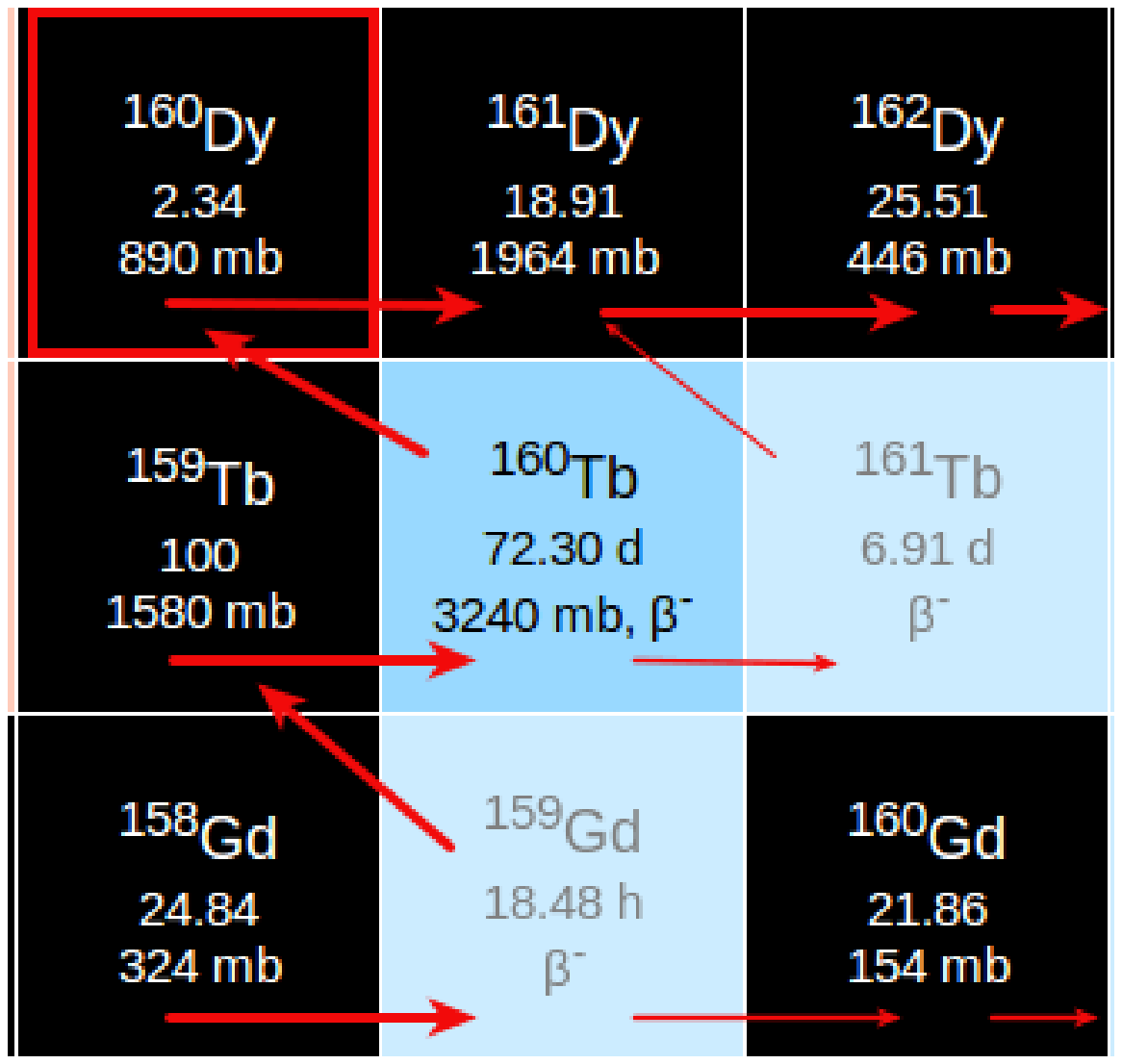}
\caption{Schematic representation of the chart of nuclei in the atomic mass region
close to the $s$-only isotope $^{160}$Dy (red square). Thick lines represent the $s$-process
nucleosynthesis during the \Can neutron irradiation, while thin lines correspond
to the neutron capture channels.
The terrestrial half-life of $^{160}$Tb ($t_{1/2}$ = 72.3 d) is reduced to 1.45 d at 
$T_8$ = 2 and to 11 h at $T_8$ = 3 (Takahashi \& Yokoi 1987). 
Thus, neutron captures are marginally
activated during the \Nean neutron burst, and $^{160}$Dy is almost unbranched.
(\textit{This and the following Figures are adapted from http://www.kadonis.org/. See the 
electronic edition of the Journal for a colour version of the Figures.})}
\label{branch160}
\end{figure}

The $s$-only isotope $^{160}$Dy (2.34\% of solar Dy) is mainly synthesised during the $^{13}$C-pocket 
phase and remains almost unbranched during TPs (Fig.~\ref{branch160}). The related branching points
$^{159}$Gd and $^{160}$Tb are both short-lived at $s$-process temperatures. The
decay of $^{159}$Gd ($t_{1/2}$ = 19 h) is constant under AGB conditions, while
 the terrestrial half-life of $^{160}$Tb 
is strongly reduced at $T_8$ = 3 (from $t_{1/2}$ = 72.3 d to 11 h; Takahashi \& Yokoi 1987).
Thus, less than 2\% of the s flow is bypassing $^{160}$Dy.

The $s$ contribution to solar $^{160}$Dy is 90\%.
\\
The solar abundances of Dy and Sm are both uncertain by 5\% (Lodders et al. 2009).
The $^{160}$Dy MACS is very well determined ($\sigma$[$^{160}$Dy(n, \g)] = 890 $\pm$ 12 mbarn 
at 30 keV, 1.4\%).
However, its large stellar enhancement factor 
(SEF = 1.12 at 30 keV) suggests that the neutron capture
rate of $^{160}$Dy may be affected by an additional theoretical
uncertainty.
\\
Given the negligible activation of the branches at $^{159}$Gd and $^{160}$Tb, the $s$
contribution to $^{160}$Dy is only marginally affected by the MACS uncertainties of these
branching points (e.g., $\sim$2\% variation for a 16\% change of the $^{160}$Tb MACS, 
3240 $\pm$ 510; KADoNiS).
Similarly, the large $^{160}$Tb $\beta$-decay uncertainty (up to 
a factor of three at $T_8$ = 3; Goriely 1999) has negligible effects on $^{160}$Dy. 
Accordingly, an increase of the recommended \Nean rate by factor of four  
affects the $s$ contribution of $^{160}$Dy by less than 2\%.

In conclusion, the prediction of the solar
abundance of $^{160}$Dy is mainly affected by the theoretical evaluation
of MACS of $^{160}$Dy (see also Rauscher et al. 2011).


\subsubsection{The $s$-only isotope $^{198}$Hg}

\begin{figure}
\includegraphics[angle=0,width=5cm]{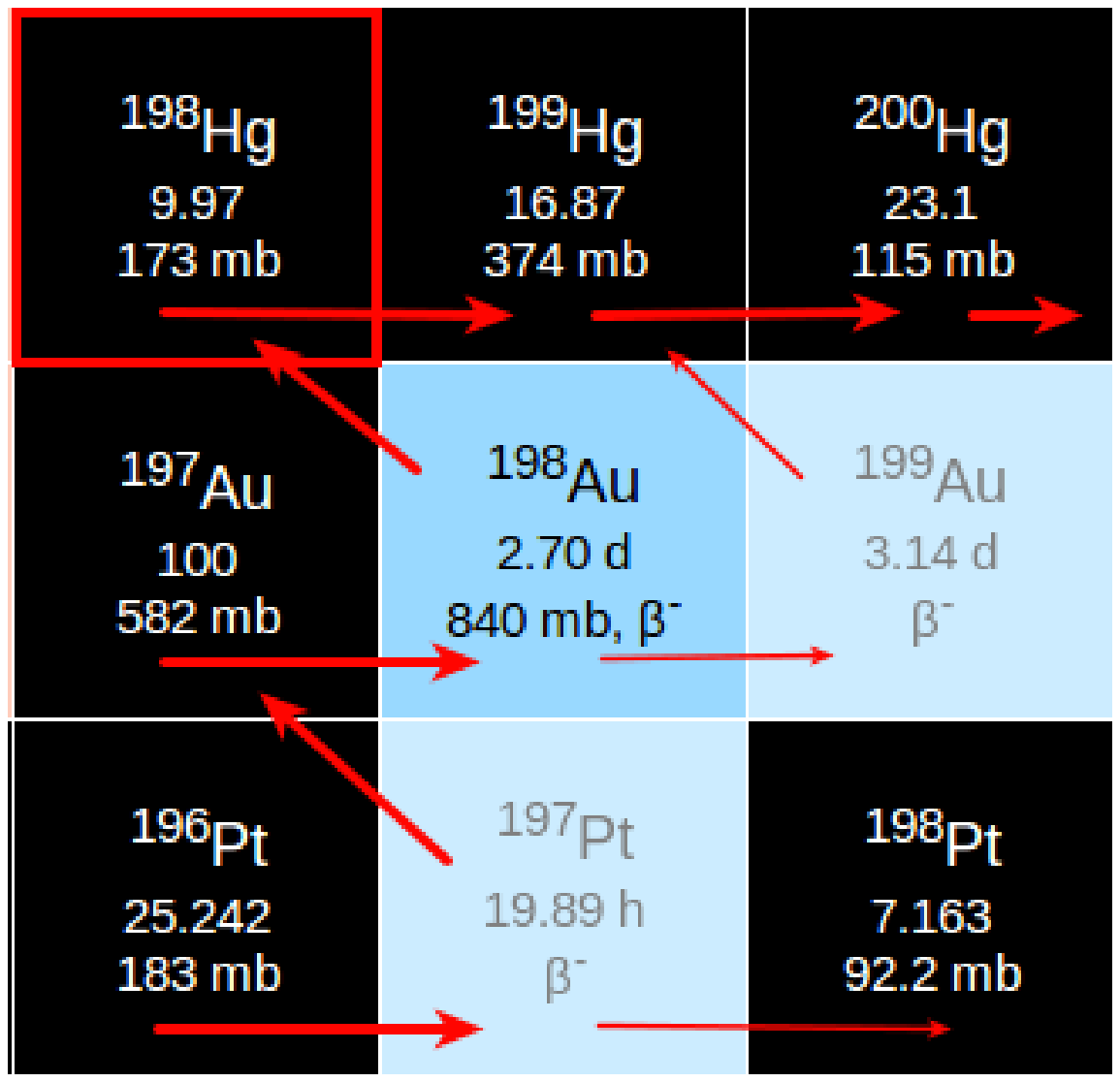}
\caption{The same as Fig.~\ref{branch160}, but for the $s$ path close to the 
$s$-only isotope $^{198}$Hg (red square). }
\label{branchhg198}
\end{figure}

The $s$-only isotope $^{198}$Hg is almost unbranched (Fig.~\ref{branchhg198}), because 
the short half-lives of $^{197}$Pt and $^{198}$Au strongly favour the $\beta$-decay channels
compared to neutron captures. The half life of $^{197}$Pt ($t_{1/2}$ = 19.89 h)
is rather constant under stellar 
conditions, while the terrestrial half-life of $^{198}$Au ($t_{1/2}$ = 2.7 d)
decreases by a factor of four during TP (Takahashi \& Yokoi 1987).
\\
Accordingly, the $s$ prediction for $^{198}$Hg varies by less than 2\% with increasing
the recommended \Nean rate by a factor of four.
\\
The major nuclear uncertainty derives from the $^{198}$Hg MACS
(8.7\%, 173 $\pm$ 15 mbarn; KADoNiS).

 About 84\% of solar $^{198}$Hg is produced by the main component.
We remind that the solar Hg abundance is largely uncertain for this very volatile
element. Originally, it was estimated using the $s$-process
systematics via the relation $\sigma$N$_s$ = constant (N(Hg) = 0.34 $\pm$ 0.04, 
Si $\equiv$ 10$^6$, Beer and Macklin, 1985). While this value was adopted in the
abundance compilation of Anders and Grevesse (1989), Lodders
(2003) determined the Hg abundance as a weighted average of 
measurements from two meteorites (Orgueil and Ivuna) with 50\% of uncertainty
due to the difficulty of finding clean samples.
Recently, Lodders et al. (2009) improved the measured Hg abundance, claiming an
uncertainty of 20\% (N(Hg) = 0.458 $\pm$ 0.092, Si $\equiv$ 10$^6$). This higher 
solar Hg abundance explains the lower $s$ contribution of $^{198}$Hg (84\%) 
compared to previous evaluations ($\sim$100\% using the solar Hg abundance of 
Anders and Grevesse 1989; see e.g., K{\"a}ppeler
et al. 2011, their Fig.~15).
The 20\% difference is, however, compatible with the respective solar 
 and nuclear uncertainties.


\subsection{Additional branches sensitive to the neutron density}


\subsubsection{The $s$-only isotopes $^{86,87}$Sr (the branch points at $^{85}$Kr and $^{86}$Rb)}\label{brkr85}

\begin{figure}
\includegraphics[angle=0,width=7cm]{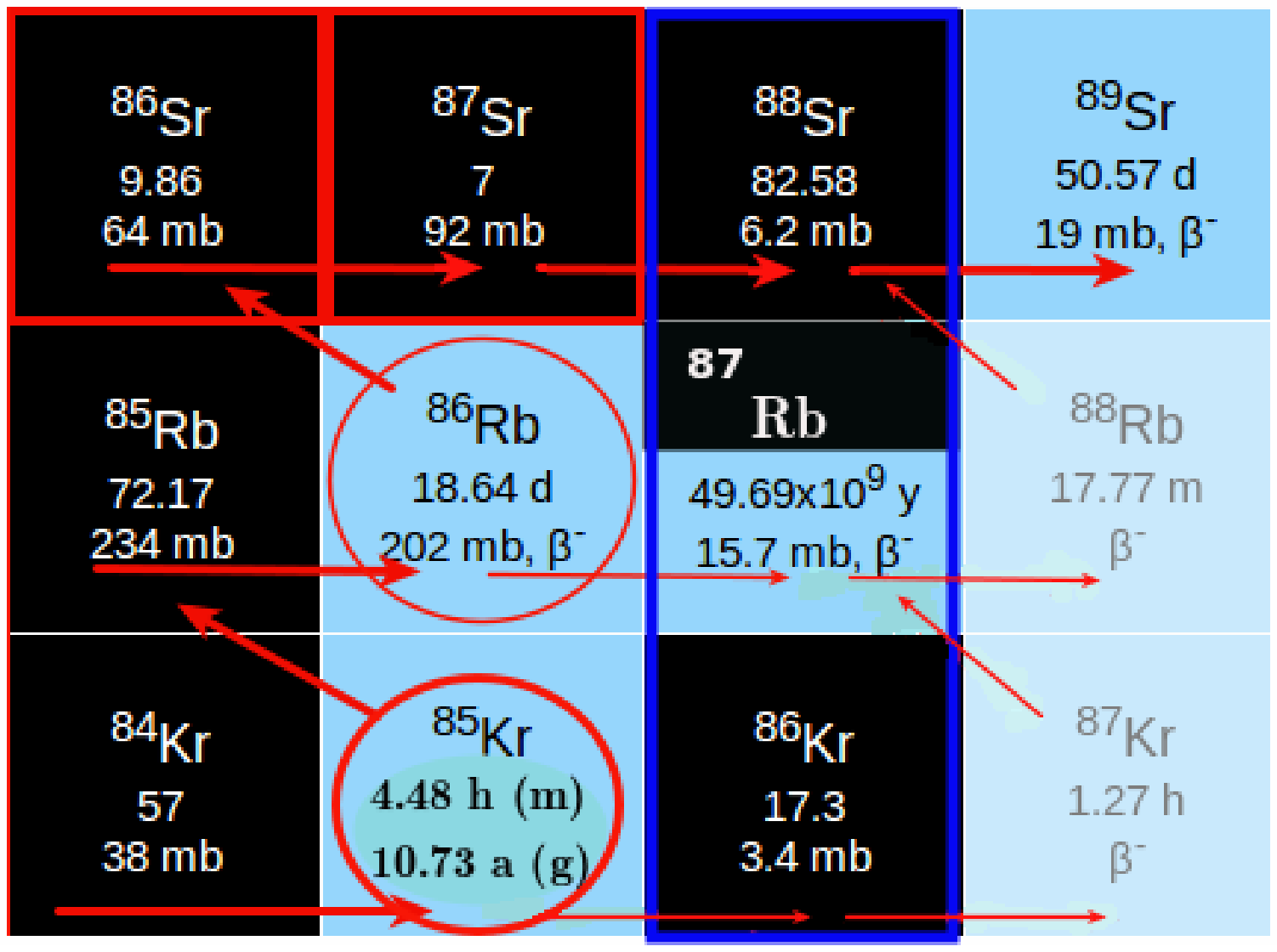}
\caption{The same as Fig.~\ref{branch160}, but for the $s$ path close to the 
$s$-only isotopes $^{86,87}$Sr. The branch at $^{85}$Kr is represented by a 
red circle.
The neutron-magic nuclei at N = 50 are indicated by a blue box.   }
\label{branch85kr}
\end{figure}

The abundances of the $s$-only isotopes
are sensitive to the branching at $^{85}$Kr and to a lesser extent to that
at $^{86}$Rb (see Fig.~\ref{branch85kr}).
\\
In fact, the branching starts already with neutron captures on $^{84}$Kr, which 
are feeding either the short-lived isomeric state of $^{85}$Kr$^{\rm m}$ ($t_{1/2}$ = 4.48 h)
or the long-lived ground state of $^{85}$Kr$^{\rm g}$ ($t_{1/2}$ = 10.73 yr).
The two $^{85}$Kr states are not thermalised and have to be treated
independently (Ward 1977). At least $\sim$80\% of the short-lived 
$^{85}$Kr$^{\rm m}$ $\beta^-$ decay 
to $^{85}$Rb, while the remaining 20\% decay 
to $^{85}$Kr$^{\rm g}$ by internal transitions.
Therefore, about 40\% of the $s$ flow coming from $^{84}$Kr feed 
$^{85}$Rb by $\beta^-$ decay of $^{85}$Kr$^{\rm m}$, and about 60\% produce 
$^{85}$Kr$^{\rm g}$.
\\
Due to the short half-life of $^{86}$Rb ($t_{1/2}$ = 18.6 d), this additional 
branching is only partially open at high neutron densities. The combined effect of
both branchings is completely determined by the neutron density,
because the half-lives of the branch point isotopes are insensitive
to the temperatures during the AGB phase (Takahashi and Yokoi 
1987).

\paragraph{During the $^{13}$C pocket. }

The neutron densities typical of the $^{13}$C pocket are too low to activate 
the neutron captures branch at $^{85}$Kr efficiently. The isomer 
with a half-life of the order of a few hours immediately $\beta^-$ decays to $^{85}$Rb,
and the ground state mainly $\beta^-$ decays as well in spite of its longer half-life 
($f_n$ $<$ 0.1 even at the maximum neutron density of $\sim$2$\times$10$^7$ cm$^{-3}$).
This means that less than $\sim$10\% of the $s$-process flow proceed 
towards $^{86}$Kr, which receives a correspondingly small $s$-process contribution 
during the $^{13}$C pocket ($\sim$10\% of solar $^{86}$Kr).
\\
The second branch point at $^{86}$Rb is not active during the $^{13}$C pocket
and the $s$-flow directly feeds $^{86,87,88}$Sr.
Only $\sim$7\% of solar $^{87}$Rb is produced during the \Can neutron irradiation.

\begin{table}
\caption{Updated solar $s$ contributions to $^{86}$Kr 
and $^{87}$Rb using our \Can rate by Denker et al. (1993; ``D93'', see Table~A1) 
and the \Can rate by La Cognata et al. (2013; ``L13'').
The present $s$ contributions are compared with 
those evaluated by changing the \Can rate by factors of two and three.}
 \label{tab1}
\resizebox{8.6cm}{!}{\begin{tabular}{lcccc}
\hline   
           &  D93  &  L13   &  D93$\times$2    &  D93/3  \\
\hline
 $^{86}$Kr &  16.4 &  18.6  &   19.6 (+19.5\%) &  10.7 ($-$34.7\%)  \\
 $^{87}$Rb &  18.3 &  19.8  &   20.5 (+12.0\%) &  13.7 ($-$25.1\%)  \\
\hline   
\end{tabular}}
\end{table}

The present contributions of the main component to
$^{86}$Kr and $^{87}$Rb are 16\% and 18\%, respectively (Table~A1). 
By including the recent \Can rate by La Cognata et al. (2013), 
these values increases to 19\% and 20\%, respectively, 
(see Table~\ref{tab1}), and are given with less than 3\% uncertainty. 
By assuming an extended range of uncertainty based on analyses given in literature,
variations up to 20--35\% are obtained by changing the \Can rate by factors of 
two and three.

All stable Br, Kr and Rb isotopes have well known (n, \g) cross sections (Mutti et al. 2005, 
Heil et al. 2008a).
Consequently, the major uncertainties affecting the $s$-process predictions in this
mass region are due to the branch point nuclei $^{85}$Kr and $^{86}$Rb, for
which only theoretical estimations are available.
The $\sim$80\% uncertainty of the $^{85}$Kr(n, \g)$^{86}$Kr MACS recommended by KADoNiS
(55 $\pm$ 45 mbarn at 30 keV; Bao et al. 2000) has been recently reduced:
Raut et al. (2013)  have measured photo-disintegration and (\g, \g') cross sections 
on $^{86}$Kr for improving the prediction of the $^{85}$Kr(n, \g) cross section.
Their recommended MACS of 83 $^{\rm +23}_{\rm -38}$ mb
is about 50\% higher than in Bao et al. (2000) with a two times
smaller uncertainty.
Nevertheless, the $^{85}$Kr MACS by Bao et al. (2000) 
and Raut et al. (2013) are in agreement within the respective uncertainties.
\\
The MACS uncertainty of $^{85}$Kr of Raut et al. (2013) results in variations of 
16--30\% and 10--20\% in the contribution of the main component to $^{86}$Kr and 
$^{87}$Rb during the $^{13}$C-pocket phase, respectively.

\paragraph{During thermal pulses. }

The neutron density reached during TPs by the \Nean reaction 
(up to a few 10$^{10}$ cm$^{-3}$) allows the activation of both branches 
at $^{85}$Kr and $^{86}$Rb, largely overcoming the uncertainties affecting 
$^{86}$Kr and $^{87}$Rb discussed above.
\\
About 40\% of the $s$ flux coming from $^{84}$Kr feeds $^{85}$Rb by $\beta^-$ decay 
of $^{85}$Kr$^{\rm m}$ ($t_{1/2}$ = 4.48 h), and 60\% produces $^{85}$Kr$^{\rm g}$ 
($t_{1/2}$ = 10.73 yr).
The longer-lived $^{85}$Kr$^{\rm g}$ easily captures neutrons, and the $s$ path proceeds 
towards $^{86}$Kr, $^{87}$Rb (a very long-lived isotope with $t_{1/2}$ = 
4.8$\times$10$^{10}$ yr), and $^{88}$Sr, bypassing $^{86,87}$Sr.
At the end of the convective instability, all the $^{85}$Kr$^{\rm g}$ stored during TP
decays into $^{85}$Rb (see Fig.~\ref{stamperb85}).
At $^{86}$Rb ($t_{1/2}$ = 18.7 d), neutron capture and $\beta^-$ decay compete 
for neutron densities larger than $\sim$10$^{10}$ cm$^{-3}$, further increasing 
the $^{87}$Rb abundance. 
The neutron-magic isotopes $^{86}$Kr, $^{87}$Rb and $^{88}$Sr (at N = 50) operate 
as bottlenecks of the $s$ path, owing to their very small neutron capture cross sections,
thus enhancing the effect of the branches at $^{85}$Kr$^{\rm g}$ and $^{86}$Rb.
\\
$^{86}$Kr and $^{87}$Rb are largely affected by the \Nean rate (see Table~A1):
the $s$-contributions to $^{86}$Kr and $^{87}$Rb increase by a factor of 1.6 and 2.3
by increasing the recommended \Nean rate by a factor of four.
The $^{86,87}$Sr $s$-predictions are less affected than $^{86}$Kr and $^{87}$Rb, with decreases
of $\sim$30\%.
The overabundances of $^{88}$Sr, which has been accumulated during the \Can neutron irradiation (MACS 
at 30 keV is 6.13 $\pm$ 0.11 mb; KADoNiS), is almost unchanged. Owing to the normalisation
effect to $^{150}$Sm (Section~5.1.1), the $s$ contribution to $^{88}$Sr shows variations of 
12.5\%.

Additional uncertainties affecting $^{86}$Kr and $^{85,87}$Rb derive from the  
theoretical MACS values of $^{85}$Kr and $^{86}$Rb (Raut et al. 2013; KADoNiS).
In Table~\ref{tab7} we evaluate the effects of the MACS uncertainties discussed above 
on the $s$ predictions of the isotopes from $^{84}$Kr to $^{88}$Sr.
While the MACS uncertainty of $^{86}$Rb influences only $^{87}$Rb up to 20\%, 
the one of $^{85}$Kr affects $^{86}$Kr as well as $^{87}$Rb (variations up to 40 and 25\%, 
respectively). Also $^{85}$Rb changes by 5--10\%.
The $s$-only isotopes $^{86,87}$Sr show smaller variations ($\la$2\%). 

\begin{figure}
\vspace{5 mm}
\includegraphics[angle=-90,width=9cm]{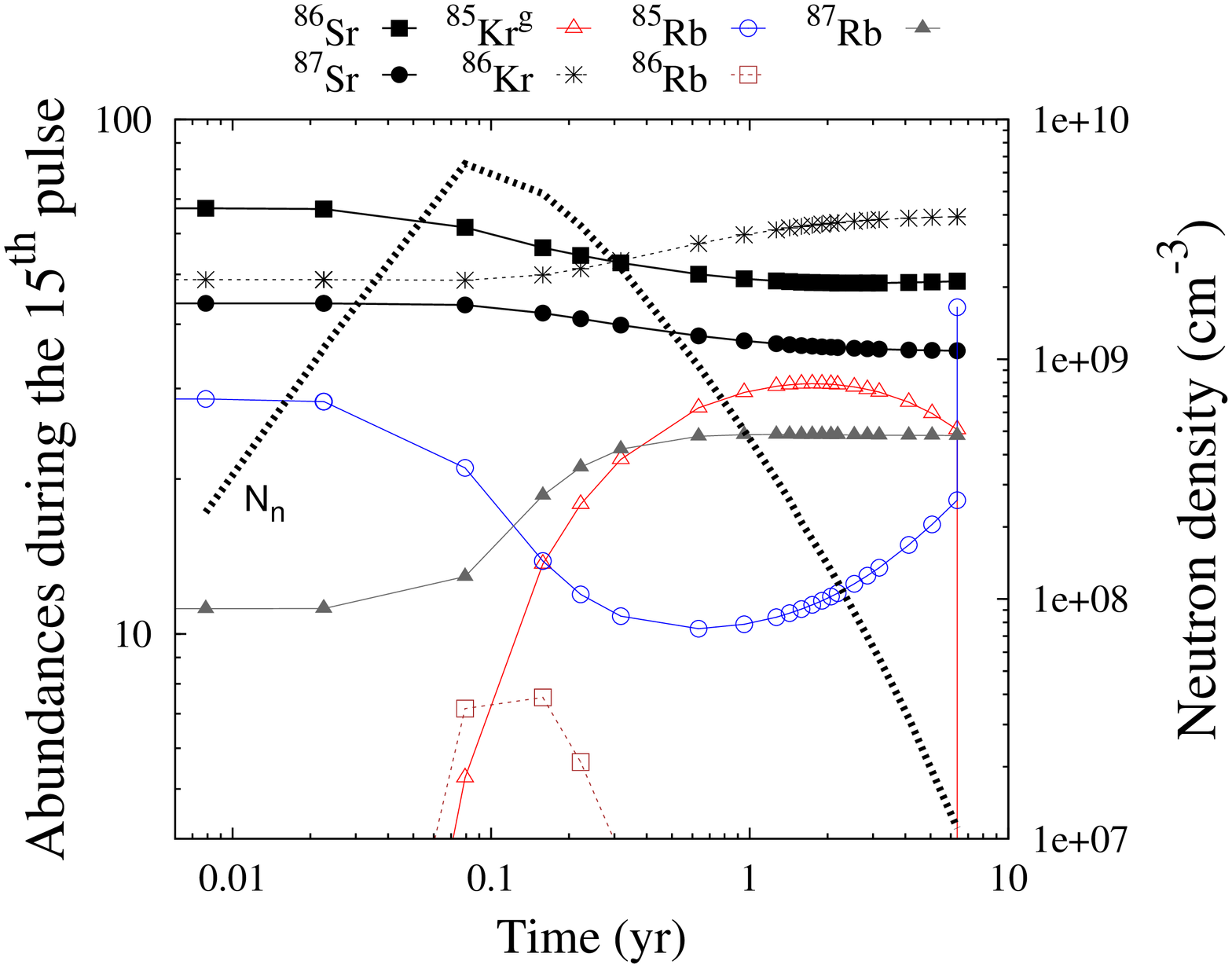}
\caption{Temporal evolution of the neutron density, and the isotopic abundances of
$^{86,87}$Sr, $^{85}$Kr$^g$, $^{86}$Kr and $^{85,86,87}$Rb during the 15$^{\rm th}$ He shell flash 
in an AGB star with 1.5 
$M_\odot$ and half-solar metallicity. The abundance values are given as number fractions normalised 
to $^{150}$Sm at the end of the He shell flash. The time-scale starts when the temperature at the 
bottom of the convective TP reaches $T_8$ = 2.5, which corresponds to the onset of the \Nean reaction. }
\label{stamperb85}
\end{figure}

\begin{table}
\caption{The abundance contributions by the main component (in \%) to the
isotopes from $^{84}$Kr to $^{88}$Sr compared with the corresponding values 
obtained with the upper and lower MACS limits of $^{85}$Kr and $^{86}$Rb 
(Raut et al. 2013, "R13"; KADoNiS, "KAD").
Variations higher than 10\% are listed between brackets.}
 \label{tab7}
\resizebox{6.5cm}{!}{\begin{tabular}{l|ccc}
\hline    
    & \multicolumn{3}{l}{Tests of the $^{85}$Kr(n, \g)$^{86}$Kr MACS}  \\
          \hline
Isotope   & [R13]   & [R13]LL  & [R13]UL   \\
$^{84}$Kr &  10.5  &  10.5        &  10.6      \\ 
$^{86}$Kr &  16.4  &   9.5 (0.58) &  20.4 (1.24)\\ 
$^{85}$Rb &  12.0  &  13.4 (1.12) &  11.4 (0.95)\\ 
$^{87}$Rb &  18.3  &  13.8 (0.75) &  21.0 (1.15)\\ 
$^{86}$Sr &  58.4  &  59.2        &  58.6      \\ 
$^{87}$Sr &  58.5  &  59.1        &  58.7      \\ 
$^{88}$Sr &  82.6  &  83.3        &  83.1      \\                                          
\hline
    & \multicolumn{3}{l}{Tests of the $^{86}$Rb(n, \g)$^{87}$Rb MACS}  \\
          \hline
Isotope   & KAD   & KAD LL  & KAD UL    \\
$^{84}$Kr &  10.5  &  10.5        &  10.5        \\
$^{86}$Kr &  16.4  &  16.5        &  16.6        \\
$^{85}$Rb &  12.0  &  12.0        &  12.0        \\
$^{87}$Rb &  18.3  &  15.7 (0.86) &  21.9 (1.20) \\
$^{86}$Sr &  58.4  &  60.3        &  56.8       \\
$^{87}$Sr &  58.5  &  59.8        &  57.7       \\
$^{88}$Sr &  82.6  &  83.2        &  83.2       \\
\hline
\end{tabular}}
\end{table}

 Note that the solar $s$ abundances of isotopes with $A$ $<$ 90 receive additional contributions.
\\
Firstly, the weak $s$ process in massive stars synthesises isotopes from Cu up to Kr-Rb-Sr
(Rauscher et al. 2002; Pignatari et al. 2013; Chieffi \& Limongi 2013). In this case, 
the $s$ predictions are largely 
affected by the $^{12}$C $+$ $^{12}$C rate (see, e.g., Pignatari et al. 2013).
\\
Second, intermediate-mass AGB stars contribute to solar $^{86}$Kr and $^{87}$Rb.
The $s$ contribution is difficult to be assessed because of the large uncertainty
affecting the treatment of mass loss in IMS AGB stars. Bisterzo et al. (2014) estimate
that the predicted contribution to the solar $^{86}$Kr abundance may increase from 15\% up 
to $\sim$+35\% and a still larger effect is found for $^{87}$Rb (from $\sim$25\% up to $\sim$70\%).
In Section~C (Supplementary Information), 
we discuss the result of a $M$ = 5 $M_\odot$ model at half-solar metallicity,
in the light of the neutron source uncertainties.


\subsubsection{The $s$-only isotope $^{142}$Nd (the branches at $^{141}$Ce and $^{142}$Pr)}\label{142nd}

\begin{figure}
\includegraphics[angle=0,width=8cm]{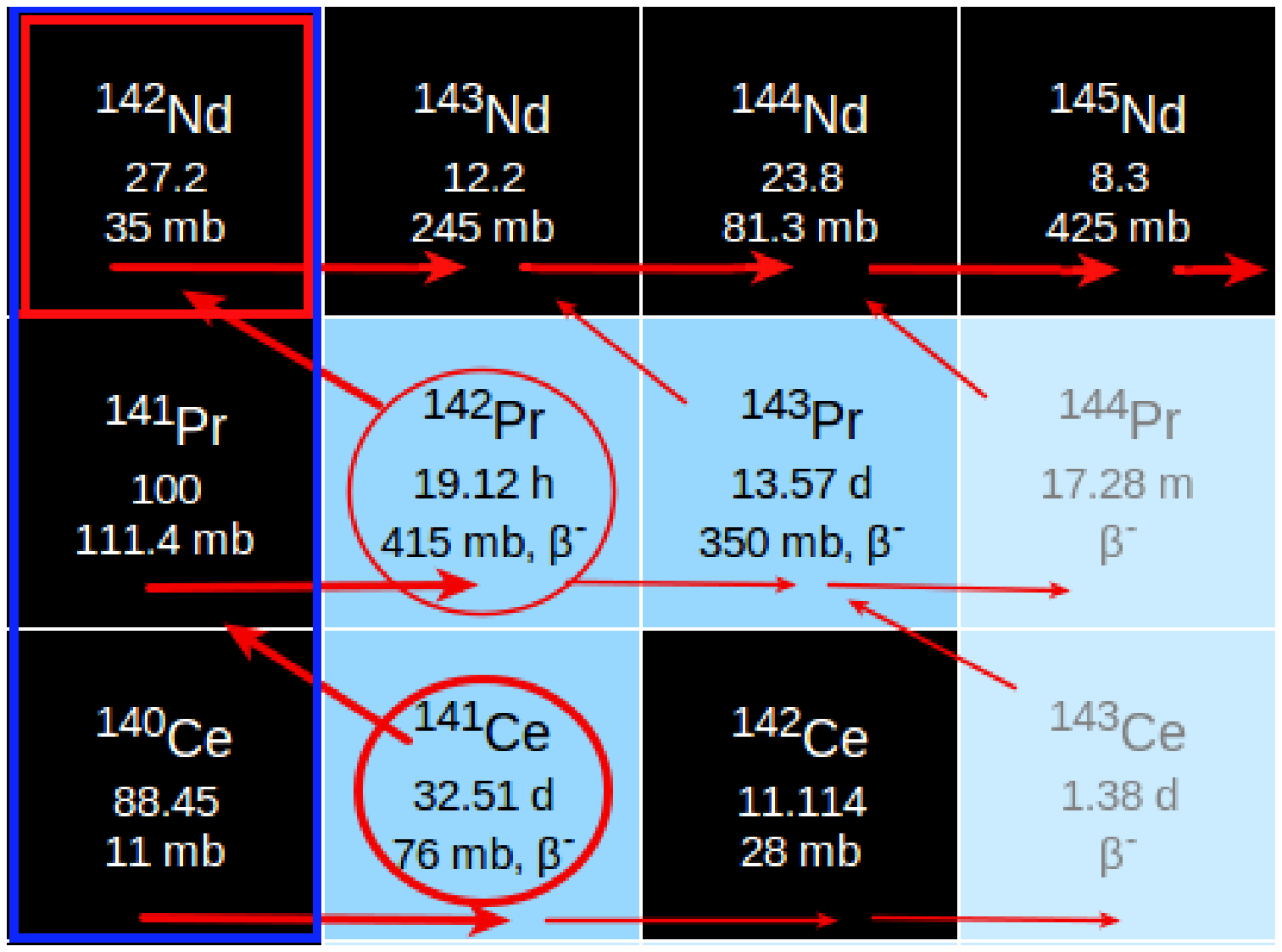}
\caption{Schematic representation of the chart of nuclei in the atomic mass region
close to the
$s$-only isotope $^{142}$Nd (red square). 
The neutron-magic nuclei at N = 82 are indicated by a blue box.}
\label{branchce141}
\end{figure}

$^{142}$Nd belongs to the magic nuclei at N = 82, and has a fairly small MACS (35.0 $\pm$ 0.7 mbarn; 
KADoNiS).
The reproduction of the solar $^{142}$Nd abundance was the first evidence that the classical model 
was not adequate to describe the main component close to the neutron-magic 
numbers (N $\sim$ 50, 82, and 126), located at the deep steps of the $<$$\sigma$$>$N$_s$ curve 
(Seeger et al. 1965, Clayton 1968, K{\"a}ppeler et al. 1989).
The success of the stellar model in reproducing the $s$-process pattern of several branchings
(e.g., $^{142}$Nd, and $^{116}$Sn, $^{134,136}$Ba as well) was demonstrated by Arlandini et al. (1999).

The $s$-only isotope $^{142}$Nd is shielded against the $r$-process by $^{142}$Ce. 
Most of solar $^{142}$Nd is synthesised during the $^{13}$C-pocket phase where the $s$ path
is restricted to the neutron-magic isotopes $^{140}$Ce, $^{141}$Pr, and $^{142}$Nd
(Fig.~\ref{branchce141}) because of the low neutron density. The branchings at
$^{141}$Ce and $^{142}$Pr, which open at higher neutron densities, are
affecting the $s$ abundance of $^{142}$Nd only during the TPs.

In spite of the relatively long half-life of $^{141}$Ce ($t_{1/2}$ = 32.51 d, 
constant at $s$-process temperatures), this branching opens only at high neutron 
densities ($f_n$ = 0.2 for $N_n$ = 4$\times$10$^{9}$ cm$^{-3}$) because of
the small MACS of 76 $\pm$ 33 mb at $kT$ = 30 keV. About 8\% of the
reaction flow proceed via $^{141}$Ce, contributing $\sim$20\% to the solar
abundance of $^{142}$Ce.
\\ 
The second branch at $^{142}$Pr is characterized by a shorter
half-life (19.25 h, increasing by a factor of four at $T_8$ = 3; Takahashi
and Yokoi 1987), only partially compensated by a larger MACS of
415 $\pm$ 178 mb (KADoNiS). Therefore, this branching remains weak
even during TPs, reaching $f_n$ $\ga$ 0.1 only for neutron densities 
$N_n$ $\ga$ 3$\times$10$^{9}$ cm$^{-3}$.

The present contribution of the main component to the $^{142}$Nd abundance is 104.5\%. 
Note that this value includes the small contribution from the $\alpha$ decay of the long-lived
$^{146}$Sm ($t_{1/2}$ = 103 Myr).

The uncertainties of the theoretical MACS values of $^{141}$Ce and $^{142}$Pr (43\% each) 
are causing variations of about 4\% in the $^{142}$Nd abundance. 
A similar effect of 5\% is found if the \Nean rate is increased by a factor of four. 
The $s$ contribution to solar $^{142}$Nd shows up to 15\% variations (owing to the      
 additional effect of the \Nean tests of the $^{150}$Sm $s$-prediction; Section~5.1.1).


\subsubsection{The $s$-only isotope $^{186}$Os (the branches at $^{185}$W and $^{186}$Re)}\label{w185}

\begin{figure}
\includegraphics[angle=0,width=7cm]{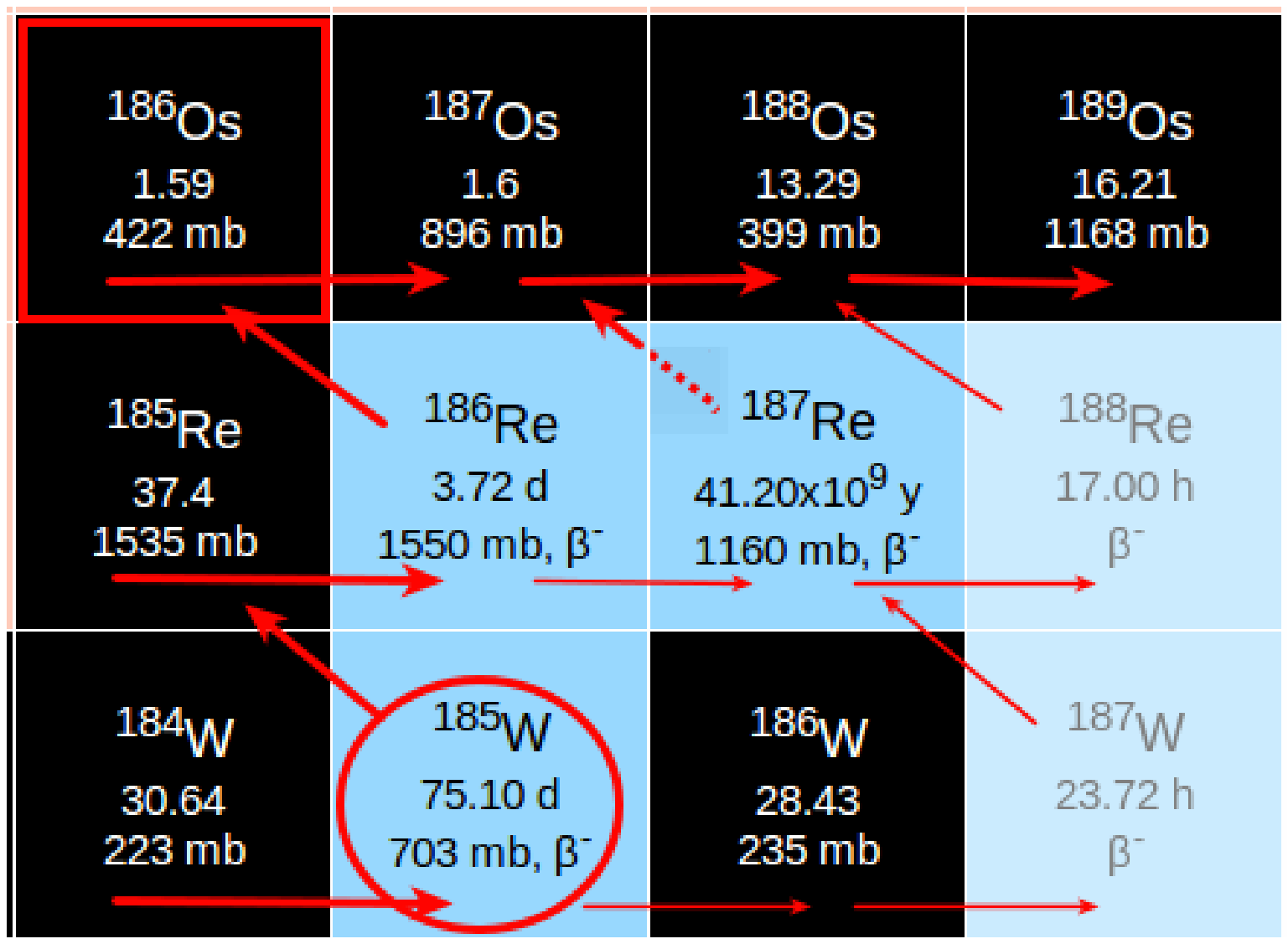}
\caption{Schematic representation of the chart of nuclei in the atomic mass region
close to the $s$-only isotope $^{186}$Os (red square). 
The branch at $^{185}$W, mainly activated during TPs, is represented with a red circle.
$^{187}$Os is shielded from the $r$-process by its long-lived isobar $^{187}$Re
($t_{1/2}$ = 4.12$\times$10$^{10}$ yr), which contributes to its cosmoradiogenic 
enrichment. }
\label{branchw185}
\end{figure}

\begin{figure}
\vspace{5 mm}
\includegraphics[angle=-90,width=9cm]{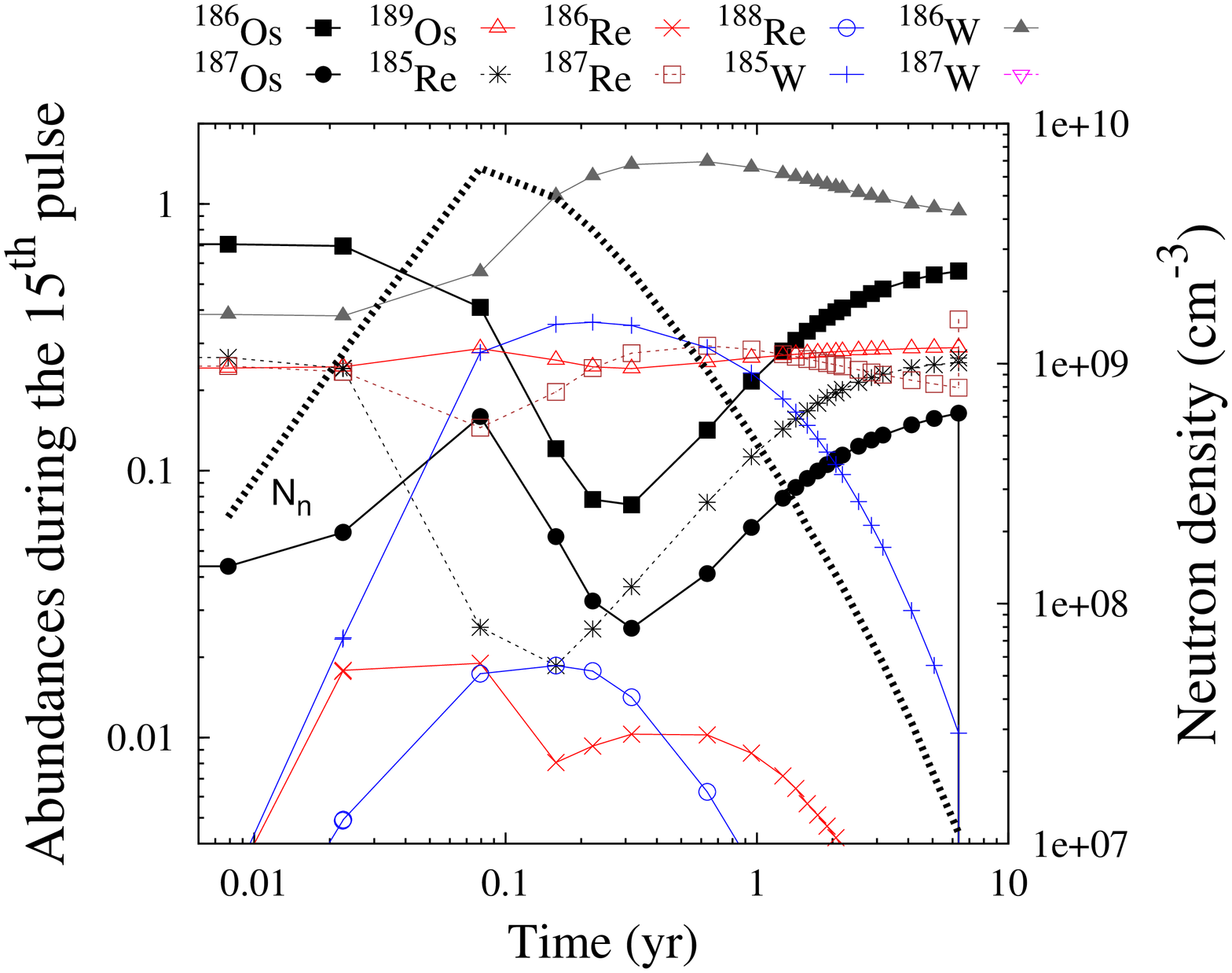}
\caption{The same as Fig.~\ref{stamperb85}, but for the isotopic abundances of 
$^{186,187,189}$Os, $^{185,186,187,188}$Re and $^{185,186,187}$W. }
\label{stampew185}
\end{figure}

The abundance of $^{186}$Os is mainly regulated by the branch at $^{185}$W,
which is efficiently activated during TPs (Fig.~\ref{branchw185}).
\\
Actually, the main component overestimates the solar abundance of $^{186}$Os by
30\%, essentially due to the recommended MACS for the branch point nucleus $^{185}$W 
(KADoNiS). This recommendation represents a weighted average between two measurements 
of the inverse $^{186}$W(\g, n)$^{185}$W reaction by Sonnabend et al (2003; 
$\sigma$($^{185}$W) = 687 $\pm$ 110 mbarn at 30 keV) and Mohr et al. (2004;
 $\sigma$($^{185}$W) = 553 $\pm$ 60 mbarn at 30 keV).
\\
This problem was firstly emphasized by Sonnabend et al. (2003), who found a 20\% 
excess of $^{186}$Os. By adopting the upper limit of this experimental value, the $s$
abundance of $^{186}$Os is still overestimated by 14\%, incompatible with
the 8\% uncertainty of the solar Os abundance (Lodders et al 2009).

The half-life of $^{185}$W decreases slightly during TPs, from $t_{1/2}$ = 75.1 d to $\sim$53 d 
at $T_8$ = 3 (Takahashi \& Yokoi 1987, depending on electron density). 
The 10\% uncertainty of the stellar half-life estimated by Goriely
(1999) is also not sufficient to explain the predicted overabundance
of $^{186}$Os.

The neutron cross sections of the Os isotopes are well determined by Mosconi et al. (2010): 
the $^{186}$Os MACS is known at 4.1\% (Mosconi et al. 2010, 430.6 $\pm$ 17 mbarn), and produces 
marginal variations of the $^{186}$Os $s$-contribution (2\%).
Possible uncertainties related to the SEF estimation have been substantially reduced 
by Fujii et al. (2010).

On the other hand, the second branch point at $^{186}$Re is only marginally activated and
can also be excluded as an explanation of the $^{186}$Os problem.
Indeed, $^{186}$Re decays quickly into $^{186}$Os ($t_{1/2}$ = 4.09 d) and 
neutron captures are favoured only at very high neutron density ($f_n$ $>$ 0.5 for 
$N_n$ $\ga$ 4$\times$10$^9$ cm$^{-3}$; see Fig.~\ref{stampew185}). 
The $^{186}$Re $\beta^+$-decay channel is negligible.
Moreover, $^{186}$Re does not build up to a significant abundance because of its large 
calculated MACS (1550 $\pm$ 250 mb, 16\% estimated uncertainty) and has, therefore, a negligible 
effect on the $^{186}$Os $s$-abundance ($\sim$1\% ). 
\\
Also the long-lived $^{186}$Re isomer does not provide any appreciable contribution at $s$-process
temperatures, and the branch at $^{186}$Re remains small (Mohr et al. 2008, 
Hayakawa et al. 2005).
Finally, the experimental photodisintegration measurement of the inverse $^{187}$Re(\g, n)$^{186}$Re 
reaction carried out by M{\"u}ller et al. (2006)
provides a $^{186}$Re MACS similar to that given by Bao et al. (2000).

In order to reproduce the solar $^{186}$Os abundance, we suggest that
the MACS of $^{185}$W should be 50\% higher than the first value
recommended by Bao et al. (2000). This guess is based on the fact that the uncertainty
of the $^{185}$W MACS may be underestimated because only the inverse ($\gamma$, n) reaction
has been measured. Rauscher (2014) estimates that our guess may be plausible. However,
specific theoretical analyses on the $^{185}$W(n, $\gamma$) reaction should be carried out 
to provide a more realistic uncertainty. 
\\
The $^{186}$Os $s$-contribution, computed by including the above guess on the $^{185}$W MACS,
is overestimated by 6\%, in agreement within the uncertainties.
By increasing the recommended \Nean rate by a factor of four, the solar $s$ $^{186}$Os is reduced to 
$\sim$98\%.

$^{187}$Os is shielded from the $r$-process by its stable isobar $^{187}$Re. 
Apart from neutron captures on $^{186}$Os, $^{187}$Os is also produced by 
the cosmoradiogenic $\beta^-$ decay of $^{187}$Re. The half-life of the ground
state ($t_{1/2}$ = 4.35 $\times$ 10$^{10}$ yr) is reduced in stellar interiors 
(Takahashi \& Yokoi 1987; Bosch et al. 1996), but
remains quasi stable within the time scale of TPs. The stable
daughter $^{187}$Os becomes unstable under stellar conditions
(Takahashi and Yokoi 1987), but behaves also like a stable
isotope during the 6 yr duration of a TP. Therefore, its temporal
evolution follows essentially that of $^{186}$Os, which is depleted during
the peak neutron density and re-established towards the end of the
TP (see Fig.~\ref{stampew185}).
A comprehensive investigation of the $^{187}$Os/$^{187}$Re ratio
with chemical evolution models
will be the topic of a future work.


\subsubsection{The $s$-only isotope $^{192}$Pt (the branch at $^{192}$Ir)}\label{pt192}

\begin{figure}
\includegraphics[angle=0,width=8.5cm]{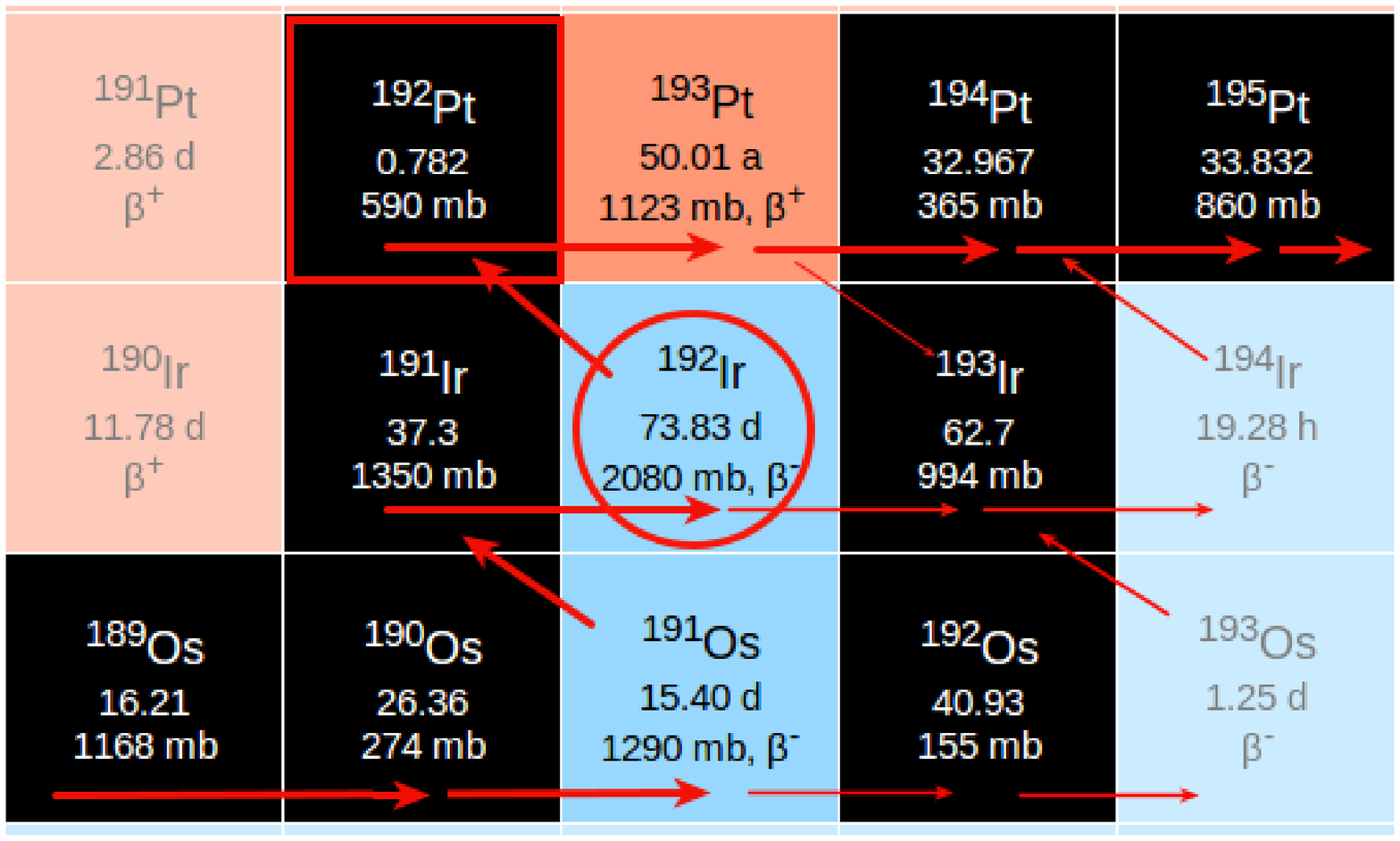}
\caption{Schematic representation of the chart of nuclei in the atomic mass region
close to the $s$-only isotope $^{192}$Pt (red square). 
The branch at $^{192}$Ir, mainly activated during TPs, is represented with a red circle.}
\label{branchpt192}
\end{figure}

\begin{figure}
\vspace{5 mm}
\includegraphics[angle=-90,width=9cm]{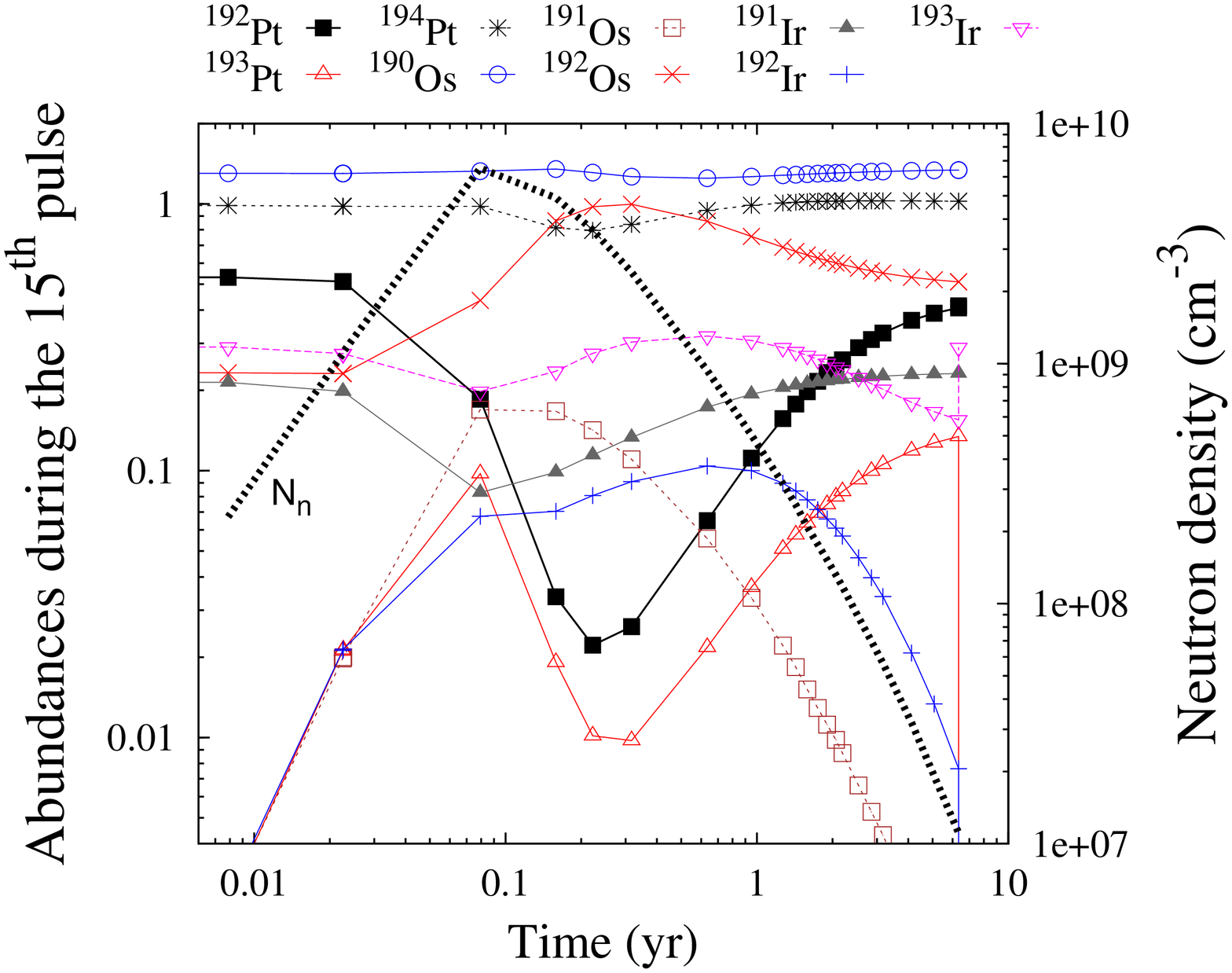}
\caption{The same as Fig.~\ref{stamperb85}, but for the isotopic abundances of 
$^{192,193,194}$Pt, $^{190,191,192}$Os and $^{191,192,193}$Ir.}
\label{stampept192}
\end{figure}

The abundance of $^{192}$Pt is mainly determined by the branch at $^{192}$Ir 
(Fig.~\ref{branchpt192}).
At neutron densities higher than 2$\times$10$^8$ cm$^{-3}$, $^{192}$Pt is mainly 
bypassed by neutron captures on $^{192}$Ir (the terrestrial half-life of 77.54 d is 
reduced to 55 d at $T_8$ = 3), to be partially restored 
(78\% of the $^{192}$Pt abundance produced during the previous 
$^{13}$C pocket; see Fig.~\ref{stampept192}) at the end of TP.
\\
The branch at $^{191}$Os has a smaller influence on $^{192}$Pt, because of the 
shorter half-life (the terrestrial value of 15.04 d is reduced by a
factor of two at $T_8$ = 3, Takahashi and Yokoi 1987). Therefore,
neutron captures on $^{191}$Os (MACS = 1290 $\pm$ 280 mbarn at 30 keV; KADoNiS) 
can compete only at peak neutron densities ($N_n$ $>$ 3.6$\times$10$^9$ 
cm$^{-3}$; $f_n$ $>$ 0.5).
 The $s$-process prediction underestimates the solar $^{192}$Pt by about 20\%. 
\\
Previously, the missing $s$ contribution was compatible with the 20\% uncertainties 
that were estimated for all theoretical $^{192}$Pt, $^{191}$Os and $^{192}$Ir MACS
($\sigma$($^{192}$Pt) = 590 $\pm$ 120 mbarn, $\sigma$($^{191}$Os) = 1290 $\pm$ 280 mbarn, and
$\sigma$($^{192}$Ir) = 2080 $\pm$ 450 mbarn at 30 keV; KADoNiS).
The $^{192}$Ir $\beta$-decay rate has a marginal impact on the $^{192}$Pt abundance.
\\
Recently, Koehler et al. (2013) measured the neutron capture cross sections of Pt 
isotopes with much improved accuracy (e.g., $\sigma$($^{192}$Pt) = 483 $\pm$ 20 mbarn
at 30 keV; 4\% uncertainty), and used their experimental results to provide
a new theoretical estimation of the $^{192}$Ir MACS (3220 $\pm$ 720 mbarn at 30 keV).
With the new MACS by Koehler et al. (2013) the uncertainty was significantly reduced 
but the $s$ contribution of 81\% remains too small. Allowing for a 2$\sigma$ uncertainty 
of the theoretical MACS of $^{192}$Ir, the $s$ contribution of $^{192}$Pt can be increased 
to 95\%, consistent with the 8\% uncertainty of the solar Pt abundance (Lodders 
et al. 2009).
\\
More detailed analyses concerning the MACS of $^{192}$Ir would help to improve the
understanding of this branching point and to reproduce the solar abundance of $^{192}$Pt more
accurately.

The impact of the \Nean neutron source is marginal in this case. An increase
of the recommended \Nean rate by a factor of four is affecting the $s$ abundance of $^{192}$Pt by 
less than 4\%.


\subsection{Additional branches strongly sensitive to stellar
temperature (and/or electron density) and neutron density}


\subsubsection{The $s$-only isotopes $^{152,154}$Gd (the branches at $^{151}$Sm and $^{154}$Eu)}\label{152154gd}

\begin{figure}
\includegraphics[angle=0,width=8.5cm]{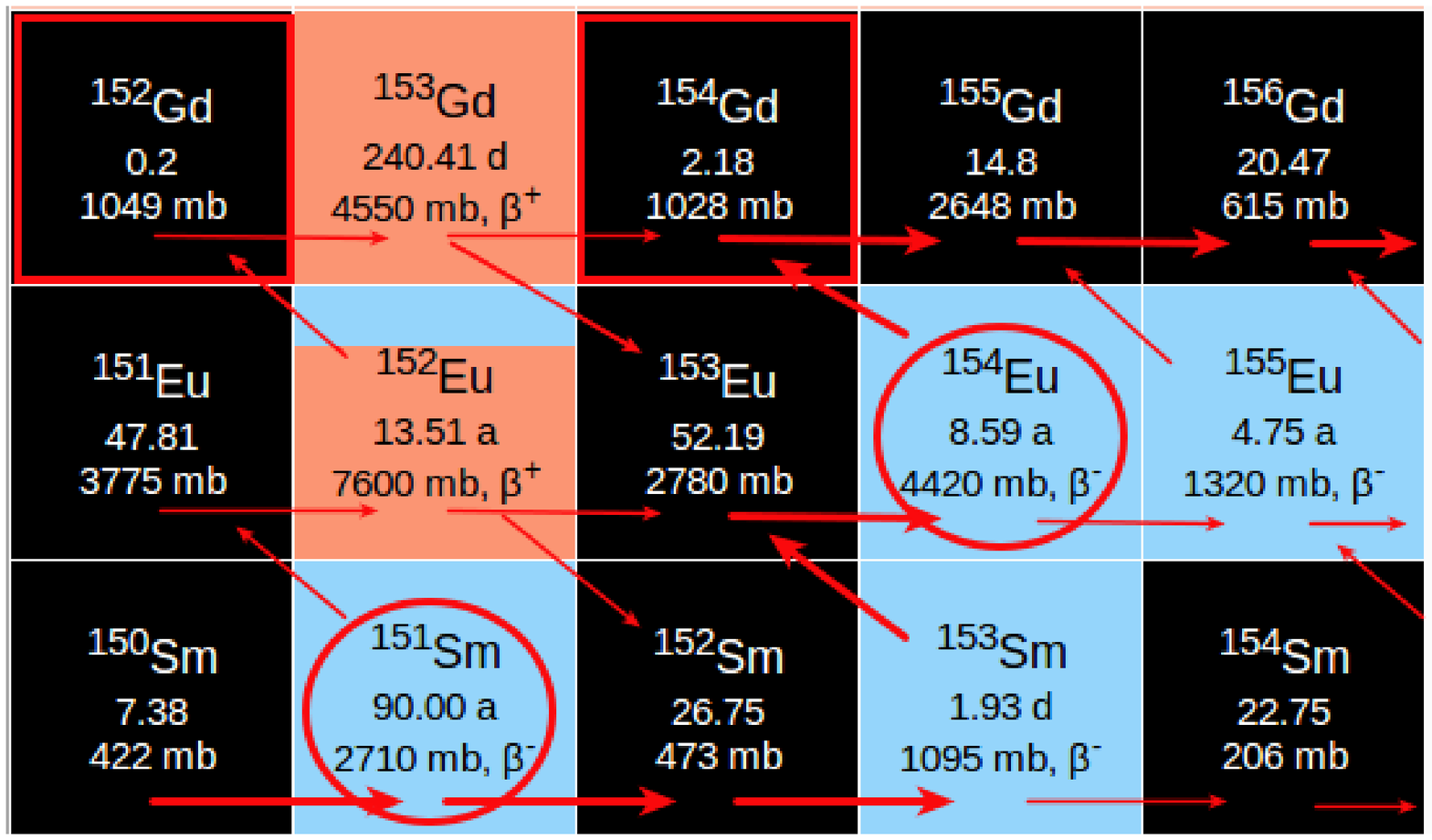}
\caption{Schematic representation of the chart of nuclei in the atomic mass region
close to the
$s$-only isotopes $^{152,154}$Gd (red squares). 
The terrestrial half-life of $^{151}$Sm ($t_{1/2}$ = 90 yr), which depends 
strongly on temperature and neutron density (Takahashi and Yokoi 1987), is reduced 
to a few years at $T_8$ = 3. From $^{152}$Eu, the $s$-process flow proceeds directly 
to $^{152}$Gd, because the half-life of $^{152}$Eu is reduced to about 5 hours at stellar
temperatures. }
\label{branchsmeugd}
\end{figure}

\begin{figure}
\vspace{5 mm}
\includegraphics[angle=-90,width=9cm]{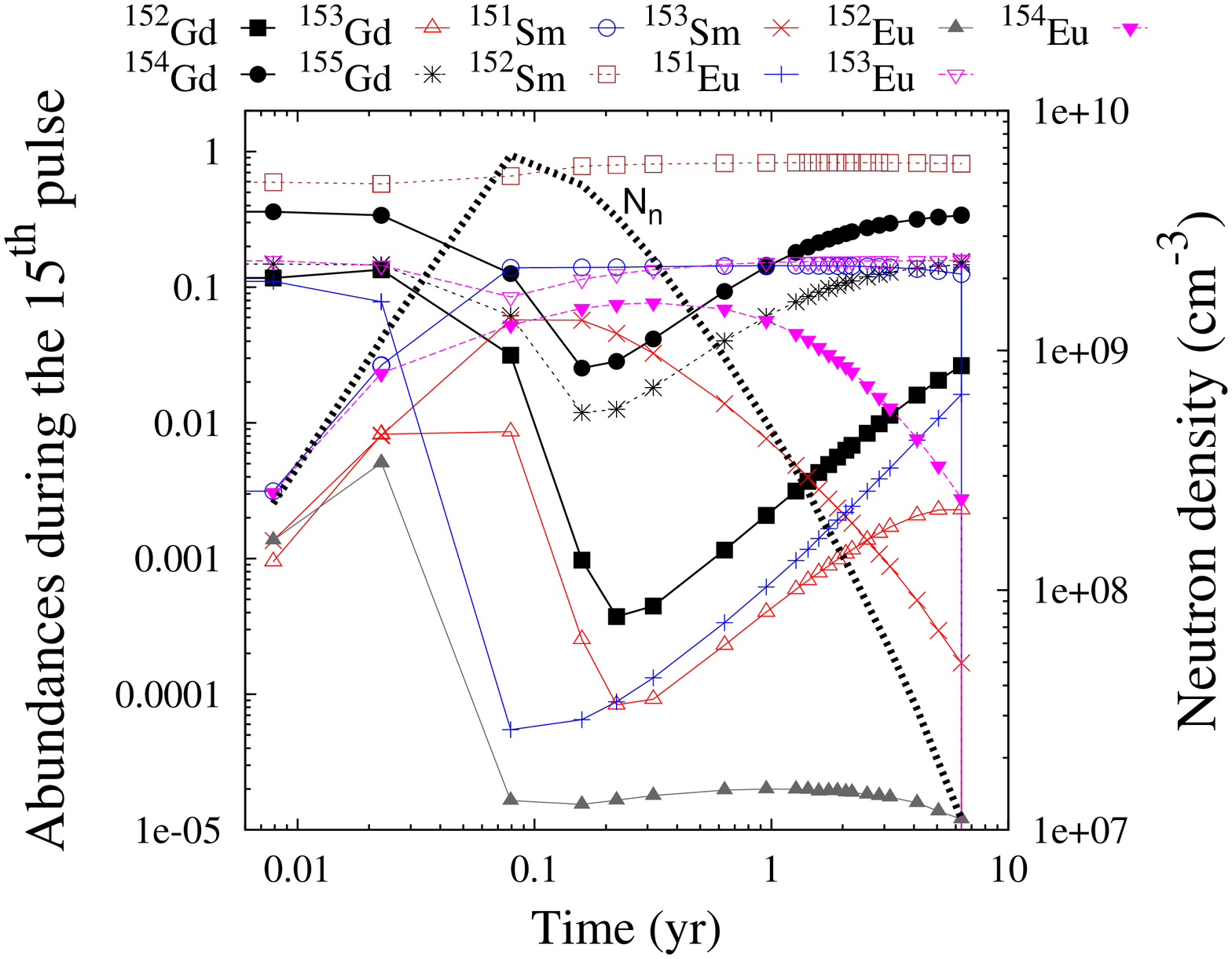}
\caption{The same as Fig.~\ref{stamperb85}, but for the isotopic abundances of
 $^{152,153,154,155}$Gd, $^{151,152,153}$Sm and $^{151,152,153,154}$Eu. }
\label{stampesm151}
\end{figure}

The production of the $s$-only isotopes $^{152,154}$Gd is affected by two main branches
at $^{151}$Sm and $^{154}$Eu, respectively (Fig.~\ref{branchsmeugd}). 

During the $^{13}$C-pocket phase most of the $s$ path feeds the Sm isotopes from A = 
150 to 153, mainly bypassing $^{152}$Gd ($\sim$0.2\% of solar Gd).
Although the terrestrial half-life of $^{151}$Sm ($t_{1/2}$ = 90 yr) may decrease 
to $t_{1/2}$ = 55 yr (at $T_8$ = 1 and an electron density $n_e$ = 30$\times$10$^{26}$ 
cm$^{-3}$, Takahashi \& Yokoi 1987), the neutron density is still sufficient to bypass
$^{152}$Gd ($N_n$ $\ga$ 1$\times$10$^6$ cm$^{-3}$; $f_n$ $\ga$ 0.5).
\\
At the peak neutron densities of the TPs ($N_n$ $>$ 10$^9$ cm$^{-3}$, $f_n$ $>$ 0.2) 
when the half-life of $^{151}$Sm is reduced by a
factor of 30 ($t_{1/2}$ $\sim$ 3 yr at $T_8$ = 3), the branching at $^{151}$Sm has a
strong effect on $^{152}$Gd. Accordingly, $^{151}$Eu and $^{152}$Gd are depleted
near the maximum of $N_n$, before they are partially restored during
the freeze-out phase. As shown in Fig.~\ref{stampesm151} the $^{151}$Eu abundance is
even increasing ($\sim$20\%) above the value after the $^{13}$C-pocket
phase, because of the additional feeding by the decay of $^{151}$Sm
after the end of the TPs. In contrast, $^{152}$Gd, which feels the strong variations 
of neutron density and temperature in the He flash, reaches only about 20\% of the 
abundance produced in the $^{13}$C pocket.
 
Partial neutron captures on $^{151}$Sm are allowed at the bottom of the advanced TPs when 
peak neutron densities are reached ($N_n$ $>$ 10$^9$ cm$^{-3}$, $f_n$ $>$ 0.2), and $^{151}$Eu and 
$^{152}$Gd are mostly destroyed (see Fig.~\ref{stampesm151}).
At the decreasing tail of the neutron density, most of $^{151}$Sm present in the hot TP layers 
$\beta^-$ decays into $^{151}$Eu, feeding again $^{152}$Gd. The $^{151}$Sm abundance stored at the 
end of the TP decays into $^{151}$Eu during the interpulse phase, increasing the $^{151}$Eu amount 
present in the previous $^{13}$C pocket ($\sim$+20\%). 
As shown in Fig.~\ref{stampesm151}, $^{152}$Gd feels the strong variations of neutron density
and temperature in the He flash, being destroyed at peak neutron density and partially 
rebuilt at the end of the TP: only $\sim$20\% of $^{152}$Gd yielded during the previous $^{13}$C pocket 
is restored.

The branch at $^{154}$Eu has important effects on the $s$-prediction of $^{154}$Gd.
\\
The terrestrial $^{154}$Eu half-life ($t_{1/2}$ = 8.8 yr) strongly decreases during convective 
instabilities: from a few years down to a few days at the bottom layers of the advanced TPs 
($t_{1/2}$ = 10.67 d at $T_8$ = 3, Takahashi \& Yokoi 1987). Given the large probability for $^{154}$Eu 
to capture neutrons (MACS($^{154}$Eu) = 4420 $\pm$ 670 mb at 30 keV; KADoNiS), the $s$ path mainly 
proceeds towards $^{155}$Eu starting from $N_n$ $\sim$9$\times$10$^8$ cm$^{-3}$ ($f_n$ $\ga$ 0.5),
and bypasses $^{154}$Gd.
When the neutron density decreases, the abundance of $^{154}$Gd produced during the previous $^{13}$C pocket 
is almost completely restored ($\sim$94\%, Fig.~\ref{stampesm151}).

The additional branchings at $^{153}$Sm, $^{152}$Eu, and $^{153}$Gd are relatively weak. 
Almost the entire $s$-process flow runs from $^{152}$Sm to $^{153}$Eu due to the short 
half-life of $^{153}$Sm ($t_{1/2}$ = 1.1 d at $T_8$ = 3). While the rather slow terrestrial 
decay of $^{152}$Eu (half-lives of 13.51 and 47.84 yr for the $\beta^+$ and EC
channels, respectively), the $\beta^-$ decay dominates under stellar
conditions ($t_{1/2}^-$ = 20 and 5 h at $T_8$ = 2 and 3, Takahashi and Yokoi
1987), directing the $s$-process flow from $^{151}$Eu directly to $^{152}$Gd.
\\
The next unstable isotope of the $s$ path is $^{153}$Gd, which slowly $\beta^+$ decays with 
$t_{1/2}$ = 240 d (rather constant under stellar conditions, Takahashi \& Yokoi 1987). 
Due to the large MACS recommended for $^{153}$Gd (4550 $\pm$ 700 mb at 30 keV; KADoNiS) 
the reaction flow through $^{152}$Gd continues to $^{154}$Gd already at low neutron densities
($N_n$ $\ga$ 4$\times$10$^7$ cm$^{-3}$; $f_n$ $\ga$ 0.5).

\vspace{2mm}

 Previous AGB models (see $filled$ $circles$ in Fig.~5) predict
that 81\% 
of solar $^{152}$Gd and 87\% of solar $^{154}$Gd are produced by the main component. 
These results were computed by adopting constant $\beta$-decay rates for $^{151}$Sm and $^{154}$Eu. 
 These values increase to 85\% and 92\%, respectively, by including an improved 
treatment of the $\beta$-decay rates of unstable isotopes close to $^{152,154}$Gd 
($^{151}$Sm, $^{154}$Eu, $^{153}$Sm, $^{152}$Eu, $^{153}$Gd)
over the convective TPs (see $filled$ $diamond$ in Fig.~5).
\\
 However, the $s$ predictions of $^{152,154}$Gd are largely influenced by the 
uncertainties of the decay rates of $^{151}$Sm and $^{154}$Eu. According to
Goriely (1999) the $\lambda^-$ rate of $^{151}$Sm is increasing by factors of $\sim$2 and
3 at $T_8$ = 2 and 3, respectively, and a similar uncertainty is estimated for the $^{154}$Eu $\beta$-decay 
rate.
This may produce up to 10-15\% variations of the $^{154}$Gd $s$-contribution, and an
extreme impact on the $^{152}$Gd $s$-prediction (up to a factor of 2).
\\
Consequently, an appropriate treatment of the $\beta$-decay rates close to $^{152,154}$Gd
improves the $s$ predictions, but the uncertainty remains large.
Investigation of the $^{151}$Sm and $^{154}$Eu half-lives at stellar temperature 
is suggested to assess the $s$ contribution to $^{152,154}$Gd.

Marginal effects on $^{152,154}$Gd are produced by the MACS uncertainties.
Wisshak et al. (2006) and Marrone et al. (2006) have provided important
constraints on the branch at $^{151}$Sm (3031 $\pm$ 69 mbarn and 3080 $\pm$ 150 mbarn, respectively).
In our calculations we adopt a weighted average of the two measurements.
The MACS of $^{152,154}$Gd are well determined 
with less than 3\% of uncertainty (KADoNiS; Bao et al. 2000).
The 15\% uncertainties of the measured MACS of the stable isotopes $^{152,154}$Eu and the theoretical
MACS of $^{153}$Gd are somewhat larger, but have no significant effect on
the predictions for $^{152,154}$Gd ($<$1--2\%, respectively).
 
An additional uncertainty comes from 
the \Nean reaction: increasing the recommended rate by a factor of four, the $s$ contribution to 
solar $^{152}$Gd decreases to 69\%, whereas $^{154}$Gd is affected by only $\sim$4\%.


\subsubsection{The $s$-only isotopes $^{176}$Lu and $^{176}$Hf}\label{176lu}

\begin{figure}
\includegraphics[angle=0,width=6cm]{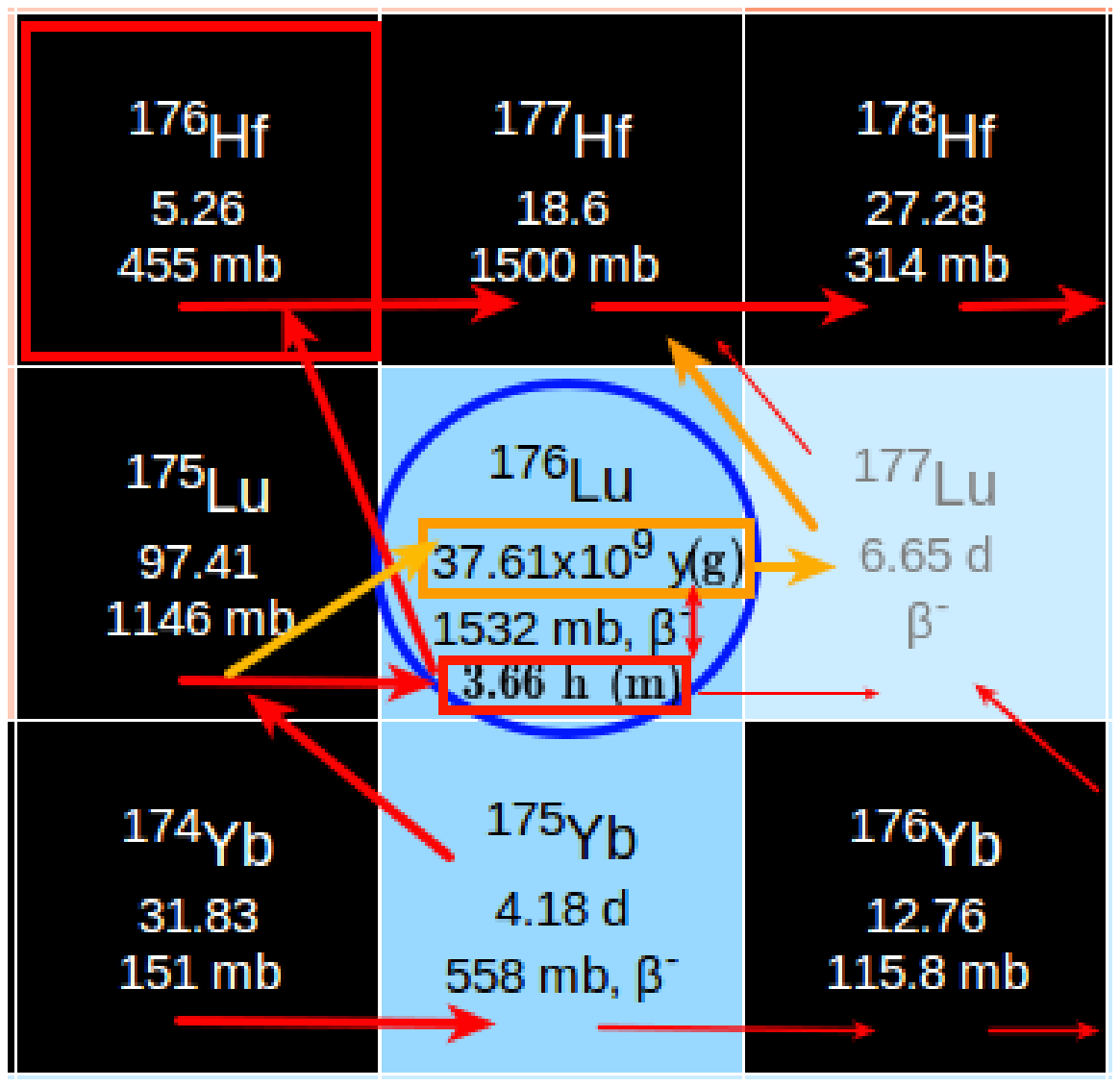}
\caption{Schematic representation of the chart of nuclei in the atomic mass region
close to the $s$-only isotope $^{176}$Lu (red rectangle). $^{176}$Lu has a short-lived isomeric 
state (m) with $t_{1/2}$ = 3.66 h and a long-lived ground state (g) with $t_{1/2}$ = 
38 Gyr. During the \Can neutron irradiation these states are 
treated independently; during \Nean neutron burst thermally induced transitions
between isomer and ground states are allowed, largely affecting the $^{176}$Lu/$^{176}$Hf
ratio (see text).}
\label{branchluhf}
\end{figure}

\begin{figure}
\vspace{5 mm}
\includegraphics[angle=-90,width=9cm]{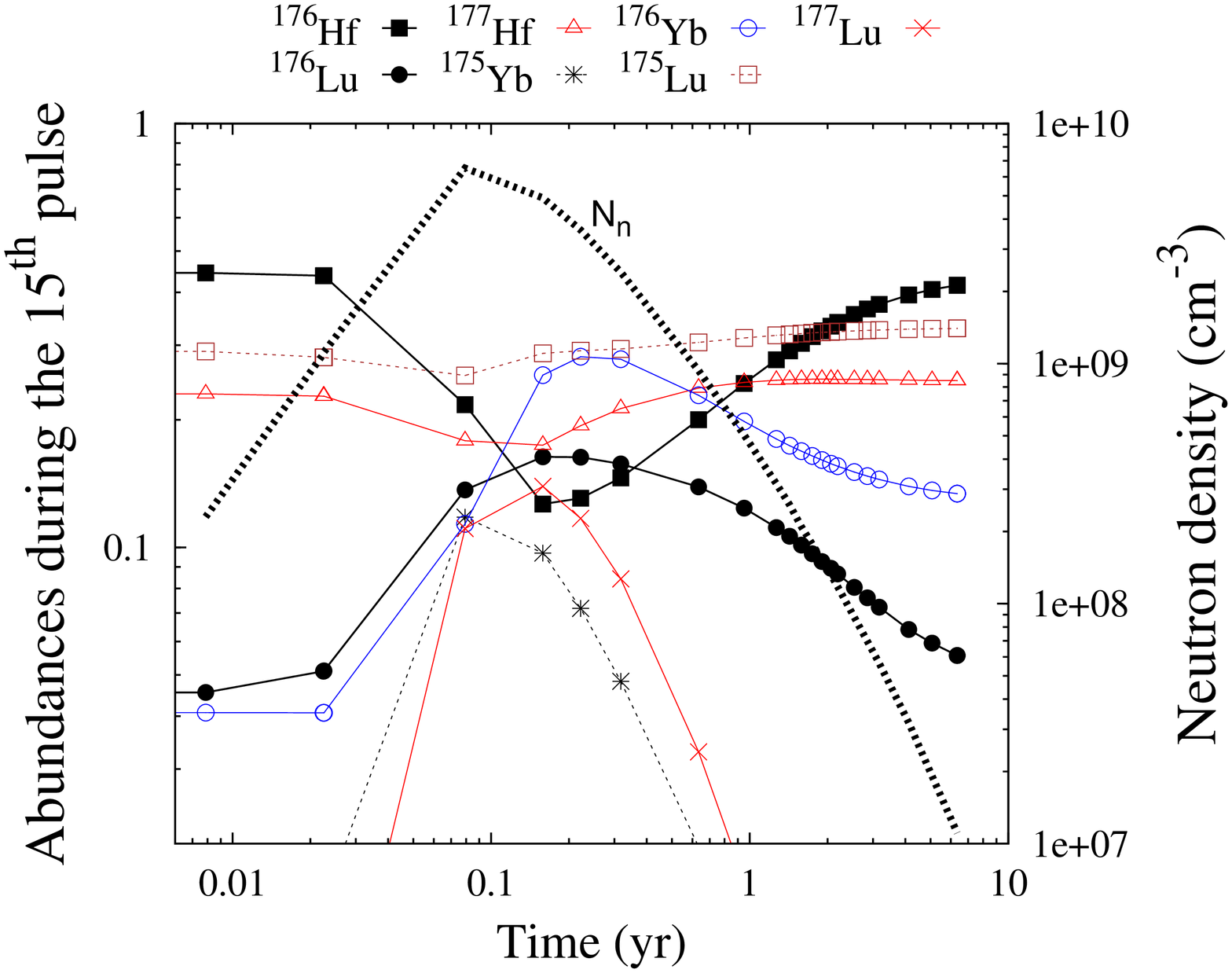}
\caption{The same as Fig.~\ref{stamperb85}, but for the isotopic abundances of 
$^{176,177}$Hf, $^{175,176,177}$Lu and $^{175,176}$Yb. }
\label{stampeertmyb}
\end{figure}

The $^{176}$Lu/$^{176}$Hf ratio is largely affected by the branch at $^{176}$Lu, which
is strongly sensitive to the temperature during TPs 
(Takahashi \& Yokoi 1987; Klay et al. 1991; Doll et al. 1999).
\\
$^{176}$Lu has a short-lived isomer ($t_{1/2}^m$ = 3.66 h) and an extremely long-lived 
ground state ($t_{1/2}^g$ = 38 Gyr; see Fig.~\ref{branchluhf}). In this case, internal
transitions are highly forbidden by nuclear selection rules and any coupling can only 
be provided by states with intermediate quantum numbers and higher excitation energies. 
Because the low temperatures during the $^{13}$C-pocket phase are not sufficient for that
coupling, the $s$-process flow via both states has been treated independently. Accordingly, 
the production of $^{176}$Lu$^g$ is determined by the partial cross section of $^{175}$Lu 
(1 $-$ IR = 1 $-$ 0.84; Wisshak et al. 2006a).
\\
At temperatures above $T_8$ = 1.5, thermally induced transitions by the hot stellar 
photon bath start to reach the first known mediating states at an excitation energy 
of 838.6 keV. As this mediating state can decay to the long-lived ground state as
well, this fraction contributes to the production of $^{176}$Lu$^g$ and to an
enhancement of the (n, \g) branch bypassing $^{176}$Hf. Above $T_8$ = 3, ground and 
isomeric states are in thermal equilibrium, and internal 
transitions are much faster than the timescales for $\beta^-$ decay and neutron capture.
\\
$^{176}$Lu$^g$ is actually produced via thermal transitions in the bottom layers 
of the advanced TPs, where temperatures are $T_8$ $\sim$ 3. 
Once produced, the long-lived $^{176}$Lu$^g$ survives in the cooler external layers 
of the convective flashes, outside of the burning zone. 
Accordingly, the detailed neutron density and temperature profiles in the TPs have 
to be considered for the $s$-process calculations. For an extensive discussion of the 
treatment of the branch at $^{176}$Lu in our AGB models see Heil et al. (2008c).
In the advanced TPs, the temperature reached at the bottom layers is high enough to provide 
additional feeding of the $^{176}$Lu long-lived ground state via thermal population of the 
mediating state.
Consequently, at the beginning of the TP $^{176}$Hf is destroyed, while 
$^{176}$Lu is produced (+23\%; see Fig.~\ref{stampeertmyb}). 
The abundance of $^{176}$Hf produced in the previous $^{13}$C pocket is restored to 93\% at the
end of the TP (Fig.~\ref{stampeertmyb}).

Note that the branch at $^{175}$Yb ($t_{1/2}$ = 4.18 d, reduced up to a factor of three at stellar
temperature) has marginal effects on the production of $^{176}$Hf,
because almost all the $s$ path coming from $^{174}$Yb $\beta^-$decays into $^{175}$Lu (99\%).

Both $^{176}$Lu and its daughter $^{176}$Hf are of pure $s$-process origin, because
they are shielded against the $r$ process by their stable isobar $^{176}$Yb.
About 10\% of the long-lived $^{176}$Lu ground state $\beta^-$ decays in the interstellar 
medium prior to the formation of the solar system, with a negligible increase of $^{176}$Hf, 
which is about ten times more abundant.
Therefore, $s$ contributions between 100\% and 110\% were found acceptable for $^{176}$Lu, 
and between 95\% and 105\% for $^{176}$Hf (Heil et al. 2008c).

With respect to the results of Heil et al. (2008c), 
the MACS of $^{176}$Hf has been reduced by 5\% due to a new measurement 
($\sigma$$_{\rm ^{176}Hf}$) = 594 $\pm$ 16 mbarn at 30 keV, Wisshak et al. 2006b) and 
to a revision of the SEF value (KADoNiS).
Also the adopted solar abundance ratio by Lodders et al. (2009) is $\sim$4\% higher than
previous assumed (Anders and Grevesse 1989).
 Moreover, the recommended \Nean rate (about a factor of two smaller than previously adopted)
increases the $s$ abundance of $^{176}$Lu from 101\% to 112\%, and reduces the $s$
prediction of $^{176}$Hf from 104\% to 101\%. 
The increase found for $^{176}$Lu is clearly exceeding the 5\% uncertainty of the 
solar abundance given by Lodders et al. (2009). 
Otherwise, $^{176}$Lu is particularly sensitive to the \Nean rate during TP:
an increase of the recommended \Nean rate by a factor of four is underproducing the solar 
$^{176}$Lu abundance by about 8\%, while the $s$ abundance of $^{176}$Hf remains almost 
unchanged. This highlights that large uncertainties affect the $s$ contribution to $^{176}$Lu, 
despite the MACS of the Lu and Hf isotopes are very well known, with uncertainties of 1\% to 
3\% (KADoNiS).

Note that Mohr et al. (2009) reported a significantly stronger coupling between 
isomer and ground state than adopted here. Therefore, the longer-lived ground state is largely
populated, thus strongly increasing the overproduction of $^{176}$Lu.
As  evidenced by Mohr et al. (2009), it is not possible to find a consistent solution with our
AGB models within the given experimental errors of the neutron capture cross sections of the Lu 
and Hf isotopes, and the uncertainty of the thermal coupling.
Mohr et al. (2009) suggested that a possible solution of the problem may originate from rather 
small modifications of the temperature profile or convective mixing during the helium shell flashes.
Cristallo et al. (2010) show that convective velocities have to be reduced by
a factor of 1000 with respect to the value estimated by means of the mixing length theory, to
lead to a lower Lu production and to a lower Hf destruction. This strong discrepancy seems incompatible
with 3D hydrodynamical calculations of shell He flashes, which justify uncertainties of a factor 2-3 
in the determination of convective velocities.
The solution may be found in the nuclear coupling scheme between the populated levels 
in $^{176}$Lu (Gintautas et al. 2009; Dracoulis et al. 2010; Gosselin et al. 2010).
At present, the discrepancy between experimental data and astrophysical determination 
remains to be solved.


\subsubsection{The $s$-only isotope $^{204}$Pb (the branch at $^{204}$Tl)}\label{pb204}

\begin{figure}
\includegraphics[angle=0,width=7cm]{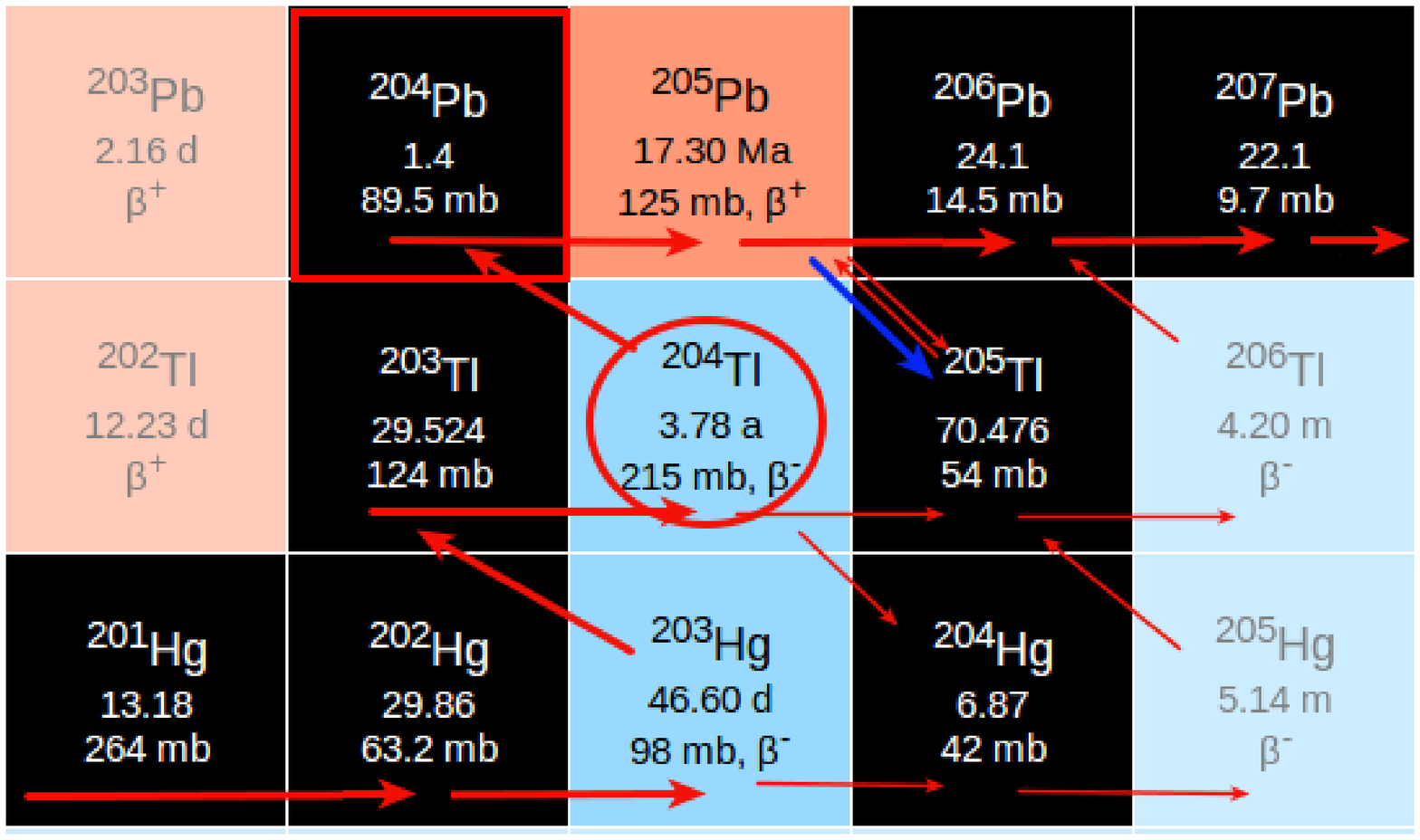}
\caption{Schematic representation of the chart of nuclei in the atomic mass region
close to the $s$-only isotope $^{204}$Pb
(red rectangle). 
The branch at $^{204}$Tl, mainly activated during TPs, is represented with a red circle.
The $^{205}$Tl abundance is influenced by the radiogenic $\beta^+$ decay of the long-lived
$^{205}$Pb ($t_{1/2}$ = 17.3 Myr; blue arrow). 
At $T_8$ = 3, the stable $^{205}$Tl becomes unstable, while the half-life of the long-lived $^{205}$Pb
is strongly reduced (see text). }
\label{branchtl204}
\end{figure}

\begin{figure}
\vspace{5 mm}
\includegraphics[angle=-90,width=9cm]{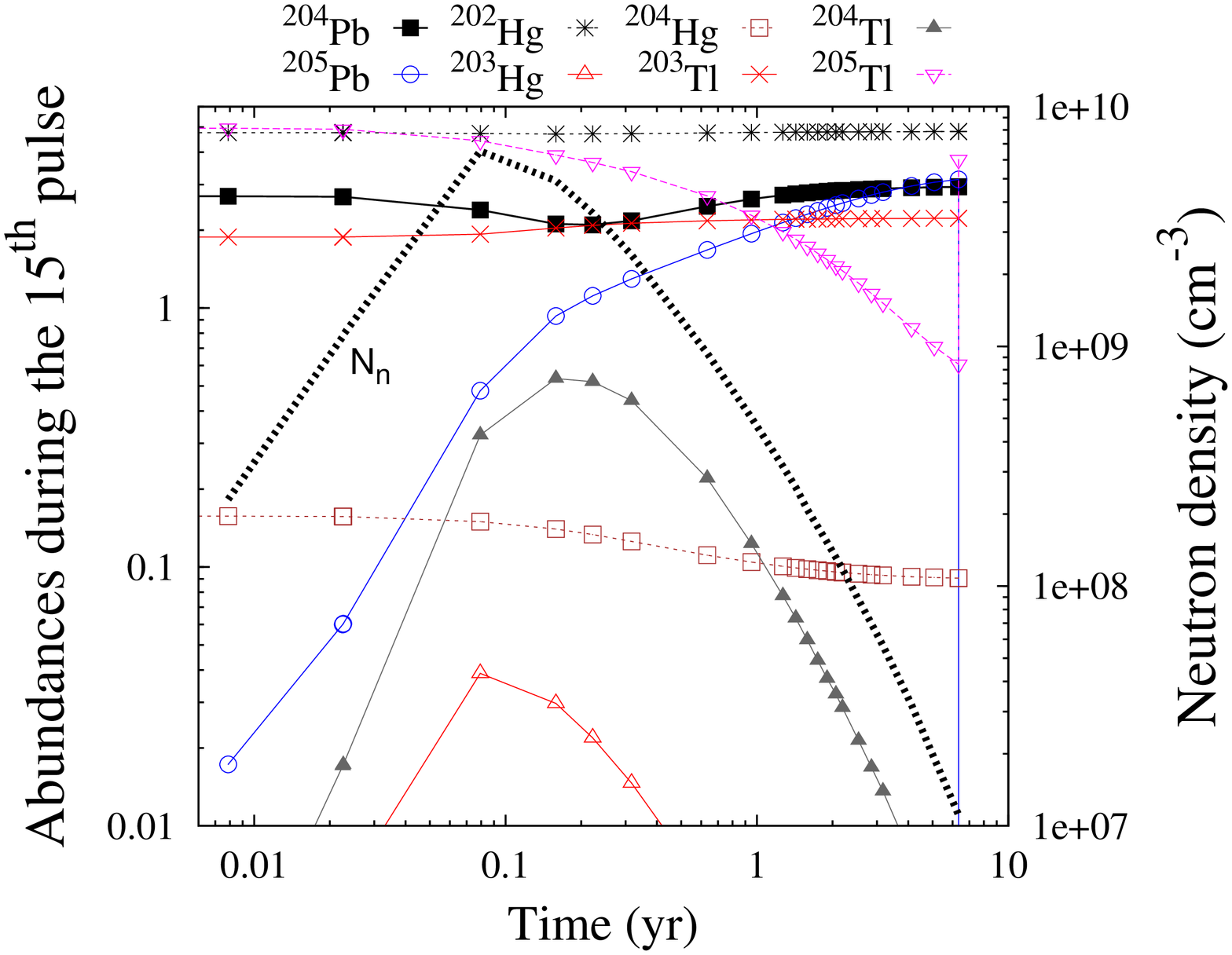}
\caption{The same as Fig.~\ref{stamperb85}, but for the  isotopic abundances of
$^{204,205}$Pb, $^{202,203,204}$Hg and $^{203,204,205}$Tl. }
\label{stampepb204}
\end{figure}

The most important branch point affecting the $s$-only nuclide $^{204}$Pb is the unstable isotope
$^{204}$Tl with a temperature-sensitive decay rate (Takahashi \& Yokoi 1987; Fig.~\ref{branchtl204}).
Although $^{204}$Pb is partly produced during the \Can irradiation, most of the AGB $s$ contribution 
derives from the \Nean neutron source (Ratzel et al. 2004; Domingo-Pardo et al. 2007). 
\\
During the $^{13}$C-pocket phase at $T_8$ $\sim$ 0.9, only about 25\% of $^{204}$Pb are produced 
because of the branching at $^{204}$Tl ($t_{1/2}$ = 3.78 yr; $N_n$ $\geq$
10$^7$ cm$^{-3}$, $f_n$ $\geq$ 0.1).
\\
With increasing stellar temperature, the $^{204}$Tl half-life is strongly reduced 
($t_{1/2}$($^{204}$Tl) = 7 d at $T_8$ = 3; Takahashi \& Yokoi 1987), and most of the 
$s$ path feeds $^{204}$Pb. At the peak neutron density $^{204}$Pb is slightly depleted 
by neutron capture ($N_n$ $\geq$ 2$\times$10$^9$ cm$^{-3}$, $f_n$ $\geq$ 0.2), but is 
recovered again as soon as the neutron density decreases (Fig.~\ref{stampepb204}). 
At the end of the TP, the abundance of $^{204}$Pb has increased by 9\%.

The branch at $^{203}$Hg has no significant impact on the $^{204}$Pb abundance, 
because the half-life is strongly reduced to less than one day during TPs (Takahashi 
\& Yokoi 1987).

 Previous AGB results (see $filled$ $circles$ in Fig. 5)
predict that about 87\% of solar $^{204}$Pb is produced by the main component.
This value increases to 91\% by including an improved treatment of the $\beta$-decay rates
of nearby unstable isotopes ($^{203}$Hg and $^{204}$Tl) over the convective TPs (see $filled$ 
$diamonds$ in Fig. 5). Accordingly, 
it is well compatible with the solar Pb abundance, which is given with a 10\%
uncertainty (Lodders et al. 2009).
\\
The MACS of $^{204}$Pb is well known (81.0 $\pm$ 2.3 mbarn with $\sim$3\% uncertainty at 30 keV; KADoNiS).
The major uncertainties affecting the $^{204}$Pb $s$-prediction derive from the branch at
$^{204}$Tl: the $^{204}$Tl MACS is evaluated theoretically with $\sim$18\% uncertainty (215 $\pm$ 
38 mbarn); moreover, Goriely (1999) estimated that the $^{204}$Tl $\beta^-$-decay rate may increase
up to a factor of two, increasing the $s$ contribution to $^{204}$Pb by a few percentages. 
\\
 Recently, Gonzalez (2014) suggests that the current solar log(Pb/Sm) ratio may be overestimated 
by +0.1 dex, owing to a systematic error (related to a trend with condensation temperature) that affects
the evaluation of the Pb/Sm solar abundance.
This would reconcile the present $s$ prediction of $^{204}$Pb.

In spite of its long terrestrial half-life of 17.3 Myr, $^{205}$Pb acts as an 
additional branching point because of the strong dependence on temperature and 
electron density ($t_{1/2}^+$($^{205}$Pb) = 5 yr at $T_8$ = 3; Takahashi \& Yokoi 
1987). Similarly, the stable daughter isotope $^{205}$Tl becomes unstable
during TPs ($t_{1/2}^-$($^{205}$Tl) = 0.55 yr at $T_8$ = 3; Takahashi \& Yokoi 1987) 
and its $\beta^-$ decay is competing with the $\beta^+$ decay of $^{205}$Pb. 
As the peak neutron density decreases, the $^{205}$Pb decay starts to prevail, 
however, and at the end of the TP-AGB phase the radiogenic decay of $^{205}$Pb is 
responsible for the production of $^{205}$Tl (Fig.~\ref{stampepb204}; see also
Mowlavi et al. 1998, Busso et al. 1999).


\subsection{Additional branches mainly sensitive to stellar
temperature (and/or electron density)}


\subsubsection{The $s$ contribution to $^{164}$Er 
(the branches at $^{163}$Dy and $^{163,164}$Ho)}\label{er164}

\begin{figure}
\includegraphics[angle=0,width=7cm]{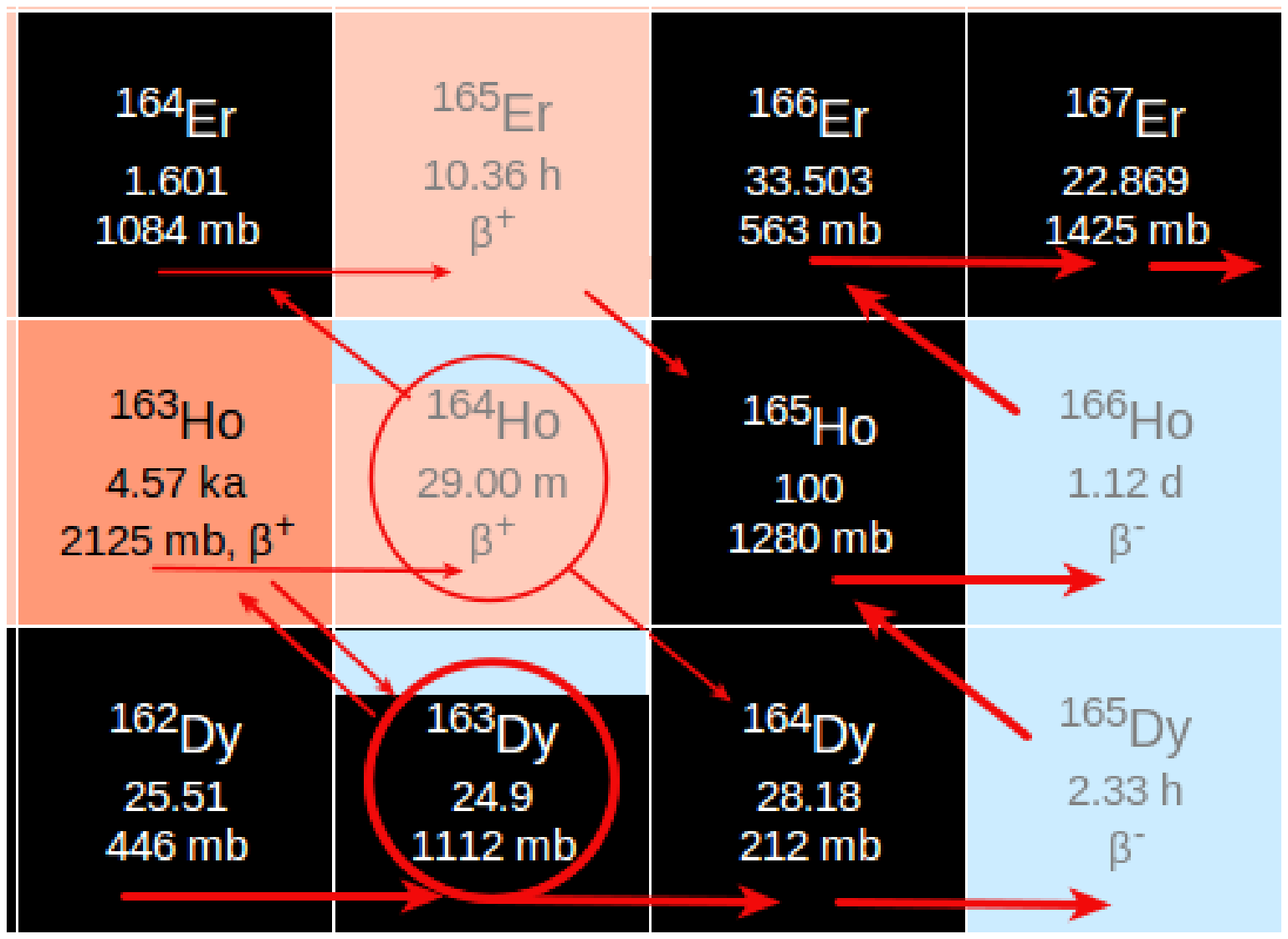}
\caption{Schematic representation of the chart of nuclei in the atomic mass region
close to $^{164}$Er. The stable isotope $^{163}$Dy becomes unstable during TPs and 
acts, therefore, as the branch point isotope responsible for the $s$ production of 
$^{164}$Er (see text).}
\label{branchdyho}
\end{figure}

\begin{figure}
\vspace{5 mm}
\includegraphics[angle=-90,width=9cm]{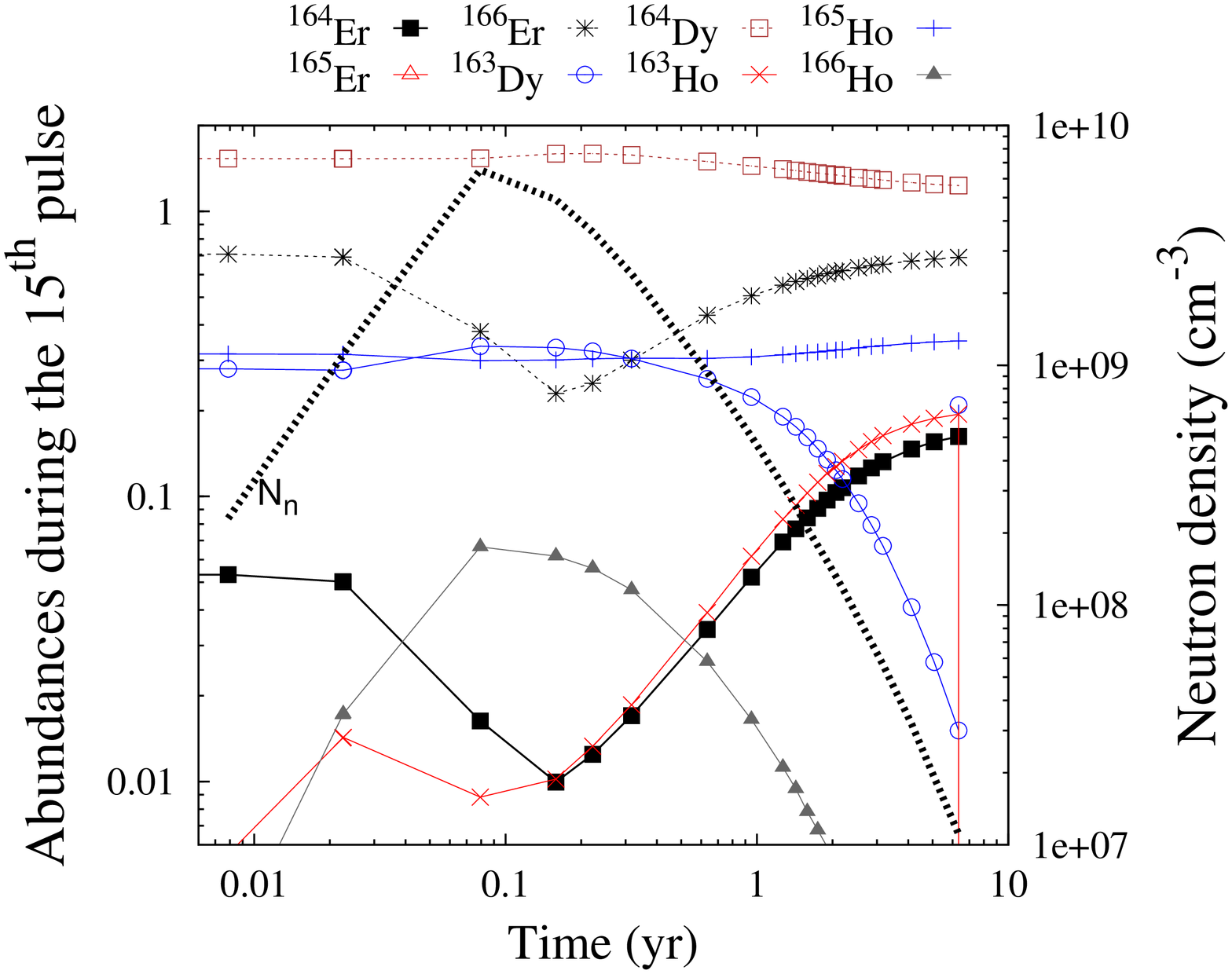}
\caption{The same as Fig.~\ref{stamperb85}, but for the isotopic abundances of 
$^{164,165,166}$Er, $^{163,165,166}$Ho and $^{163,164}$Dy.}
\label{stampeer164}
\end{figure}

The origin of $^{164}$Er was first understood by Takahashi and Yokoi (1987), 
who showed that the stable isotope $^{163}$Dy becomes unstable under stellar conditions.
\\
$^{164}$Er is completely bypassed during the $^{13}$C-pocket phase when the \Can operates
radiatively at $T_8$ $\sim$ 0.9, because $^{163}$Dy is still stable at that temperature.
During TPs, however, $^{163}$Dy becomes unstable and decays to $^{163}$Ho with a rapidly 
decreasing half-life, down to 40 -- 120 d (at $T_8$ = 3), partially feeding
the unstable $^{163}$Ho (Fig.~\ref{branchdyho}).  
Neutron captures on $^{163}$Ho are efficiently activated during TPs. Although the half-life 
of $^{163}$Ho decreases from 4.57 kyr to values of 7 to 20 yr at $T_8$ = 3 (Takahashi
and Yokoi 1987) it remains larger than the time scale for neutron captures during the \Nean 
irradiation so that the $s$ path runs almost completely from $^{163}$Ho to $^{164}$Ho, 
which decays via $\beta^-$ to $^{164}$Er and via $\beta^+$ to $^{164}$Dy (Fig.~\ref{stampeer164}). 
At the end of the neutron irradiation, the $^{163}$Ho abundance stored during a TP $\beta^+$ decays
into $^{163}$Dy during the following interpulse period, when the $s$ flow bypasses $^{164}$Er.

 The improved treatment of the half-life of $^{163}$Dy and $^{163,164}$Ho 
over the convective TP provides an increase of solar $^{164}$Er from 81\% to 87\% (see filled circle
and diamond in Fig.~5).
\\
This contribution is however uncertain.
Although the $^{164}$Er MACS is rather well determined, with 4.7\% uncertainty 
(1084 $\pm$ 51 mbarn at 30 keV; KADoNiS), it is affected by a non-negligible 
theoretical stellar enhancement factor, SEF = 1.08 at 30 keV (KADoNiS;  see also Rauscher 
et al. 2011). 
\\
Moreover, the competition between the $^{164}$Ho $\beta^-$ and $\beta^+$-decay rates is 
influenced by the large uncertainty of the $\beta^-$-decay rate under stellar conditions. 
Goriely (1999) estimated that the $^{164}$Ho $\beta^-$-decay rate may change
up to $\sim$30--40\% at $T_8$ = 3, producing {changes from 75\% up to 105\%} of the
$^{164}$Er $s$-prediction. 
The $\beta^+$ decay of $^{164}$Ho has only a marginal effect, because there are no
unknown transition probabilities in this case, and the $\beta^-$ decay of $^{163}$Dy 
is also of minor importance (a 30\% uncertainty of the decay rate produces 
up to 4\% variations of $^{164}$Er).

This holds for the \Nean rate as well. An increase of the recommended \Nean rate by a factor
of four produces up to 4\% variations in the $s$ abundance of $^{164}$Er.


\subsubsection{The $s$-only pair $^{128,130}$Xe (the branch at $^{128}$I)}\label{128130Xe}

\begin{figure}
\includegraphics[angle=0,width=8cm]{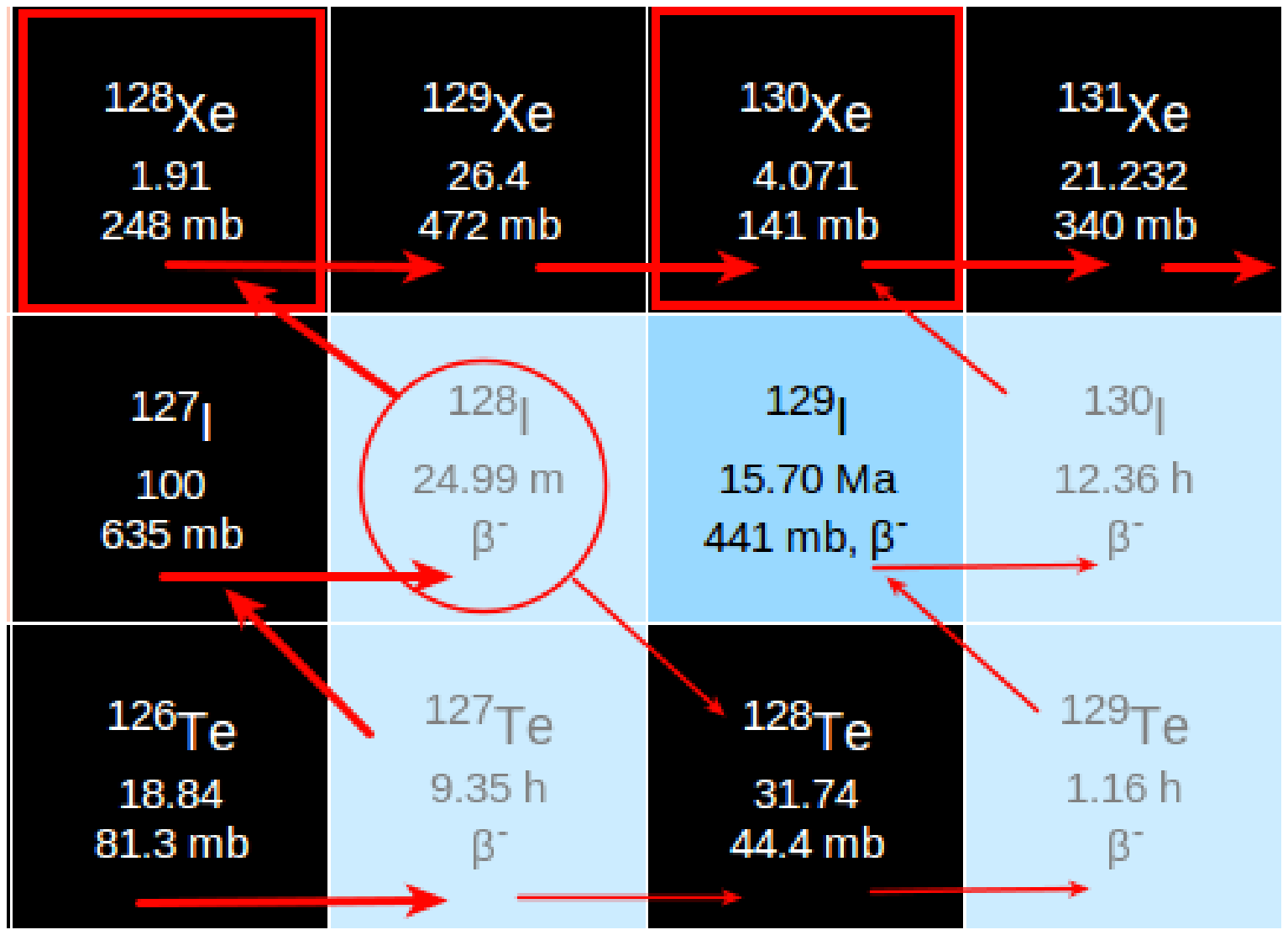}
\caption{Schematic representation of the chart of nuclei in the atomic mass region
close to the $s$-only isotopes $^{128,130}$Xe. The $s$ abundance of $^{128}$Xe
is determined only by the competition between $\beta^-$ decays and electron captures 
of $^{128}$I (red circle), completely independent of the $s$-process neutron density.}
\label{branchi128}
\end{figure}

The $s$ predictions of $^{128,130}$Xe depend mainly on the branch at $^{128}$I,
whereas the short $^{127}$Te half-life ($t_{1/2}$ = 9.34 h) prevents neutron captures
towards $^{128}$Te (Fig.~\ref{branchi128}).
\\
$^{128}$I partly $\beta^-$ decays into $^{128}$Xe and partly feeds $^{128}$Te
via electron captures ($t_{1/2}^{\beta^-}$ = 26.55 m; $t_{1/2}^{EC}$ $\sim$ 7 h).
While the $\beta^-$ half-life shows negligible variations under stellar conditions 
($t_{1/2}^{\beta^-}$ = 42 m at $T_8$ = 3), electron captures on $^{128}$Xe are strongly 
temperature and electron density dependent (e.g., at $T_8$ = 3, $t_{1/2}^{EC}$ increases to 10 d
for density of $\rho$ = 1000 g/cm$^3$; Takahashi \& Yokoi 1987). 
Note that the decay of $^{128}$I is dominated by the $\beta^-$ channel, both during the 
$^{13}$C pocket at $T_8$ $\sim$ 0.9 and during the convective TP at $T_8$ $\sim$ 3, so that
only 6\% of the $s$ path bypasses $^{128}$Xe.
\\
The branching at $^{128}$I is unique for the fact that it is completely independent of the 
neutron density and that the $s$ abundance of $^{128}$Xe is only determined by the competition
between $\beta^-$ decay and electron captures.

As outlined by Reifarth et al. (2004), the branch at $^{128}$I is a good indicator of
the convective mixing timescale during He shell flashes.
Although $\beta^-$ decays dominate over electron captures, the EC half-life of $^{128}$I
decreases to $\sim$6 h in the cooler layers of the TP (at $T_8$ = 0.5), directing about 
5 to 6\% of the $s$ path to $^{128}$Te, thus bypassing $^{128}$Xe. By assuming a 
sufficiently fast time scale for turnover mixing in the TP, freshly produced $^{128}$I is
rapidly transported from the hot bottom layers to the cooler external layers of the 
convective zone, allowing a partial activation of the EC channel. 
The resulting ratio of the $s$ abundances of $^{128}$Xe and $^{130}$Xe
is in agreement with the $^{128}$Xe/$^{130}$Xe ratio observed in SiC grains.

According to the present $s$ predictions 89\% of solar $^{128}$Xe and 98\% of solar $^{130}$Xe
are produced by the main component. 
\\
The MACS of $^{128,130}$Xe are well determined, with less than 2\% uncertainty ($\sigma$($^{128}$Xe) =
262.5 $\pm$ 3.7 mbarn and $\sigma$($^{130}$Xe) = 132.0 $\pm$ 2.1 mbarn; KADoNiS).
Both values are consistent with the uncertainty of the solar Xe abundance, which was estimated
theoretically by interpolation of measured cross-sections 
and abundances of neighbouring elements, through the relation $\sigma$N = 
constant (see Reifarth et al. 2002; Lodders et al. 2009).
The above $s$ predictions are practically independent of the \Nean rate because of 
the small influence of the stellar neutron flux: by increasing the recommended \Nean reaction
by a factor of four, $^{128,130}$Xe $s$-predictions show marginal variations ($\la$6\%).
Accordingly, the uncertainties of the theoretical MACS values of $^{127}$Te and $^{128}$I
of up to a factor of two are affecting the $^{128}$Xe/$^{130}$Xe ratio by less than 2\%.

\clearpage
\newpage

\section{}

The aim of this paper is to provide an overview of the major uncertainties that affect 
the main component:
we have focused the analysis on two AGB models ($M$ = 1.5 and 3 $M_\odot$ at [Fe/H] = $-$0.3),
selected in the mass and metallicity ranges useful to reproduce 
the solar $s$ distribution of isotopes between 90 $\leq$ $A$ $\leq$ 204. 
However, the study of intermediate-mass or low-metallicity AGB models is essential to 
understand the $s$-process nucleosynthesis in different 
environments (e.g., globular clusters, dwarf spheroidal galaxies, intrinsic or extrinsic 
peculiar stars showing $s$ enhancement; see Section~1). 
\\
To this purpose, we analyse in this Section two AGB models chosen as illustrative of extended mass and metallicity
ranges: a half-solar metallicity 5 $M_\odot$ model and a 3 $M_\odot$ model at [Fe/H] = $-$1.

The post-process 3 and 5 $M_\odot$ models are based on stellar input data (e.g., temperature during 
TPs, TDU, and He-intershell masses) of full evolutionary FRANEC models (Straniero et al. 2003, 2000).
The maximum temperature at the bottom of the intershell grows slightly as the mass of the
H-exhausted core increases (Iben \& Renzini 1983): larger core masses are attained by more
massive stars and, for a given mass, by more metal-poor stars (see, e.g., Straniero et al. 
2006). This allows an efficient activation of the \Nean neutron source in these stars.

In IMS stars, the mass of the He intershell is about a factor of ten lower than low-mass models,
the efficiency of the TDU decreases by about one order of magnitude or more, and the interpulse phase
is much shorter ($\sim$6.5$\times$10$^3$ yr). 
The \Can neutron source is expected to have a small or even negligible effect in these stars. 
Thus, IMS AGB stars play
a minor role in the Galactic enrichment of $s$ isotopes, but they are crucial for globular clusters
(e.g., D'Orazi et al. 2013).
The major uncertainty of IMS stars is the treatment of mass loss. In our IMS models we have adopted
an efficient mass loss, but we can not exclude different prescriptions (see Ventura \& Marigo 2010,
Karakas et al. 2012, Straniero et al. 2014).
\\
Our 5 $M_\odot$ model at [Fe/H] = 0.3 is characterised by an He-intershell mass of $\sim$0.003 
$M_\odot$, a TDU mass of $\sim$0.001 $M_\odot$ and experiences 24 TDU episodes.
The maximum temperature at the bottom of the TPs reaches $T_8$ $\sim$ 3.6. 
The dominant neutron source is the \Nean reaction, which is efficiently activated and 
produces high peak neutron density ($N_n$ $\sim$ 10$^{11}$ cm$^{-3}$).

In low-mass models of close-to-solar metallicity, the maximum temperature at the bottom
of TPs, although gradually increasing with the pulse number, barely reaches $T_8$ = 3. 
The 3 $M_\odot$ models at [Fe/H] = $-$1 is instead characterised by a rapidly increasing
temperature at the bottom of the convective zone, whose maximum reaches $T_8$ = 3.5 in the 
advanced TPs inducing a considerable production of the nuclei up to zirconium (and of
the few neutron-rich isotopes involved in those $s$-process branchings) via the \Nean neutron 
burst. As for the half-solar metallicity models, the mass loss is considered through the Reimers
formula ($\eta$ = 1), which allows the model to experience 35 TDU episodes.   
\\
We refer to Bisterzo et al. (2010, 2014) for more details on these two AGB models.

\begin{figure}
\includegraphics[angle=-90,width=8cm]{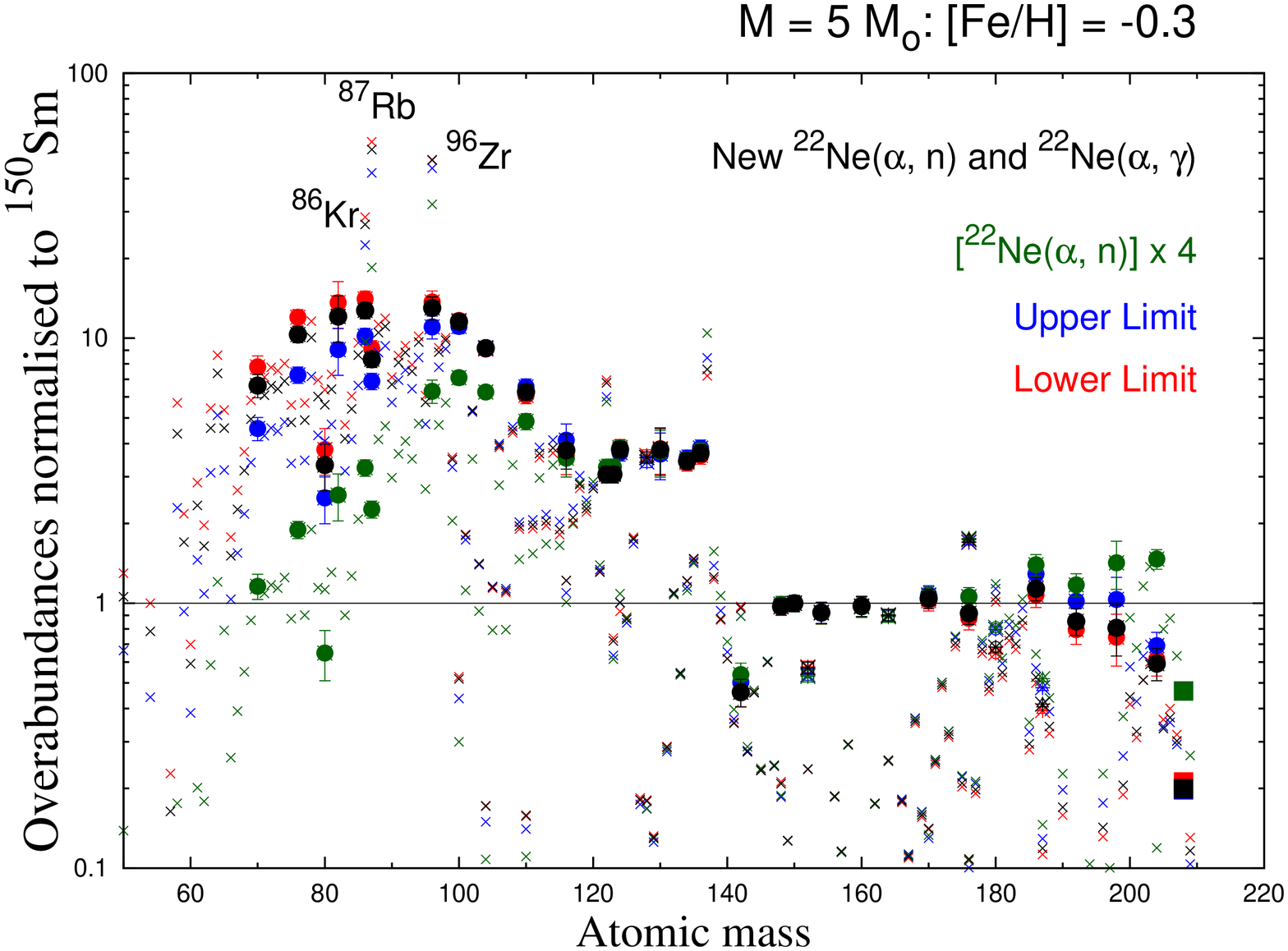}
\caption{The $s$ distribution of a $M$ = 5 $M_\odot$ model at half-solar metallicity 
($top$ $panel$) obtained with the recommended 
\Nean and \Neag rates (black symbols), compared to the results computed with 
the lower and upper limits of the \Nean rate (red and blue symbols) and with the 
recommended \Nean$\times$4 (green symbols), while the \Neag reaction is unchanged. }
\label{FigM5}
\end{figure}

\begin{figure}
\includegraphics[angle=0,width=8cm]{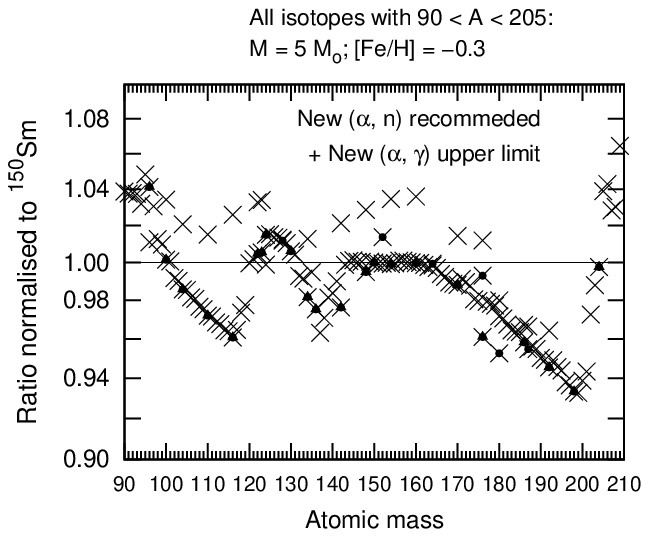}
\vspace{-5cm}
\caption{Ratios between the $s$ distribution of a $M$ = 5 $M_\odot$ model obtained with the recommended 
\Nean and \Neag rates (black symbols) compared to the results computed with the \Neag rate
upper limit (blue symbols).}
\label{FigM5ratio}
\end{figure}

\vspace{2cm}

In Fig.~\ref{FigM5}, the $s$ distribution of a 5 $M_\odot$ model at [Fe/H] = $-$0.3 is shown.
The results obtained with the recommended \Nean and \Neag rates (black symbols) are compared with those 
computed with the lower and upper limits of the \Nean rate (red and blue symbols) and with the 
test \Nean$\times$4 (which corresponds to Test A in Section~4.1; green symbols). 
No changes are applied to the \Neag reaction.
All values are normalised to $^{150}$Sm to overcome the major uncertainties of AGB models. 
\\
Larger variations are seen for isotopes with $A$ $<$ 100. However, the whole $s$ distribution 
is widely affected: the $s$-production factors increase by up to a factor of 1.7 with the
upper limit of the recommended \Nean rate, and by one order of magnitude with Test A.
The \Nean neutron burst mostly feeds isotopes between 80 $\la$ $A$ $\la$ 100 (with
abundances about a factor of ten higher than that of $^{150}$Sm), and with a minor extent up to 
the neutron magic isotopes with $N$ = 82. 
Note that the $^{150}$Sm abundance (in mass fraction) produced by the 5 $M_\odot$ model
(with the recommended \Nean rate) is about 20 times higher than its initial (solar-scaled) value. 
In general, the $s$-predictions obtained by IMS stars are almost negligible with respect to
the $s$ contribution predicted by low-mass AGB models, which produce values of one thousand or more. 
Exceptions are a few neutron-rich isotopes, $^{86}$Kr, $^{87}$Rb, $^{96}$Zr: they are
efficiently produced by the branches at $^{85}$Kr and $^{95}$Zr, which are essentially 
open during TPs.

The competition between the \Nean and \Neag rates affects mostly the Mg isotopes, while the $s$ isotopes
heavier than $A$ = 90 show variations smaller than $\sim$6\% (Fig.~\ref{FigM5ratio}).

\vspace{2cm}

In Fig.~\ref{FigM3}, the $s$ predictions (normalised to $^{150}$Sm) of a 3 $M_\odot$ model are shown. 
We compare the half-solar metallicity model adopted for the main component with a 1/10 solar metallicity
model ($top$ and $bottom$ $panels$, respectively). The results obtained by adopting the recommended \Nean and 
\Neag rates are represented by black symbols. 
The $s$ predictions computed with the lower and upper limits of the \Nean rate 
(red and blue symbols) and with the recommended \Nean$\times$4 (Test A in Section~4.1; green symbols)
are also displayed for comparison. 
\\
Focusing on black symbols, the large dependence on metallicity of the $s$ distribution is evident:
because the neutron exposure 
increases with decreasing the iron seeds, the abundances of isotopes with neutron magic numbers $N$ = 50 
and 82 are overcome (thus reducing the whole distribution between 90 $\la$ $A$ $\la$ 130 and 140 $\la$ 
$A$ $\la$ 204, respectively) and $^{208}$Pb is progressively produced.
\\
As discussed for the main component, the \Nean neutron source does not substantially 
modify the $s$ abundances in the half-solar metallicity 3 $M_\odot$ model, except in a few branchings.
In the low-metallicity model, not only branches are affected, but the whole
$s$ distribution is widely modified (see green symbols).

The competition between the \Nean and \Neag rates affects the Mg isotopes. Marginal variations
are found for $s$ isotopes heavier than $A$ = 90 ($<$2\%) by adopting the upper limit of the \Neag rate
(Fig.~\ref{FigM3ratio}).

The \Can rate operates efficiently in the $M$ = 3 $M_\odot$ model at [Fe/H] = $-$1, providing a substantial
contribution to $^{208}$Pb.  The present uncertainty associated to the \Can reaction has
marginal effects on $s$-only nuclei (see Fig.~\ref{FigM3reazc13an}).

 \begin{figure}
\includegraphics[angle=-90,width=8cm]{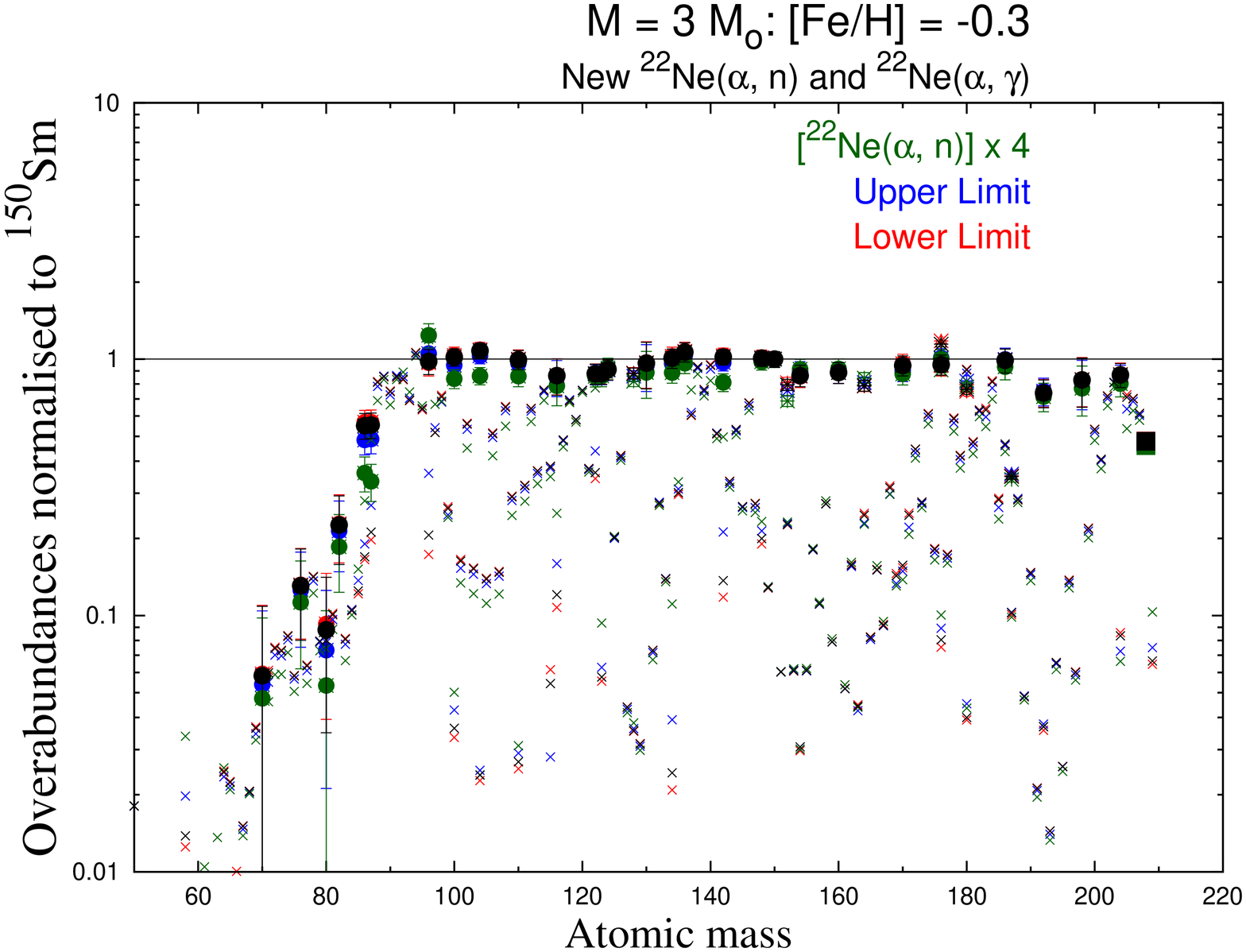}
\includegraphics[angle=-90,width=8cm]{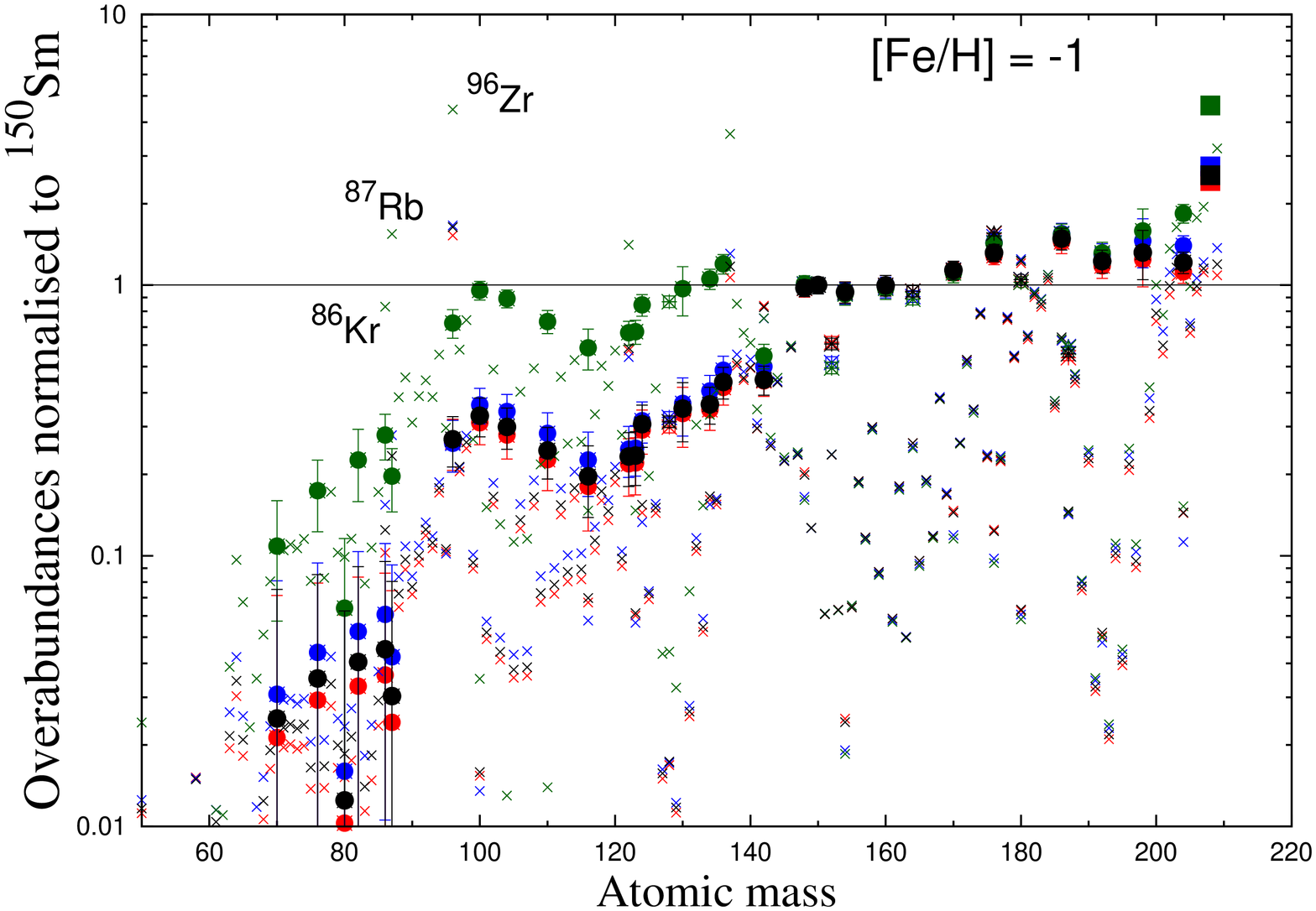}
\caption{The $s$ distribution of a $M$ = 3 $M_\odot$ model at [Fe/H] = $-$0.3 
and [Fe/H] = $-$1 ($top$ and $bottom$ $panels$ respectively) obtained with the recommended 
\Nean and \Neag rates (black symbols), compared to the results computed with 
the lower and upper limits of the \Nean rate (red and blue symbols) and with the 
recommended \Nean$\times$4 (green symbols), while the \Neag reaction is unchanged.}
\label{FigM3}
\end{figure}

 \begin{figure}
\includegraphics[angle=0,width=8cm]{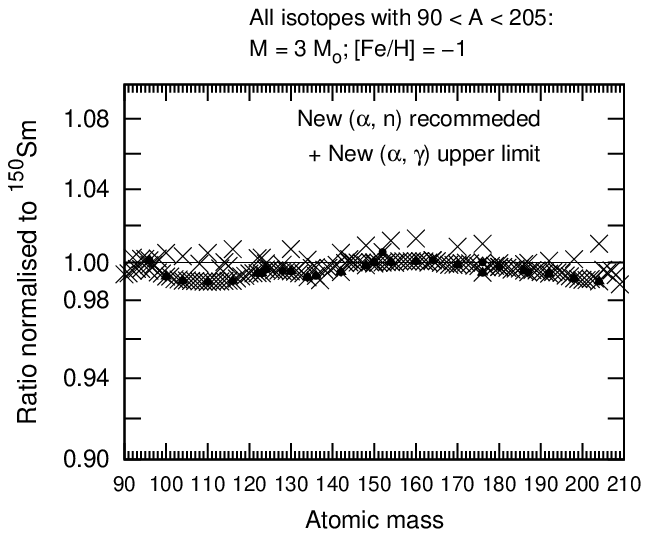}
\vspace{-5cm}
\caption{Ratios between the $s$ distribution of a $M$ = 3 $M_\odot$ model 
obtained with the recommended \Nean and \Neag rates (black symbols) compared to the results
computed with the \Neag rate upper limit (blue symbols).}
\label{FigM3ratio}
\end{figure}

 \begin{figure}
 \vspace{1cm}
\includegraphics[angle=-90,width=8cm]{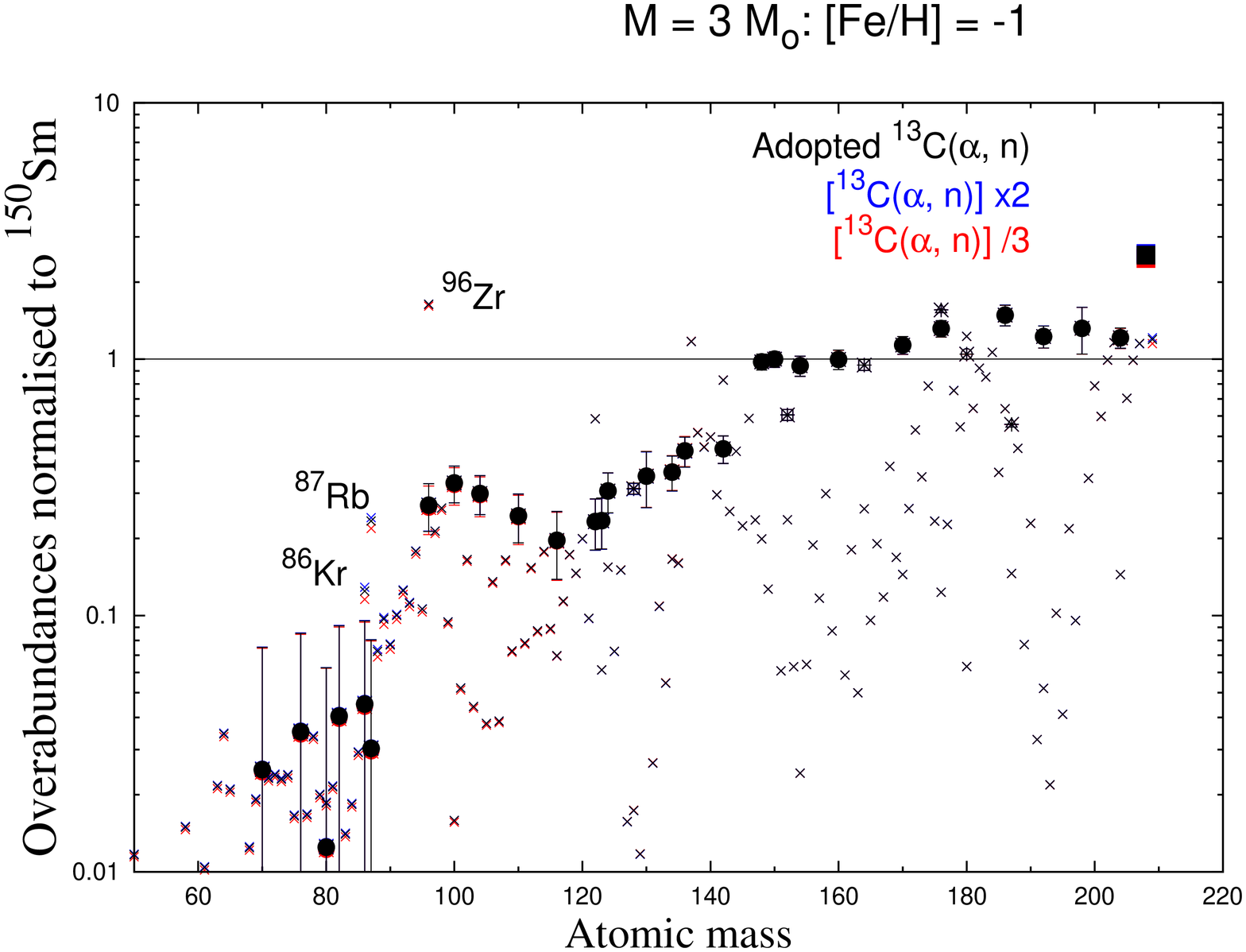}
\caption{The $s$ distribution of a $M$ = 3 $M_\odot$ model at [Fe/H] = $-$1
computed with our adopted \Can rate by Denker et al. (1995; black symbols),
with the test [\Can]$\times$2 (blue symbols) and [\Can]/3 (red symbols).}
\label{FigM3reazc13an}
\end{figure}


\clearpage
\newpage

\twocolumn

\bsp

\label{lastpage}

\end{document}